\documentclass[sigconf]{acmart}

\usepackage{url}
\usepackage{graphicx}
\usepackage{hyperref}
\usepackage{booktabs}
\usepackage{makecell}
\usepackage{longtable}
\usepackage{array}
\usepackage{xcolor}
\usepackage[normalem]{ulem}

\newif\ifdraft

\newcommand{\newtext}[1]{\ifdraft\textcolor{blue}{#1}\else\textcolor{black}{#1}\fi} 
\newcommand{\strike}[1]{\ifdraft\textcolor{red}{\sout{#1}}\fi}

\AtBeginDocument{%
  }

\begin{document}

\renewcommand{\arraystretch}{1.5}
\title[OSINT Clinic]{OSINT Clinic: Co-designing AI-Augmented Collaborative OSINT Investigations for Vulnerability Assessment}


\author{Anirban Mukhopadhyay}
\orcid{0009-0003-0925-1084}
\affiliation{%
  \institution{Department of Computer Science, Virginia Tech}
  \city{Blacksburg}
  \state{VA}
  \country{USA}
}
\email{anirban@vt.edu}

\author{Kurt Luther}
\orcid{0000-0003-1809-6269}
\affiliation{%
  \institution{Department of Computer Science, Virginia Tech}
  \city{Arlington}
  \state{VA}
  \country{USA}
}
\email{kluther@vt.edu}

\renewcommand{\shortauthors}{Mukhopadhyay et al.}

\begin{abstract}
Small businesses need vulnerability assessments to identify and mitigate cyber risks. Cybersecurity clinics provide a solution by offering students hands-on experience while delivering free vulnerability assessments to local organizations. To scale this model, we propose an Open Source Intelligence (OSINT) clinic where students conduct assessments using only publicly available data. We enhance the quality of investigations in the OSINT clinic by addressing the technical and collaborative challenges. Over the duration of the 2023-24 academic year, we conducted a three-phase co-design study with six students. Our study identified key challenges in the OSINT investigations and explored how generative AI could address these performance gaps. We developed design ideas for effective AI integration based on the use of AI probes and collaboration platform features. A pilot with three small businesses highlighted both the practical benefits of AI in streamlining investigations, and limitations, including privacy concerns and difficulty in monitoring progress.
\end{abstract}




\maketitle

\section{Introduction}

The 2019 Verizon Data Breach Investigations Report highlights that small businesses are prime targets for cybercriminals, accounting for 43\% of all breaches \cite{verizon_2019_nodate}. 
However, many small businesses are unaware of the cyber risks they face or how to mitigate them. Cyber vulnerability assessments can provide a view of their attack surface to address weaknesses before malicious actors can exploit them. Comprehensive vulnerability assessments include tasks like controls interviews, physical facility tours, and internal network scanning \cite{noauthor_nist_2020}. However, these require technical expertise and time to execute. Even if the small business has an IT staff member, they are often overwhelmed with tasks and lack specific training in cybersecurity.

\strike{Cybersecurity clinics, in which students gain valuable hands-on experience while providing free vulnerability assessments to local organizations, offer a potential solution \cite{consortium_nodate}. 
The cybersecurity clinic model has recently seen exponential growth thanks to support from private charities, government, and industry \cite{lederman_latest_nodate}. One consortium of cyber clinics reports 880 students participating in cybersecurity training and assisting at least 83 public interest organizations across U.S. \cite{srispens_university_2024}.}

\newtext{Cybersecurity clinics offer a potential solution \cite{consortium_nodate}. They provide essential cybersecurity services to community organizations — including small businesses, nonprofits, and local governments — while giving students real-world experience. Modeled after legal and medical school clinics, these programs are typically housed at colleges and universities under the direction of clinical professors \cite{clinic_uc_nodate}. Students from diverse backgrounds and disciplines receive training to offer free vulnerability assessments and cybersecurity assistance to clients who may lack the resources to secure these services otherwise \cite{consortium_nodate}. This model not only offers a valuable skills-based learning environment for students but also strengthens the cybersecurity resilience of communities. The cybersecurity clinic model has recently seen exponential growth thanks to support from private charities, government, and industry \cite{lederman_latest_nodate}.}

However, these clinics also face scalability challenges. 
Training a diverse group of students to conduct high-quality vulnerability assessments is time-consuming and resource-intensive. Even for trained students, a single investigation can take months. Further, some components of vulnerability assessments, such as controls interviews and facility tours, require significant time investments from clients.  

In this paper, we propose the concept of an \textit{OSINT Clinic}, \newtext{where students can learn and apply OSINT skills in a real-world context} to address the scalability challenges of cybersecurity clinics. Open Source Intelligence, or OSINT, refers to investigations that rely entirely on publicly available data. In a cybersecurity context, OSINT tasks include attack surface mapping for identifying all publicly accessible digital assets such as websites, servers, webcams, and social media accounts, as well as their vulnerabilities; brand monitoring for examining review sites and social media for mentions of the businesses, including negative rumors and disinformation campaigns; and breach data discovery for finding leaked credentials on the dark web or other sources \cite{hayes_open-source_2018, tabatabaei2017osint, browne_systematic_2024}. While comprehensive vulnerability assessments sometimes include an OSINT component, cyber clinics rarely, if ever, focus on OSINT. 


OSINT provides unique advantages for a student team doing vulnerability assessments. Dealing with publicly available data is a lower risk for both the students and the target businesses; this passive, “zero-touch” approach minimizes accidental damage and saves time for business owners. Second, the ethos of the field dictates that the process for finding, verifying, and documenting each piece of evidence must be transparent and repeatable, providing a rich learning environment for students. Despite the growing interest and promise of OSINT investigations \cite{hayes_open-source_2018, edwards_panning_2017, martins_generating_2022, urban_plenty_2020, riebe_values_2023}, only a few studies have taken a human-centered approach to investigate specific challenges involved in applying these techniques \cite{riebe_values_2023}. We address this gap by focusing on identifying and addressing the technical and collaborative challenges with OSINT tasks for vulnerability assessments in the clinic.   


OSINT investigations for cybersecurity are complex, requiring sifting through vast amounts of unstructured public data, which can be noisy, overwhelming, and time-consuming \cite{glassman2012intelligence, belghith_compete_2022, szymoniak_open_2024, tabatabaei2017osint}. Analysts often face difficulties in data verification due to the unreliable nature of sources and misinformation \cite{hassan2018evolution, belghith_compete_2022}. 
We argue that generative AI can be effective in addressing these challenges based on previous research on its role in narrowing the skill gap in creative and knowledge work \cite{doshi_generative_2023, suri_use_2024, wan_it_2024}. Examples from OSINT practitioners and trainers like Matt Edmondson and Chris Poulter \cite{flashpoint_generative_2024, poulter_osint_2023} also back this up. \newtext{However, reflecting Ackerman’s ``socio-technical gap''~\cite{ackerman2000intellectual}, the technical capabilities of AI currently lag behind the nuanced social requirements of trust and confidentiality for sensitive tasks like OSINT. Following a growing thread of HCI and CSCW research, our study aims to effectively integrate generative AI to augment existing workflows --- in this case, those of an OSINT clinic --- and help them scale up.}

This research is driven by the following guiding questions:
\begin{enumerate}
    \item \newtext{How can we scale up cybersecurity vulnerability assessments using OSINT?}
    \item What challenges arise when performing vulnerability assessments in the OSINT clinic?
    \item How can generative AI, particularly large language models like ChatGPT, address the technical and collaboration challenges of OSINT investigations \newtext{among non-experts}?
    \item \newtext{How can generative AI support real-world OSINT investigations for vulnerability assessments?}
\end{enumerate}

To explore these questions, we designed and launched an OSINT clinic with six undergraduate students. We trained the students on OSINT skills with regular practice sessions and workshops to prepare them for real-world investigations. Over the 2023--24 academic year, we conducted a longitudinal co-design study by adapting the Matchmaking for AI method \cite{liu2024humancenterednlpfactcheckingcodesigning} with three phases (Studies 1--3) to support the work of the students in the clinic. Study 1 identified key challenges across the Intelligence Cycle phases through workshops, revealing performance gaps in student investigations. In Study 2, we utilized generative AI and collaboration platform probes to explore how AI could help novices address the challenges and develop expertise across the stages of planning, data collection, processing, analysis, and dissemination. We analyzed the user experience and workshopped design ideas to \strike{guide the integration of generative AI into these processes} \newtext{show how generative AI can support training and be integrated into workflows}. Finally, Study 3 piloted OSINT-based vulnerability assessments with three small businesses, integrating curated AI prompts into real-world investigations. This study demonstrated AI's practical benefits and also revealed limitations, such as privacy concerns and difficulties in monitoring progress. 

Together, these studies make the following contributions:
\begin{enumerate}
    \item \newtext{We make a conceptual contribution by developing the OSINT clinic program to scale up cybersecurity vulnerability assessments based on publicly available information.}
    \item We make a methodological contribution by extending the Matchmaking for AI co-design process to include learning goals, collaboration challenges, \newtext{and real-world deployments}. 
    \item \newtext{We make an empirical contribution by identifying new challenges and learning goals in the OSINT clinic and addressing them through the use of generative AI.} 
    \item \strike{We develop design requirements based on outstanding challenges to support future research in streamlining OSINT investigations.}\newtext{We develop design considerations for practical, AI-augmented, and collaborative investigations based on the use of generative AI.}
\end{enumerate}

\section{Related Work}

\newtext{Our paper bridges two key areas of recent HCI research: Clinical Computer Security, which defines protocols for addressing targeted digital security threats (Section~\ref{sec:clinical-security}), and CSCW research focused on exploring generative AI tools to facilitate sensemaking and collaboration (Section~\ref{sec:rw-sensemaking}). By situating our work at this intersection, we highlight how AI-augmented systems can empower novice investigators in navigating complex OSINT tasks, supporting both their learning goals and the effectiveness of real-world investigations. In terms of method, we extend the Matchmaking for AI co-design framework with a longitudinal study design (Section~\ref{rw:co-deisgn}), focusing on collaboration and real-world validation. We situate the work within the context of cybersecurity vulnerability assessments using OSINT (Section~\ref{sec:vuln-assess}).}

\subsection{\newtext{Clinical Computer Security in HCI}}
\label{sec:clinical-security}

\newtext{Clinical Computer Security (CCS)~\cite{freed2019my} is an emerging approach in HCI that provides personalized, hands-on support to individuals facing technology-based security threats. CCS methods involve pairing clients with trained technologists who collaboratively investigate and address digital security vulnerabilities. This process is guided by the Understand-Investigate-Advise (UIA) framework, in which consultants conduct interviews to map a client's digital footprint, investigate potential security threats (e.g., spyware or compromised accounts), and offer tailored mitigation strategies \cite{havron_clinical_2019}. CCS draws inspiration from existing support paradigms such as the Citizen Lab \cite{citizen_lab_about_nodate} and Cybersecurity Clinics \cite{clinic_uc_nodate}, which focus on assisting victims of nation-state digital attacks. These clinics not only provide critical services but also serve as research sites for developing new security practices. For example, Tseng et al. co-designed data stewardship approaches by collaborating with consultants and clients in CCS settings \cite{tseng2024data}.} 

\newtext{The OSINT Clinic builds on CCS research by applying its foundational principles --- such as personalized, collaborative investigations and the UIA framework --- to cybersecurity vulnerability assessments for small businesses using OSINT. While CCS has primarily focused on intimate partner violence (IPV) contexts \cite{freed2019my, havron_clinical_2019, tseng2022care}, the OSINT Clinic demonstrates how these principles can be extended to broader cybersecurity challenges. Beyond tackling a new domain, our work explores the HCI and CSCW challenges of leveraging generative AI to support clinical work. We propose a socio-technical system where AI not only assists with technical tasks but also fosters learning, coordination, and leadership among student analysts. }

\newtext{OSINT investigations face numerous challenges that are amplified in the context of clinical work, where student teams analyze public information. Investigators often work with limited resources such as funds, time, and access to specialized software \cite{kang_characterizing_2014, pastor2020not}. Issues of data quality, quantity, integration, and ethical considerations further complicate the process \cite{govardhan_key_2023}. Mukhopadhyay et al. found training on technical skills and ethics was essential to get novices to collaborate successfully with OSINT experts on real-world investigations \cite{mukhopadhyay2024osint}. However, comprehensive training is difficult to scale. Effective leadership in OSINT is also challenging as it requires navigating legal and ethical boundaries while applying investigative techniques across various domains \cite{akhgar2017open}. In this paper, we explore how generative AI can help scale this clinic model by understanding and addressing these challenges.}

\subsection{CSCW Research and Generative AI Opportunities in OSINT}
\label{sec:rw-sensemaking}

OSINT investigations are inherently complex, requiring the collection and analysis of large volumes of data as part of a sensemaking process to extract meaningful insights \cite{venkatagiri2019groundtruth, dailey_dispersants_2015,alharthi_2021}. 
Collaboration can enhance these efforts by dividing up workloads and integrating diverse perspectives \cite{fisher2012distributed}. Previous HCI and CSCW research has studied various approaches to collaborative and crowdsourced investigative work, including top-down strategies \cite{alcaidinho_2017_investigation}, bottom-up explorations \cite{huang_2015_investigation}, and hybrid methods \cite{venkatagiri2021crowdsolve}. Belghith et al. studied ``social OSINT'' --- i.e., the growing landscape of competitive and cooperative OSINT-themed events --- and characterized it as a community of practice with a culture of transparency and collaboration \cite{belghith_compete_2022}. \newtext{This model with overlapping goals of training and delivering real-world value also closely relates to growing HCI research on how trained and well-supported crowds can handle complex investigations \cite{venkatagiri2021crowdsolve, venkatagiri2023cosint, mukhopadhyay2024osint}.} 
Existing systems like Newstrition \cite{news} and CAPER \cite{aliprandi_caper_nodate} facilitate expert collaboration in specific contexts like information verification and prevention of organized crime, but there is potential to expand to non-experts through the use of AI tools \cite{doroudi_toward_2016, mukhopadhyay2024osint}.

\strike{AI can support OSINT analysts' sensemaking efforts by automating parts of the collection and analysis, such as using machine learning models for data classification and anomaly detection \cite{yadav2023open, brown_reflection_2014, evangelista2021systematic, vinayakumar2018detecting}.} 
\strike{However, current AI and machine learning tools in OSINT face challenges due to the lack of focus on real-world applications and inadequate integration with existing intelligence tools \cite{browne2024systematic}. Analysts often lack the technical background to understand AI outcomes, which hinders trust and usability \cite{de2015assessing, haughey2020misinformationbeat}. Moreover, AI solutions fail to address key aspects of intelligence work, such as planning and collaboration \cite{Dorton2021CollaborativeHS, browne_systematic_2024}. Browne et al. speculate that LLMs can support multiple phases of the Intelligence cycle, but no such tools were included in their review \cite{browne2024systematic}. We address the need for developing successful tools through effective integration into the analyst’s workflow and bridging the “socio-technical gap” \cite{ackerman2000intellectual}.}

\newtext{Generative AI presents a promising solution for augmenting OSINT investigations, particularly when compared to custom-built AI solutions \cite{browne_systematic_2024}.} Current AI and machine learning tools in OSINT face challenges due to the lack of focus on real-world applications and inadequate integration with existing intelligence tools \cite{yadav2023open, evangelista2021systematic, browne2024systematic}. Moreover, AI solutions fail to address key aspects of intelligence work, such as planning and collaboration \cite{Dorton2021CollaborativeHS, browne_systematic_2024}. Following the definition by Sarkar et al., we define generative AI as “an end-user tool whose technical implementation includes a generative model based on deep learning” \cite{sarkar2023participatory}. \newtext{Unlike specialized, narrowly focused AI tools, LLMs are general-purpose and capable of performing a range of complex tasks, such as reasoning, summarization, and web searches, without the need for extensive fine-tuning. Their flexibility makes them well-suited for dynamic and high-uncertainty contexts like OSINT investigations.} Prior studies, such as PentestGPT, have demonstrated the effectiveness of LLMs in cybersecurity applications, including interpreting tool outputs and suggesting next steps during penetration testing \cite{deng2024pentestgpt}. 

\newtext{Generative AI also offers advantages in training and skill development for novice OSINT analysts \cite{chen_next_2023, suh_ai_2021}. By supporting iterative learning processes through demonstrations, feedback, and self-evaluation \cite{mukhopadhyay2024osint, dow2012shepherding}, LLMs can help bridge knowledge gaps in areas like tool usage, data collection, and verification. Unlike custom AI solutions, which often require specialized technical expertise to design and deploy, generative AI tools are accessible to a broader audience and can be seamlessly integrated into existing workflows.}

Research has begun exploring how generative AI, like LLMs, can enhance collaboration by serving as active participants in group settings \cite{he_ai_2024, zhang2023investigating}. Shifting from single human-AI interactions to multi-human and AI environments could improve performance in complex scenarios, providing more context-aware collaboration \cite{muller2022frameworks, bansal_is_2021,he_ai_2024}. Generative AI can facilitate dynamic teamwork and decision-making, improving collaborative sensemaking in high-uncertainty situations \cite{Dorton2021CollaborativeHS, suri_use_2024, he_ai_2024}. Based on related work, we posit that generative AI can significantly improve the effectiveness and scalability of OSINT investigations across the dimensions of practical integration, training, and collaboration.

\subsection{Co-Designing AI for OSINT}
\label{rw:co-deisgn}
Despite the growing importance of OSINT in cybersecurity, there is limited research on incorporating stakeholder values in developing OSINT technologies \cite{riebe_values_2023, hepenstal2021analysis}. Co-design methods applied to the design of AI tools enable stakeholders to brainstorm ideas where the design landscape is underexplored, translating high-level user needs into actionable AI features \cite{Sanders2008CocreationAT}. However, a gap remains in applying co-design methodologies with OSINT investigators to address their specific challenges and needs.

Current AI design paradigms often struggle to identify innovative and feasible design concepts that effectively integrate human values. The Matchmaking for AI approach, introduced by Liu et al. \cite{liu2024humancenterednlpfactcheckingcodesigning}, addresses this by borrowing from the earlier, more general Matchmaking method \cite{bly_design_1999, Dijk2019DiscoveringUF}. It maps AI techniques to user activities using probes, fostering mutual learning between stakeholders and AI researchers. This method extends traditional co-design practices by focusing on practical feasibility and aligning existing AI capabilities with domain-specific tasks. \strike{We propose enhancing this model by incorporating real-world OSINT investigations and collaboration, allowing participants to explore AI's role in cybersecurity risk assessment.}

\newtext{We extend the Matchmaking for AI method by incorporating a longitudinal aspect, allowing us to study the use of generative AI tools over an extended period and across multiple phases of co-design. This enables AI solutions to be practically integrated into workflows and adaptable to evolving team dynamics \cite{chin2009exploring, stahl2005groups}. By embedding the method within a real-world context, we validated design ideas through longitudinal observation of participants conducting vulnerability assessments with actual clients.}

\strike{Generative AI presents valuable material for co-design, enabling stakeholders to explore new ways to enhance OSINT investigations \cite{sarkar2023participatory, kulkarni_word_2023}. Large language models (LLMs) like ChatGPT and Gemini have been used as probes to identify areas for improvement and assess their domain-specific applications \cite{kulkarni_word_2023}. Such generative models can be integrated into OSINT workflows to support planning \cite{noy_experimental_2023, bhattacharjee_understanding_2023}, creative problem-solving \cite{he_ai_2024, doshi_generative_2023}, and provide real-time feedback \cite{peng_impact_2023, wan_it_2024}.}

\subsection{Cybersecurity Vulnerability Assessment using OSINT}
\label{sec:vuln-assess}

The study deals with tasks related to cybersecurity vulnerability assessment, which is a systematic process of identifying, analyzing, and prioritizing potential security weaknesses in an organization. These assessments can have many components, including internal network discovery and vulnerability scanning with tools like Nmap  \cite{nmap_nodate} and Nessus \cite{nessus_nodate}; security controls interviews based on frameworks like CIS \cite{cis_nodate}; and physical security reviews looking for access control, fire suppression, and water detection \cite{noauthor_nist_2020}. Passive information-gathering techniques in OSINT achieve a subset of these tasks to assess cybersecurity vulnerabilities without direct interaction with the target. Prior work informs our use of OSINT tools and techniques relevant to vulnerability assessment \cite{browne_systematic_2024, hwang2022current, pastor2020not}.

Urban et al. found that over 83\% of the analyzed companies leak enough employee attributes to enable highly sophisticated phishing attacks \cite{urban_plenty_2020}. 
From a defense perspective, organizations can leverage OSINT techniques to assess their attack surfaces or identify social engineering opportunities \cite{hayes_open-source_2018, edwards_panning_2017, pervez_towards_2023}. As the use of social media and other online resources has grown, the availability of open-source information has expanded, making it a key tool for organizations to understand the threat landscape, identify the techniques and tools used by adversaries, and mitigate potential cyberattacks \cite{williams_defining_2018, tundis2022feature, pervez_towards_2023}. OSINT-based vulnerability assessments therefore shift the focus of cyber defense from reactive measures to proactive strategies \cite{tabatabaei2017osint, hayes_open-source_2018}. OSINT enables fast, real-time information collection, and clear sourcing, and is both convenient and cost-effective \cite{gibson2016acquisition}. 
Despite its critical need during the reconnaissance phase, there is a lack of understanding of challenges involved in effective OSINT investigations for vulnerability assessments.


\section{Study Design}

\begin{table*}[t]
\caption{OSINT tasks in cybersecurity vulnerability assessments}
\label{tab:osint-tasks-table}
\begin{tabular}{@{}p{0.2\linewidth}|p{0.35\linewidth}|p{0.35\linewidth}@{}}
\hline
\textbf{Category}                    & \textbf{Task Description  }                                                                                                                                                                 & \textbf{OSINT Tools and Techniques}                                                                                                                        \\ \hline
Network Discovery           & Identify all digital assets (websites, email, servers, webcams) associated with the business.                                                                                      & DNS enumeration, IP mapping, Reverse IP lookup, WHOIS lookup \cite{whoiscom_nodate}, Shodan \cite{noauthor_shodan_nodate}                                                                              \\ \hline
Website Vulnerabilities     & Discover vulnerabilities on the business’s website such as outdated software, insecure admin panels, open ports, files that should not be public, etc.                             & BuiltWith \cite{builtwith_2024_nodate}, CMS identification, port scanning, mail server scanner, Google dorking \cite{googledorking_nodate}                                                                \\ \hline
Physical Assets             & Identify physical assets like buildings, properties, or vehicles from online resources or images.                                                                                  & Google Maps, Google Earth, Street View, Image metadata analysis, Geolocation techniques                                                           \\ \hline
Third-Party Services        & Document third-party services (vendors, etc.) that the client does business with to identify possible supply chain vulnerabilities.                                                & External service provider evaluation based on reviews, past breaches, and terms and conditions                                                    \\ \hline
Employee Public Information & Identify key employees from a company using social media platforms, gathering information like job titles, email addresses, or sensitive personal data accidentally shared online. & LinkedIn search; Advanced searches on social media platforms like Instagram, Twitter, YouTube, TikTok; Censys \cite{censys_exposure_nodate}, other phone number and email verification tools \\ \hline
Brand Monitoring            & Find instances of negative mentions, disinformation, inappropriate content, legal filings, and settlements related to the business.                                                & Online reviews, social media monitoring, court case databases                                                                                     \\ \hline
Breach Data                 & Search for a company’s leaked passwords and data in breach data dumps and on the dark web.                                                                                         & Have I Been Pwned \cite{haveibeenpwned_nodate},  IntelX \cite{noauthor_intelligence_nodate}, other breach databases                                                                               \\ \hline
\end{tabular}%
\end{table*}

\subsection{Study Protocol}

\begin{figure*}[h]
\includegraphics[width=15cm]{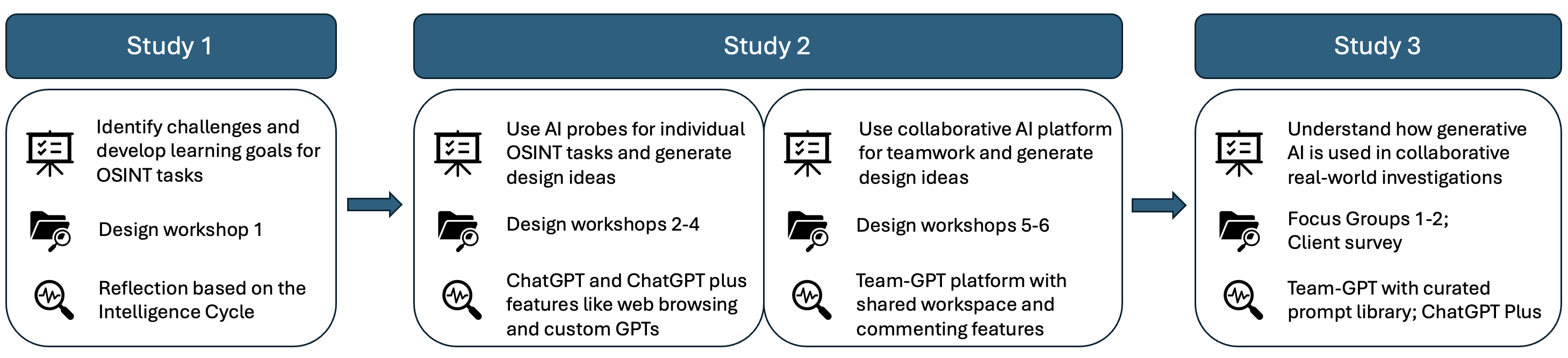}
\caption{Study Diagram: (1) Study 1 was formative and aimed at identifying the key challenges in the Intelligence Cycle phases of OSINT investigations. We asked participants to reflect on previous investigations and document the steps and challenges faced during Design Workshop 1. (2) In Study 2, we introduced generative AI probes in Design Workshops 2--4 to explore how AI could address challenges experienced earlier. During the workshops, we identified areas where generative AI was effective and also its limitations. We focused on the emergent challenges of collaboration and leadership during \strike{design workshops 4 and 5} \newtext{Design Workshops 5 and 6}, with a collaborative AI platform (Team-GPT) as a probe. (3) In Study 3, we piloted OSINT-based vulnerability assessments with three small businesses, integrating curated generative AI prompts into ongoing investigations. This real-world application highlighted the practical utility and surfaced challenges of using foundational generative AI models like ChatGPT in vulnerability assessments, especially in collaborative settings.}
\label{fig:study-diagram}
\Description{The image shows the study workflow diagram that outlines the structure of three studies focused on collaborative AI and Open-Source Intelligence (OSINT) tasks. The first study, "Study 1," is dedicated to identifying challenges and developing learning goals for OSINT tasks. This is achieved through design workshop 1 and reflection based on the Intelligence Cycle. Moving forward, "Study 2" uses AI probes for individual OSINT tasks, generating design ideas through workshops 2 to 4, and leveraging ChatGPT's advanced features such as web browsing and custom GPTs. The final part, "Study 3," investigates the application of generative AI in real-world investigations through a combination of focus groups, client surveys, and a Team-GPT platform equipped with curated prompts to enhance the collaborative process. This description was generated using ChatGPT.}
\end{figure*}

We adopted the Matchmaking for AI method \cite{liu2024humancenterednlpfactcheckingcodesigning} to conduct our co-design study. The method allows stakeholders --- students in our case --- to actively contribute to the design and development of effective tools for OSINT investigations. The iterative workshops and use of AI probes provide a structured yet flexible framework for identifying user needs and aligning them with existing AI capabilities. This method ensures that the solutions we develop are not only technically feasible but also usable in practical scenarios, particularly in the context of cybersecurity vulnerability assessments. We discuss our methodological contributions to the framework by supporting real-world investigations and collaboration among students in Section~\ref{reflection-MMFAI}.

We ran the co-design study with a group of six undergraduate students during the 2023--24 academic year at our institution, a large public university in the US. Previous research has demonstrated that groups of student analysts can help characterize intelligence investigations \cite{chen2021supporting, kang_characterizing_2014}, effectively experiment with new crowdsourcing methods, and provide valuable process recommendations \cite{xu2015classroom, williams2016axis}. Students have also securely engaged in real-world OSINT investigations, including cybersecurity vulnerability assessments and human rights investigations, as part of experiential learning \cite{MITCybersecurityClinic, HRC_Berkeley}. 

As a part of the clinic, the authors, who have relevant industry experience and certifications, trained students on different aspects of OSINT investigations like (1) practicing operational security (OPSEC) and being aware of privacy, legal, and ethical issues; (2) mining different public sources of information like social media for brand monitoring and employee profiles; (3) satellite imagery for geolocating physical assets; (4) identifying network infrastructure for identifying open ports, IoT devices; (5) inspecting website and network infrastructure to identify vulnerabilities and unpatched systems; and (6) looking for breach data and leaked credentials. Table~\ref{tab:osint-tasks-table} shows the seven key areas for vulnerability assessments along with relevant OSINT tools and techniques.

Our longitudinal study spanned two semesters. Study 1 (Design Workshop 1) was conducted after three months of OSINT training. Right after, we ran Study 2 (Design workshops 2--6) over two months during the end of the first semester and the beginning of the second. Finally, Study 3 involved two focus group interviews conducted over a span of one month at the end of the second semester. \newtext{The longitudinal design of this study was essential for us to observe how the use of generative AI evolved over time and across different phases of the Intelligence Cycle. This extended engagement provided insights into how AI is integrated, how collaboration challenges develop, and how leadership roles evolve in response to AI support.}
Figure \ref{fig:study-diagram} represents the flow of the research study with the objective, data source(s), and probe(s) for each of Studies 1-3. 

\subsection{Participant Recruitment}
This study was approved by our university’s IRB. Six undergraduate students (P1--P6) were recruited based on their interest in gaining practical OSINT skills as part of a team focused on cybersecurity vulnerability assessments. The group represented a range of academic backgrounds, including majors in Cybersecurity, Computer Science, National Security, and Engineering. The participants’ ages ranged from 18 to 22. Three participants identified as women, two as men, and one as non-binary. Some students had prior experience with OSINT-related activities, such as capture-the-flag (CTF) challenges, OSINT counter-disinformation projects, and research involving fake identity generation. Others were newer to the field but eager to learn.  

The participants met twice a week for two-hour sessions to learn and practice OSINT skills. They were paid an hourly wage of \$18 for a maximum of 10 hours per week to attend the OSINT clinic. The design workshops were integrated into the clinic to help them reflect and augment their workflows with generative AI. 


\strike{Our study involved only six participants over two semesters, which may limit the diversity of needs and design ideas explored. Previous research in intelligence analysis, including studies with 2--10 participants \cite{kang_characterizing_2014, chin2009exploring, riebe_values_2023}, often face similar constraints due to the sensitive nature of the work and client involvement.} 

\strike{In Study 2, short sessions (10--20 minutes) for trying out prompts and features may have restricted idea generation. Future studies could expand on ideas by including longer workshops with power users of generative AI tools or OSINT practitioners.} 

\strike{Generative AI probes can produce incorrect responses or "hallucinations," potentially limiting the reliability of disseminated results. While our focus on verification and cross-referencing addresses some issues, further research is needed to develop accountable human-AI collaboration frameworks \cite{chen_next_2023}.} 

\section{Study 1: Identifying Challenges \newtext{and Learning Goals} with OSINT Tasks}

\subsection{Method}

In this study, we sought to understand participant workflows related to the OSINT tasks in Table~\ref{tab:osint-tasks-table} during a recorded 2-hour long co-located workshop (design workshop 1). We adapted the Domain Speciality Canvas \cite{liu2024humancenterednlpfactcheckingcodesigning} from Step 1 of the Matchmaking for AI process and added another column for collaboration challenges to facilitate data collection (shown in Figure~\ref{fig:figjam-study1}). Before this study, participants were trained on and practiced common OSINT tools and techniques mentioned in Table \ref{tab:osint-tasks-table}. They also worked in teams of three to investigate a few local businesses. 
To reflect on the work, we asked participants to discuss their most recent investigation, one successful team investigation, and one challenging team investigation. Corresponding to the normative workflow used in Step 1 by Liu et al.~\cite{liu2024humancenterednlpfactcheckingcodesigning}, we presented the Intelligence Cycle (IC) \cite{browne_systematic_2024}, widely used to structure investigations, to guide participants to add notes. We mapped out the different challenges on a digital whiteboard (FigJam) and asked follow-up questions to clarify participants' responses. 

\begin{figure}[]
\includegraphics[width=8.5cm]{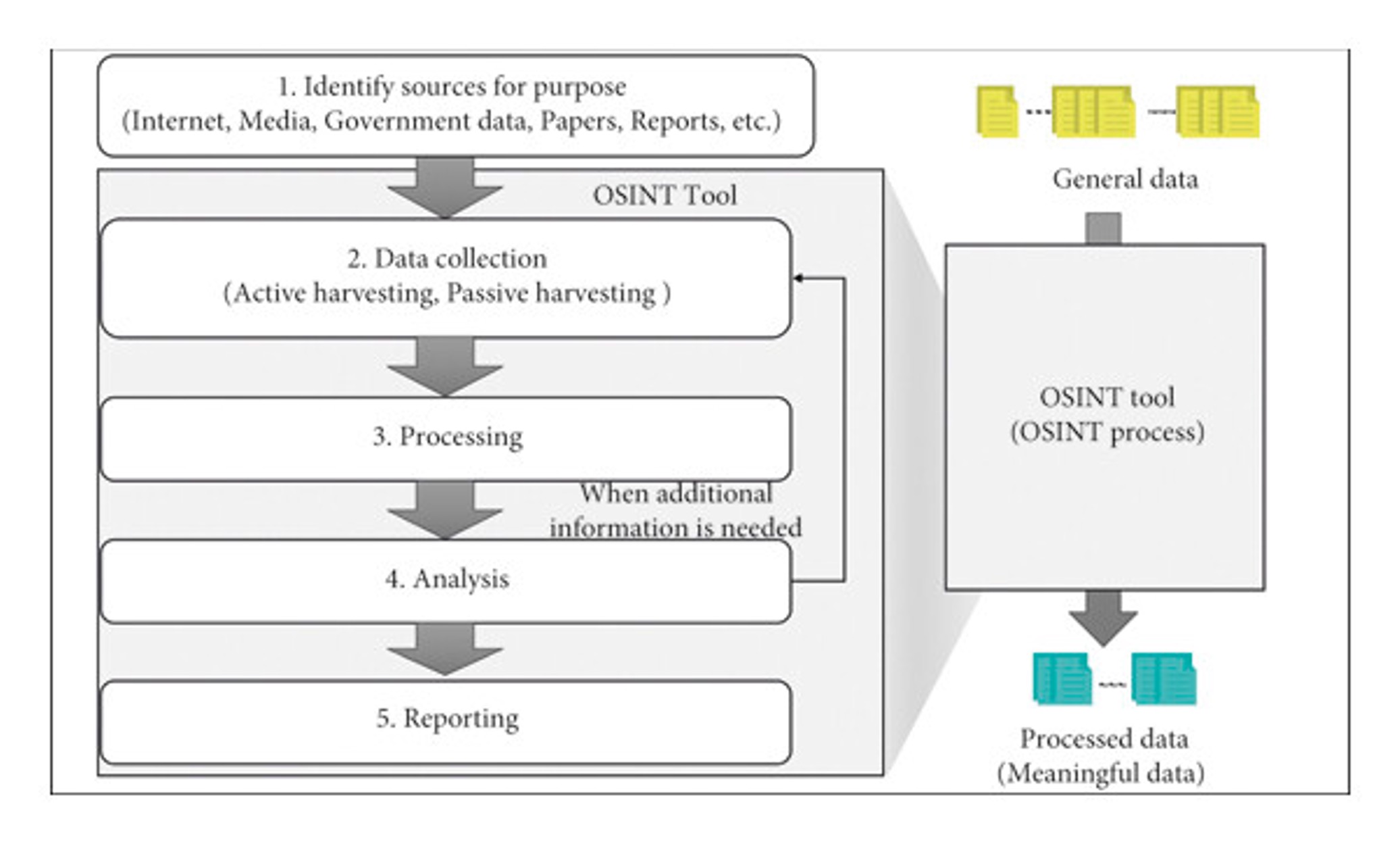}
\caption{OSINT Intelligence Model in the context of cybersecurity as described by Hwang et al. \cite{hwang2022current}}
\label{fig:OIM}
\Description{This image shows the OSINT Intelligence Model in the context of cybersecurity (Fig. 2). This flowchart illustrates the OSINT Intelligence Model in cybersecurity, detailing the steps involved in gathering and processing information. The model shows sequential phases: identifying sources, data collection (active and passive harvesting), processing, analysis, and reporting. It highlights the use of OSINT tools for converting general data into meaningful processed data for cybersecurity purposes. This description was generated using ChatGPT.}
\end{figure}

\subsection{Analysis}

The first author transcribed the recorded workshop and, in collaboration with the rest of the research team, conducted an inductive thematic analysis \cite{braun2006thematic} of the transcript to identify the challenges. We aligned the identified challenges with the five phases of the OSINT Intelligence Model (OIM) presented by Hwang et al. \cite{hwang2022current} (see Figure~\ref{fig:OIM}), which closely follows the IC. \newtext{Hwang et al's model focuses on continuous iteration and practical applications to cybersecurity \cite{hwang2022current}.} We denote the first stage of ``identifying sources for purpose'' as ``planning'' to match the IC, as it entails the same task of figuring ``where and how to get information.''


\subsection{Results}

We identified major challenges and learning goals across the five phases of planning, collection, processing, analysis, and dissemination in OSINT investigations. In the following sections, we describe the challenges students faced while completing the tasks, and develop learning goals based on existing guidelines \cite{tabatabaei2017osint, mukhopadhyay2024osint} and our own experiences as OSINT researchers. Table~\ref{tab:challenges-table} collates the challenges and learning goals across the five phases.

\begin{table*}[h]
\caption{Challenges and Learning Goals across phases of OSINT investigations}
\label{tab:challenges-table}
\begin{tabular}{@{}p{0.12\linewidth}|p{0.4\linewidth}|p{0.4\linewidth}@{}}
\hline
\textbf{OSINT Cycle Step} & \textbf{Challenges}                                                      & \textbf{Learning Goals}                                                                                 \\ \hline
Planning         & Navigating ethical and legal boundaries                         & Develop appropriate trust on OSINT data, tools, and techniques; Implement operational security \\ \cline{2-3} 
                 & Defining investigation objectives and overlapping work          & Breakdown investigations into subtasks with data sources                                       \\ \hline
Collection       & Skill gaps in data collection                                   & Identify and utilize the most effective OSINT tools                                            \\ \cline{2-3} 
                 & Lack of clarity in information presentation                     & Develop and utilize standardized templates to systematically collect and organize information  \\ \hline
Processing       & Handling high volume of raw information                         & Structure raw information into parsable format and improve documentation                       \\ \cline{2-3} 
                 & Verifying public information                                    & Find multiple sources for OSINT verification                                                   \\ \hline
Analysis         & Identifying vulnerabilities                                     & Interpret results from tools and identify vulnerabilities                                      \\ \cline{2-3} 
                 & Developing recommendations from vulnerabilities identified & Generate actionable recommendations                                                            \\ \hline
Dissemination    & Not knowing “what a good end product will look like”            & Iterate on recommendations to tailor it to business needs      \\                          \hline 
\end{tabular}%
\end{table*}

\subsubsection{Planning}


\paragraph{Navigating the ethical and legal boundaries}
Participants were trained on the ethics and safety of investigations, but they often faced dilemmas around the use of tools and techniques. In OSINT investigations, active techniques commonly used beyond the reconnaissance phase of penetration testing are typically avoided. P6 highlighted the need to understand the limitations and restrictions on using such tools. P1 thought investigations were complicated by the need to stay clearly within legal boundaries while investigating vulnerabilities, particularly when trying to gain access to leaked data. P6 also highlighted practical concerns such as private browsing on platforms like LinkedIn to avoid alerting the target of the investigation.

We argue that understanding which OSINT data, tools, and techniques are both private and trustworthy can help alleviate the challenge. Participants also need to implement strong operational security practices, such as creating and using sock puppet accounts, to protect both the investigation and the investigator. These two aspects can form a learning goal to address the ethical challenges.

\textbf{Learning Goal 1 (LG1):} Develop appropriate trust for OSINT data, tools, and techniques; Implement operational security

\paragraph{Defining investigation objectives and overlapping work}

Participants found it difficult to define clear objectives and scoping for the OSINT investigations, which sometimes led to inefficiencies and miscommunication. P1 emphasized the need for better procedural guidelines, stating, "... just more of an understanding of format and kind of a procedure that we should follow as to what things should look like". There were instances where a lack of communication led to redundant efforts, and P1 found multiple members focusing on a specific platform reduced efficiency. P3 and P4 reported that working on a shared document to highlight information collected wasn’t effective in minimizing overlap. 

These insights suggest that clearer division of work and structured guidelines could enhance the effectiveness of team-based OSINT investigations. A learning goal could be to develop the ability to effectively break down investigations into specific subtasks. This includes identifying relevant data sources and facilitating a structured approach for task division.

\textbf{Learning Goal 2 (LG2):}  Break down investigations into subtasks with data sources

\subsubsection{Collection}


\paragraph{Skill gaps in data collection}
  
Participants faced significant challenges in identifying effective techniques for data collection, compounded by varying skill levels among team members in using advanced OSINT tools and platforms. Many struggled with the technical aspects, such as understanding code, creating effective queries for search, and navigating tools like the Linux command line interface, which are not always user-friendly for beginners. P2 highlighted the steep learning curve: "I don’t really know coding languages, so I’m trying to mess with an inspect element, and it just looks like a bunch of text to me." Participants also struggled to identify the right OSINT tool(s) for the job among potentially dozens of possibilities. P3 and P4 mentioned the difficulty of staying updated with changing techniques.

Participants need to select appropriate tools among the many available options to improve performance. Finding alternate sources of information becomes essential when stuck.

\textbf{Learning Goal 3 (LG3):} Identify and utilize the most effective OSINT tools

\paragraph{Lack of clarity in information presentation}

Participants mentioned difficulties in collecting and organizing data due to the lack of a clear structure and formatting for tracking different intelligence sources. Inconsistencies in format and overall disorganization of documents where they collected data often hindered the utility of the gathered information. For example the “data dumps” or raw results from tools made it hard for other members of the team to parse the information. As P6 put it, "It's been told we have to interpret the information, analyze it, and then put it in a format to be understandable. But I often notice it comes out very variable, just because some people put it in some ways, or it has to be very messy in the [working] document." P1 emphasized the need for a clearer structure, stating, "Just more of an understanding of format and kind of a procedure that we should follow as to what things should look like”. 

Implementing templates can significantly improve the organization and clarity of OSINT investigations. Templates provide a structured approach to document and cover multiple sources of information, ensuring consistency and comprehensiveness.

\textbf{Learning Goal 4 (LG4):} Develop and utilize standardized templates to systematically collect and organize information

\subsubsection{Processing}


\paragraph{Handling high volume of raw information}

The sheer volume of unstructured information often made it difficult for participants to identify and focus on the most pertinent details. The team spent a lot of time summarizing, sorting, and categorizing information and eliminating irrelevant data. Additionally, there was a need for better documentation practices, like P5 wanted to, ``\ldots document better how I found certain information so somebody else can follow the steps I took''. The effectiveness of collaboration was often hindered when team members failed to properly document their work or explain their processes to others, leading to potential oversights like duplication of work or inaccurate findings.

To tackle the overwhelming amount of data, participants emphasized the importance of identifying relevant details that can be verified and used for further analysis. P4 thought, ``AI could be very useful, especially if you have a large data set that you wanna maybe format in the best way''. Improved documentation could help with getting other members to leverage the collected information.

\textbf{Learning Goal 5 (LG5):} Structure raw information into parsable format and improve documentation

\paragraph{Verifying public information}

A key challenge for participants was accurately verifying information, especially in relation to smaller companies where multiple sources were harder to find. Participants wanted to follow best practices and ensure the accuracy of important details by cross-referencing multiple sources. P2 highlighted the importance of "getting at least two sources confirming something," such as verifying an employee's identity. This process often involved checking social media profiles and company websites or finding official state records to double-check the accuracy of the information. However, the difficulty of finding multiple sources, especially for verifying website vulnerabilities, made this task more challenging.

Developing the ability to thoroughly verify information by consistently finding and cross-referencing multiple sources is essential for the accuracy of findings. Supporting this process, particularly when dealing with less transparent or hard-to-verify sources, can improve the investigation results.

\textbf{Learning Goal 6 (LG6):} Find multiple sources to enable OSINT verification 

\subsubsection{Analysis}


\paragraph{Identifying vulnerabilities}

Participants lacked a deep understanding of common vulnerabilities and how different data sources can be combined for vulnerability assessments. There were difficulties in interpreting and analyzing complex technical information from tools. P6 wanted to be ``able to pull the information that you have readily available, and be able to interpret that in a way that can help the small business.'' Participants highlighted the challenge of not only gathering relevant data, but also interpreting it in a meaningful way that aligns with the investigation's objectives.

Supporting the ability to interpret results from tools and accurately identify vulnerabilities is crucial in overcoming the challenges. By helping bridge the skill gaps, participants can more effectively analyze complex data and generate reliable vulnerability assessments.

\textbf{Learning Goal 7 (LG7):} Interpret results from tools and effectively identify vulnerabilities

\paragraph{Developing recommendations from vulnerabilities identified}

Participants found it difficult to develop recommendations for mitigating the vulnerabilities based on analyzed data. For example, P1 mentioned that it was hard to identify vulnerabilities for websites that are well-maintained by third-party vendors; in this case, the recommendation could be to continue using these services. Participants needed guidance to effectively present their recommendations, integrating the identified vulnerabilities with supporting data.

Understanding different types of recommendations helps the students in iterating over their investigations and making them actionable. Providing a starting point and directions for recommendations to the students can improve this process.

\textbf{Learning Goal 8 (LG8):} Generate actionable recommendations

\subsubsection{Dissemination}


\paragraph{Not knowing ``what a good end product will look like''}

We observed that the issue of developing appropriate recommendations was exacerbated when presenting findings to non-technical clients, as they require clear explanations and simplifications. Participants also lacked clarity on the best format to use for presenting information. P1 thought their deliverables ``\ldots come out very disorganized and the way information is presented is highly variable.''

Creating standardized formats for reporting findings will ensure clarity and consistency. Iterating on recommendations to tailor it to the needs of the small business will make the results of investigations highly impactful.

\textbf{Learning Goal 9 (LG9):} Iterate on recommendations to tailor them to business needs \newline

Study 1 findings highlighted performance gaps in OSINT investigations. To address these challenges and achieve the learning goals (LG 1--9), Study 2 explores the use of generative AI probes across the five phases of IC.
\section{Study 2: Using Generative AI Probes to \strike{Explore Design Ideas for OSINT Tasks} \newtext{Achieve Learning Goals for OSINT Investigation}}

\subsection{Method} 
In Study 2, we conducted three co-located design workshops (Design workshops 2--4) to address the challenges identified in Study 1 and achieve the learning goals. These two-hour workshops allowed participants to individually explore how AI could enhance various phases of OSINT investigations. Here, we closely followed Step 2 (playing with AI probes) and Step 3 (brainstorming AI design elements) of the Matchmaking for AI method \cite{liu2024humancenterednlpfactcheckingcodesigning}. Participants spent time working on OSINT tasks using generative AI probes before coming up with design ideas during the workshops. The ideas were represented on a digital whiteboard (FigJam) using the ``Co-Design AI Canvas'' from the Matchmaking for AI method (see Figure~\ref{fig:figjam-study2}). We included columns for generative AI ideas, challenges faced, and function and form, to gather what added customizations or controls they would like beyond the probe. Participants were asked to copy over the purpose of the prompt and the prompt used into a shared prompt library document for all the workshops.

Design Workshop 2 focused on planning and collection-related challenges. These two phases go hand-in-hand, as there needs to be a clear plan before starting with data collection. Participants performed two sets of tasks or two activities; after each activity, there was a focus group to brainstorm design ideas and potential improvements on a digital whiteboard. The activities and brainstorming lasted 25 minutes each with a break of 10 minutes between the sets. Participants created subtasks and assigned them to their team members, and also identified the right tools and techniques for the tasks during the first activity. They developed templates using ChatGPT and did some preliminary data collection to fill out the template during the second activity. This workshop utilized the standard out-of-the-box ChatGPT 3.5 with no web browsing, as this was the latest version at the time.

Design Workshop 3 targeted challenges in the processing, analysis, and dissemination phases. Similar to the previous workshop, there were two sets of tasks. In the first activity, the tasks included verifying information, documenting the investigation process, and interpreting the results as to how they could be vulnerabilities. For the second activity, participants had to create a report based on their findings so far that could be sent to small businesses. Participants could use the templates developed in the earlier workshop and the data collected. This workshop also relied on the standard ChatGPT 3.5 without additional enhancements.

Design Workshop 4 addressed all identified challenges using a new set of AI probes, including web browsing, image analysis, custom GPTs, and plugins (now discontinued) made available through ChatGPT Plus \cite{chatgpt_plus}. These features are relevant to OSINT investigations as they help utilize publicly available information from websites and provide access to external tools for analysis. Participants were first presented with specific goals related to the challenges. They played around with each of the four AI probes for 10 minutes and then brainstormed how these tools could be used to help overcome each challenge. This workshop aimed to explore more advanced AI capabilities to support participants in each phase of OSINT investigations.

We used deductive thematic analysis \cite{braun2006thematic} to code transcripts from the workshops and notes from the FigJam boards, corresponding to each learning goal. We coded (1) how participants used ChatGPT; (2) what worked well and what challenges faced; and (3) desired features to improve the experience. We organized our findings for each challenge into three main areas to provide a clear understanding of how generative AI could be leveraged in OSINT investigations. First, we discuss the tasks performed and prompting methods used. Second, we explore the utility of generative AI, highlighting what worked well and the challenges encountered. Finally, we present new ideas and suggested features that participants proposed to enhance the effectiveness of AI in their investigations. 

\subsection{Results}

\subsubsection{Planning} 

\paragraph{LG1: Developing appropriate trust on OSINT data, tools, and techniques; Implementing operational security}

Participants used AI to analyze privacy measures of various tools and identify trustworthy resources. P1 sought guidance on whether their investigative methods stayed within ethical boundaries. P3 and P6 mentioned using prompting techniques to work around restrictions in ChatGPT’s responses, asking for ethical means to obtain information. P2 highlighted the importance of finding tools commonly used in real OSINT investigations and then understanding how they work before applying them. 

Participants found that while ChatGPT was useful in creating and updating sock puppet accounts and providing general guidance on privacy and trust, it often failed to provide detailed and specific information about the trustworthiness of tools. P3 noted, ``Finding specific details about tools and their trustworthiness was challenging. ChatGPT’s responses were sometimes too general to be useful." P1 echoed this idea, pointing out that the AI's comments on trust were often based on popularity rather than a deep analysis of the tool's security features. Participants realized the need to ask more targeted questions to extract useful information.

\strike{P3 and P5 suggested that ChatGPT should provide a rating or value scale to help users better understand the reputation of a tool or source before using it. Participants wanted AI to offer unbiased views on privacy and trust, with P6 stating, ``I would like AI to be able to provide comments on the privacy and amount of trust I can place in different verification and informative sources." Participants explored web browsing and custom GPT models and thought they could enhance operational security training and simulations. They also noted the need for regular updates and specificity in the AI’s knowledge base due to the rapid pace of change in the OSINT tools ecosystem.}

\paragraph{LG2: Break down investigations into subtasks with data sources}

Participants used ChatGPT to break down complex tasks into manageable subtasks tailored to their team’s capabilities. For example, P1 mentioned, "I gave it one of the OSINT tasks, and I asked it to divide the work into subtasks\ldots trying to make it adapt to our team by providing the number of members and experience levels". The templates served as conceptual models to guide the investigations. Participants also used specific prompting techniques to receive more structured and relevant answers, such as asking ChatGPT to provide references, using chain-of-thought by asking for steps to get to the subtasks and tools, and framing questions in a scenario-based manner like ``if I were to tell my friend how to \_\_\_ what should I say?'' (P2). 

Participants found generative AI to be effective in assisting with task division and resource identification, making it easier to manage team-based OSINT investigations. P5 reflected, "I basically just asked it to come up with a list of subtasks and a list of tools and techniques\ldots and it worked perfectly". However, there were challenges with the AI's limitations in providing specific details related to smaller businesses as P4 observed, ``ChatGPT provides little to no information on smaller businesses''. There were also difficulties crafting prompts to avoid ``guardrails'' against inappropriate content or eliciting new, non-repetitive information. Participants proposed custom GPT models trained with OSINT-specific databases and plugins to make the AI more tailored to the unique needs of OSINT investigations.

\strike{P2 wanted to improve the experience was for ChatGPT to link directly to tools and resources it recommends, simplifying the follow-up steps in investigations. They saw value in ChatGPT's ability to create visually appealing documents and diagrams, which would aid in team discussions and presenting findings. The use of advanced features like custom GPT models trained with OSINT-specific databases and plugins that create visual templates were also proposed as ways to make the AI more tailored to the unique needs of OSINT investigations.}

\subsubsection{Collection}

\paragraph{LG3: Identify and utilize the most effective OSINT tools}

Participants used ChatGPT to gather information sources and tools relevant to their tasks, along with descriptions of how to use them effectively. P3 tried to obtain specific ``website links to look into data breach sources'' as he found broad directions for investigation unhelpful. Prompt structures varied, like role-playing as a cybersecurity student or asking for queries for techniques like Google dorking. P1 and P5 used creative approaches to bypass ChatGPT guardrails, such as asking it to suggest websites similar to those known for posting leaked passwords rather than directly requesting such information.

Participants found ChatGPT to be particularly effective for discovering resources when they asked for specific tools. They could generate alternative options through follow-ups when the initial suggestions did not meet their needs. However, challenges were noted, particularly regarding the AI's outdated knowledge base, as it only contained information on tools up to January 2022. Furthermore, the tools suggested were not always available. P1 described the process of verifying each tool suggested by ChatGPT: ``You have to painstakingly copy and paste each one [tool] and research it yourself.'' Participants noted that while ChatGPT's web browsing capabilities could cover a wide range of knowledge, the credibility of the sources provided was questionable and emphasized the need for careful cross-verification.

\strike{A common suggestion for improvement was for ChatGPT to provide more contextual information on an OSINT tool, or as P6 explained: ``getting a comprehensible description/summary of the tool so I know what and how to use it in my work.'' Participants wanted more clarity about the tool's data sources, whether tools had a paywall, as well as how to find specific details about their trustworthiness.}

\paragraph{LG4: Develop and utilize standardized templates to systematically collect and organize information}

Participants used ChatGPT to create comprehensive templates that could be applied in future investigations. P6 said, ``I specifically used ChatGPT to help construct a table so all the information could be organized and easy to read.'' The process often involved iterative prompting, where participants refined their queries to add information like data sources and OSINT techniques. P2 and P6 mentioned that these details helped divide up work within the team.

Participants particularly appreciated the AI's ability to quickly lay out different sections and specify what should go under each. Participants faced challenges in reformatting the AI-generated templates to make them more presentable. P1 explained, ``ChatGPT is great for getting your data quickly into format but honestly it doesn’t look very good. With a little bit of editing and using a format made by a human I think it would look much better.'' Importantly, the AI's responses varied depending on the prompts, leading to inconsistencies in the templates generated across different users that impeded collaboration.
Participants proposed the idea idea of a prompt library, where successful prompts for templates could be shared and rated by others.

\strike{Participants wanted ChatGPT to generate files containing templates, such as spreadsheets or documents, which would reduce the need for additional formatting. P6 recalled, ``I tried asking ChatGPT to make the template into a spreadsheet, but it is unable to directly make files. It would be more helpful if ChatGPT could generate files.'' 
P3 highlighted the potential of using plugins to generate and convert templates more efficiently.} 

\subsubsection{Processing}

\paragraph{LG5: Structure raw information into parsable format and improve documentation}

Participants employed ChatGPT to organize raw information into readable formats and eliminate duplicate or less relevant information. For instance, P6 mentioned, ``I asked it to look for different verification sources mentioned throughout the data\ldots it was a lot of information just written, but it's hard to pick through and find all the little details.'' Participants created specific formats for documenting sources and then used ChatGPT to follow these formats, though some manual adjustments were still required.

While the AI was able to organize data into a list or by topics, it often failed to follow requests like placing data directly into tables. The organized lists were still helpful in making the raw data easier to process and document. Another issue was with ChatGPT's overwhelming amount of information, making it difficult to identify the most relevant details.

\strike{Participants reiterated the need to directly generate files in the specified format, such as spreadsheets with particular columns, to reduce manual formatting efforts. They explored the integration of automation tools like web scraping through Python libraries, though they found it frustrating that ChatGPT could not perform these tasks directly and instead provided instructions for them. These suggestions highlight the potential for ChatGPT to be more automated in the data organization and documentation process.}

\paragraph{LG6: Support OSINT verification based on multiple sources}

Participants noted that ChatGPT was useful in identifying potential verification sources and organizing data. However, tool suggestions often fell short of providing the detailed, specific information needed to trust the AI's recommendations, as sources or tools were unfamiliar. P2 cautioned about relying too heavily on AI’s web browsing capabilities, as they still needed to be double-checked for accuracy.

\strike{Participants suggested a more iterative experience for diving deeper. P1 said, "I would like AI to be more interactive and help complete the investigation with me by giving recommendations on places to find information not found". They wanted detailed ratings or community sentiment analysis of tools for reliability.} 

\subsubsection{Analysis}

\paragraph{LG7: Interpret results from tools and identify vulnerabilities}

Participants often began by attempting to interpret the results on their own and then used ChatGPT to further analyze the data. This approach helped participants gain a broader understanding of the vulnerabilities they were investigating, even if the AI's feedback was sometimes generalized.

Participants found ChatGPT's explanations to be somewhat helpful. As P3 noted, ``It was useful for elaborating on vulnerabilities and explaining in simpler terms\ldots especially for a non-technical audience.''. However, P6 found output adjustments to be hard, "Using ChatGPT for interpretation requires specific prompts for better detail. It can be helpful, but you need to be clear about what you’re asking for". P5 found that ChatGPT did not provide detailed security analysis when using web browsing capabilities, often citing limitations in its scope. \strike{P3 mentioned specialized plugins like Check Point Security, which evaluates the security status of IP addresses, URLs, and file hashes, to augment ChatGPT's general responses.} 

\paragraph{LG8: Generate actionable recommendations}
 
Participants used ChatGPT to develop tailored recommendations for non-technical small business owners. Prompting techniques included asking ChatGPT to structure recommendations into a table and develop a timeline for implementing these actions to present information professionally.

ChatGPT helped present supporting data for reports and generate recommendations. Participants found it difficult to identify the most relevant actions from a large number of recommendations. Similar to previous challenges, ChatGPT and custom GPTs often failed to generate specific and actionable recommendations tailored to the unique needs of a given situation, sometimes refusing a detailed analysis for being out of scope or unethical. \strike{Participants wanted support in creating data visualizations for recommendations, such as Gantt charts, to better illustrate timelines and implementation strategies.} 

\subsubsection{Dissemination}

\paragraph{LG9: Iterate on recommendations to tailor it to business needs}

Participants iterated on recommendations by refining them through a feedback loop. P3 described the process as ``using ChatGPT to create specific recommendations based on vulnerabilities and then simulating client feedback to further improve the recommendations.'' They also wanted outputs to have severity ratings for vulnerabilities.

Participants found that using ChatGPT to simulate client feedback was particularly useful in preparing for client interactions. P1 emphasized that iterative feedback from ChatGPT helped in making the recommendations more actionable and better tailored to client needs. However, P4 observed that the iterative process was often limited to removing, or reformatting content rather than delving into more detailed analysis. Participants also worried about privacy issues in involving the AI at this final stage, when security flaws were at their most distilled and contextualized. \strike{Some suggested that turning off chat history or using personal chat configurations could enhance privacy during the iterative process of sharing sensitive information like vulnerabilities.}  

\subsection{Follow-up Study: Collaboration and Leadership}
\label{collab-study2}

As participants kept performing OSINT investigations and incorporated generative AI in their workflow, we observed new collaboration challenges emerge:

\paragraph{Difficulty in achieving consistent formatting and style}

Participants saw despite using similar prompts, the outputs varied widely, making it difficult to maintain uniformity across documents. P6 summarized it as, "ChatGPT is pretty different depending on the prompt you give, and it always gives different answers for different people. So it is a little difficult to make our templates match". This inconsistency required additional time and effort to ensure consistency across the team’s work.

\paragraph{Difficulty in developing shared resources}

Participants found it challenging to build on each other's work and learn from prompts used by others. P6 reflected, "I didn't end up testing out anyone else's prompts, but I might do that in the future just to see what [comes] as a result of it". P3 and P4 were interested in finding successful prompts to use them as starting points. In terms of improvements, P6 said that "the collaborative process could be improved if we used a different platform to document our investigation notes and sources". This platform would help to share editable results from generative models.

\subsubsection{Method}
\begin{figure*}[t]
\includegraphics[width=14cm]{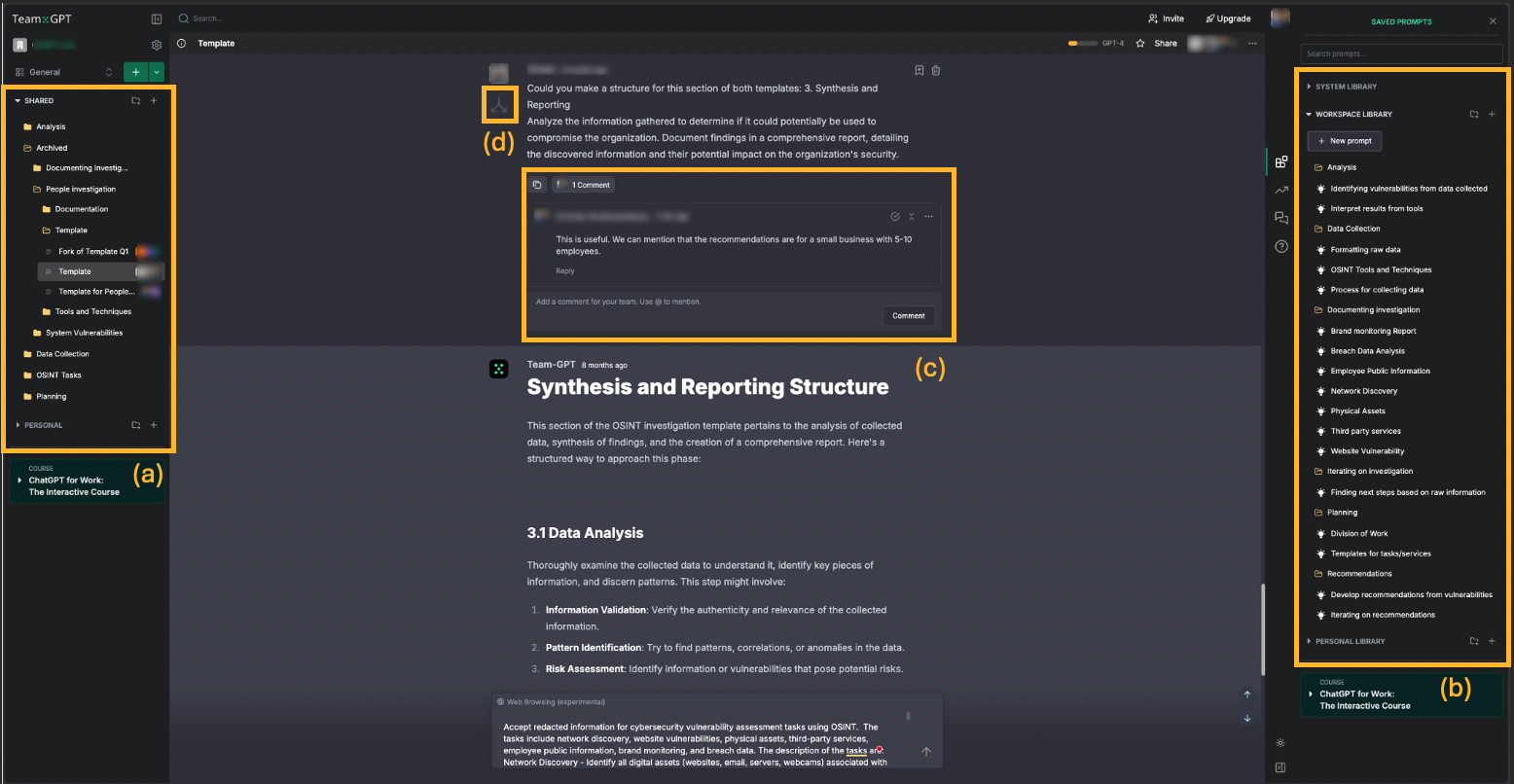}
\caption{Team-GPT platform \cite{team-gpt_enterprise_nodate} with core features: (a) personal chats for individual work and shared workspaces for team collaboration. Chats were grouped by the different phases of OSINT investigation in our case; (b) a team library to store shared prompt templates which can be invoked with a click after filling in the placeholders. We used the library to curate a set of useful prompts for the participants in Study 3; (c) directly comment on prompts and responses to provide feedback; (d) fork chats to new personal or shared chats to continue the conversation.}
\label{fig:teamgpt}
\Description{The image shows a screenshot of the Team-GPT interface, annotated to highlight its features. On the left side, there is a navigation sidebar (labeled "a") containing sections such as "Analysis," "Documenting Investigation," "Tools and Techniques," and other relevant categories to organize the workspace. Below this, a course area (labeled "b") offers an interactive course on using ChatGPT effectively in a professional setting. In the center, the document editing area (labeled "c") displays a section titled "Synthesis and Reporting Structure," which outlines steps for data analysis, synthesis of findings, and the creation of a comprehensive report. A comment feature (labeled "d") is visible, showing a comment bubble where a user has made a note regarding the report recommendations, suggesting modifications for small businesses with 5-10 employees. This description was generated using ChatGPT.}
\end{figure*}

As a follow-up study, we brainstormed design ideas around these two collaboration challenges. We conducted an online workshop (Design Workshop 5) to focus on collaboratively forming templates for data collection and dividing up work, with P1 serving as the team leader. We used Team-GPT \cite{team-gpt_enterprise_nodate}, a commercially available web-based platform that enables teams to collaboratively interact in real-time with AI models like ChatGPT, as a probe. Figure~\ref{fig:teamgpt} shows a screenshot of the Team-GPT workspace with its core features.

Effective leadership is crucial for guiding the team, assigning roles, and facilitating the efficient use of collaboration tools. P6 emphasized that having a leader helped streamline tasks and prevent overlap. We argue that while Team-GPT offers valuable collaboration features, integrating strong leadership could significantly improve the collaboration process. However, we observed during design workshop 5 and a team-based practice session, that assigned leaders found it overwhelming to help get the team started and keep track of all the information. To address the issue, we wanted students to take on the role of leaders and imagine how they would use ChatGPT to display task-oriented leadership behaviors during a 30-minute focus-group interview (Design Workshop 6) \cite{luther_redistributing_2013}.

For collaboration, we performed a deductive thematic analysis \cite{braun2006thematic} of the workshop 5 transcript based on (1) Team-GPT features used and (2) how they impacted collaboration.

For leadership, we performed deductive thematic analysis \cite{braun2006thematic} of the workshop 6 transcript based on the themes of 4 leadership behaviors (1) dividing and assigning tasks; (2) clarifying requirements by thinking about the end goal and client presentation; (3) monitoring progress and identifying areas to dig deeper; and (4) problem-solving by providing feedback and iterating on results.

\subsubsection{Results}

\paragraph{Collaboration} Participants found several of Team-GPT's collaboration features helpful. The shared workspace was particularly useful for visibility and access, allowing team members to build on each other’s contributions and avoid duplication. P4 mentioned, "I thought it was really useful for collaboration\ldots being able to just all contribute to one was really useful". However, when multiple active chats were created, it sometimes led to confusion and a lack of coordination, making it feel as though team members were working in silos. Similarly, excessive forking of chats, while flexible, resulted in a fragmented workflow that complicated team collaboration. To mitigate these challenges, participants suggested staying within the same chat more consistently and marking successful strategies within the chat for others to build upon.

Commenting on prompts and responses allowed for immediate feedback from team members, which helped refine outputs and guide the next steps based on that information. Participants found the prompt library useful for storing and reusing effective prompts but were skeptical about its practicality in diverse, case-specific scenarios. P5 said, "I'm not sure how often that would be used\ldots a lot of this is just a case-by-case basis". Participants mentioned that placeholders within prompt templates allowed for easy customization, which could increase the library's utility. 

\paragraph{Leadership} Participants suggested that AI could use team members' skills and past performance to propose optimal task assignments to leaders, but human judgment would still be crucial for considering team members' preferences. P6 thought AI-generated templates could help leaders ensure that the final outputs were thorough and aligned with client needs. P1 suggested AI could create monitoring frameworks, such as tables or templates with detailed timelines and task descriptions. Participants reiterated the need for real-time communication and human oversight along with AI for effective monitoring. 

AI could help identify gaps in the data for the leader, who can then follow up on it, as P6 stated, "I'm sure if you put like the info from a shared document to ChatGPT and ask like which areas need more research, it might tell you". 
As a leader, P1 wanted suggestions for visualization to enhance the presentation's impact. They also noted the importance of the human touch in finalizing decisions and being accountable for presenting results to clients. \newline

In conclusion, Study 2 explored how generative AI could address key challenges in OSINT investigations and developed design ideas. The findings show how AI supported learning and helped achieve the learning goals. Building on the insights from using it in different parts of the investigation, participants used generative AI in conducting OSINT-based vulnerability assessments on small businesses during Study 3.
\section{Study 3: Performing Real-World Investigations Augmented By Generative AI}
\label{study3}

\subsection{Method}
In this study, our six student participants of the OSINT Clinic offered free cyber vulnerability assessments to three small businesses in our region. These assessments provided small businesses with a view of their attack surface to address vulnerabilities before malicious actors could exploit them. Unlike a comprehensive vulnerability assessment, which might include control interviews with IT staff, physical facility tours and inspections, and internal network enumeration and scanning with tools like Nmap and Nessus / OpenVAS, ours were limited to vulnerabilities that could be detected via OSINT techniques.

\strike{At an initial meeting, the researchers described the process to the clients, set expectations around timelines and their involvement, and answered questions. Clients were asked to select from a menu of services based on the categories mentioned in Table~\ref{tab:osint-tasks-table}. After the meeting, if clients wanted to proceed, they signed a letter of agreement consenting to be investigated and provided some basic information about their digital infrastructure.}

\strike{As mentioned in Studies 1 and 2, participants had training and practice with OSINT tradecraft for over a semester, as well as experience using generative AI platforms like ChatGPT and Team-GPT. Participants worked together as a single team of six members for the first investigation, and then as teams of 2--3 members each (including a designated leader) for the next two, which ran in parallel. The teams spent around 2.5 weeks per investigation, from data collection and analysis to developing a presentation describing the vulnerabilities identified and making recommendations to mitigate them.}

\newtext{We structured the consultation process using the Understand-Investigate-Advise (UIA) framework introduced by Havron et al.\cite{havron_clinical_2019, freed2019my}. In the initial Understand phase, the researchers met with the clients to explain the consultation process, set expectations regarding timelines and involvement, and answer any questions. Clients were then asked to select relevant services from a menu of tasks based on the categories outlined in Table\ref{tab:osint-tasks-table}. Those who wished to proceed signed a letter of agreement consenting to the investigation and provided basic information about their digital infrastructure.}

\newtext{In the Investigate phase, participants drew on their semester-long training in OSINT tradecraft and their experience with generative AI platforms like ChatGPT and Team-GPT (as discussed in Studies 1 and 2). For the first investigation, all six participants collaborated as a single team. In subsequent investigations, they split into smaller teams of 2--3 members, each with a designated leader. Over approximately 2.5 weeks per investigation, the teams collected and analyzed data, identifying potential vulnerabilities in the clients’ digital systems.}

\newtext{Finally, during the Advise phase, the teams synthesized their findings into presentations. These presentations detailed the identified vulnerabilities and offered tailored recommendations for mitigation. By delivering clear and actionable advice, the teams ensured clients received practical strategies to improve their cybersecurity posture.}

We curated a prompt library on Team-GPT \cite{team-gpt_enterprise_nodate} based on highly-rated prompts from the previous study workshops across the different phases of the Intelligence Cycle. This library was provided to participants as an instrument to reuse the ChatGPT capabilities they found useful. The prompt templates allowed user customization for variables like the OSINT task type, output from tools, team size, and data collection details. We include these templates in Appendix~\ref{appendix_prompts} in Tables \ref{tab:prompts-table1}, \ref{tab:prompts-table2}, \ref{tab:prompts-table3}, and \ref{tab:prompts-table4}. In addition to the prompt library, we gave teams access to a separate group chat on Mattermost (an open-source Slack clone) and individual subscriptions to ChatGPT Plus. 

We followed best practices for security, privacy, and ethics established in previous research on student-led real-world OSINT investigations \cite{mukhopadhyay2024osint, clinic_commonly_2024}. The students signed Non-Disclosure Agreements (NDAs) to protect the confidentiality of their findings except for sharing with the customer. The students also enhanced their safety by taking operational security measures like using Virtual Private Networks (VPNs), Virtual Machines (VMs), role-based access control and encryption for storing and sharing data, and sock puppet accounts.  

We conducted two 30-minute focus group interviews, one after the first investigation and then after the final one. The goal was to understand the use of generative AI and document challenges faced during real-world investigations. We added a reflection survey for participants to rate their use of generative AI across the different phases of the investigations and for supporting leadership. We also distributed a survey to the small business clients asking them to rate the vulnerability assessment service in order to measure the real-world value of the service and how actionable the recommendations were.

In terms of analysis, first, we performed deductive thematic analysis \cite{braun2006thematic} on the focus group transcripts to understand (1) the use of generative AI, 
and (2) the challenges that emerged and how participants circumvented them. We also analyzed the shared chats on the Team-GPT workspace to monitor the use of the prompt library. Then, we aggregated the three client evaluations of vulnerability assessments to understand their utility.

\subsection{Results}
\label{study3-results}

\subsubsection{Utility of generative AI during investigations}

Participants employed AI to structure project plans, identify tasks, and allocate responsibilities. P4, the leader for the first investigation, used templates for streamlining brand monitoring and employee information management. This initial structuring helped teams set clear objectives and assign roles.

AI was particularly useful in identifying tools and generating templates to gather relevant information. P2 mentioned, ``I used the prompt library for generating report templates for brand monitoring and data breach exposure.'' These templates allowed team members to collect data in a structured manner, ensuring consistency and thoroughness across different team members' efforts. AI also assisted them in identifying gaps in data and making suggestions on how to add data from other sources for verification. 

Teams used AI to interpret data from tools and draft vulnerability mitigation strategies. They organized findings and added explanations, making it easier to present findings to non-technical stakeholders. P1 emphasized, "I used it a lot when it came to rephrasing my thoughts. So sometimes I would just dump all the information I got, and then I would say, hey, can you organize this in ChatGPT?". During dissemination, AI helped participants format their findings concisely, improve the written report, and create polished client presentations. 




There was limited usage of collaborative features of Team-GPT, as participants preferred to work separately on their personal ChatGPT instances. This was to avoid getting information ``clogged up'' on a shared workspace as P4 mentioned, "[Shared chat] helped to get started and you could take a look at others, but then you did not really want to go in and like be in a shared chat". Leaders accessed the prompt library on Team-GPT more than other members to create templates to guide the team's work and structure results for presentation.

\subsubsection{Challenges and limitations faced during investigations}

A significant challenge was the concern over privacy and data security. Participants were cautious about feeding sensitive information into AI tools, as there was uncertainty about how this data might be handled or stored. One participant reflected on this issue, stating, "We should keep this in mind while we are using ChatGPT\ldots we do not want to put in the raw data in the form that breaches the privacy of the company." The researchers provided a Python script that participants could run locally to scrub Personal Identifiable Information (PII), but it wasn’t used as it required extra setup. P4 and P5 navigated privacy concerns by avoiding including specific or sensitive data in their prompts, focusing instead on general scenarios. For example, P4 said, “like keeping it general enough, and not having specific information, but having, like the scenario where I asked, what are the possible cybersecurity risks knowing the family members of an owner of a company or high-level employee".

Generative AI in the form of a large language model like ChatGPT, showed limitations in data collection and analysis. A recurring theme in Study 2, participants again found that AI often provided broad, generalized responses that required further refinement, as P5 mentioned, "It tends to give broad responses and doesn’t always provide the specific details needed for verification". This may be partly due to the use of abstracted prompts to bypass privacy issues. Consequently, participants spent the majority of their investigation outside the Team-GPT or ChatGPT platforms, using external tools and searching websites or social media accounts.

Another challenge was the difficulty in monitoring team progress and maintaining coordination, especially when using AI in asynchronous collaboration settings. Participants rated monitoring progress as the leadership behavior where generative AI was the least useful. While AI tools were used to create structured templates and organize information, keeping track of who was doing what and ensuring that all tasks were aligned with the overall investigation goals was not straightforward.

\subsubsection{Usefulness of vulnerability assessments based on client survey}

Based on the survey responses, all clients valued the insights into their cybersecurity risks provided by the student teams from the OSINT clinic. The consistently highest possible ratings for the value and actionability of the assessments indicate that the clients found the services beneficial, not only for understanding their current vulnerabilities but also for guiding their future cybersecurity efforts. All three clients wanted to return for a similar exercise and recommended the program to other organizations.

One respondent noted that the information was "very actionable" and that they now have "a game plan to address the issues and mitigate the potential risks". Another client emphasized the usefulness of the report as they were in the growth phase and had already started implementing some of the recommendations. Table~\ref{tab:recommendations} shows some of the common vulnerabilities identified and recommendations made. 

\begin{table*}[h]
\caption{Common vulnerabilities identified and recommendations for mitigation}
\label{tab:recommendations}
\begin{tabular}{@{}p{0.45\linewidth}|p{0.45\linewidth}@{}}
\hline
\textbf{Common Vulnerabilities }                                                                    & \textbf{Recommendations }                                                                                                             \\ \hline
Out-of-date tech stacks like JavaScript libraries on client websites                       & Regularly update software versions to avoid potential security risks                                                         \\ \hline
Company social media reveals questionable personal data or connections                     & Separate personal and business social media accounts to maintain professionalism and protect privacy                         \\ \hline
Employee emails found in data breach                                                       & Regularly change passwords and use a password manager to prevent unauthorized access through compromised emails              \\ \hline
Personal home address registered as business address can be easily found in public records & Register your business at a P.O. box, or use virtual office space for registration                                           \\ \hline
Third-party services have vulnerabilities and predatory policies                           & Switch to more secure alternatives and thoroughly read any agreement before signing and research their reputation beforehand \\ \hline
\end{tabular}%
\end{table*}

\section{Discussion}

\subsection{Reflection on Co-designing via Matchmaking for AI}
\label{reflection-MMFAI}
\subsubsection{Adapting the method for OSINT investigations}

We adapted and extended the Matchmaking for AI method \cite{liu2024humancenterednlpfactcheckingcodesigning} for an OSINT context to structure design workshops and generate innovative yet feasible ideas. We followed the three key steps of the original method. First, for mapping stakeholder domain expertise and AI needs, we focused on challenges and learning goals across the OSINT Intelligence Model for OSINT investigations. For Step 2, playing with AI probes to make participants familiar with ``affordances of the technology,'' we used ChatGPT and Team-GPT as probes. For the third step, brainstorming AI design elements, we identified ways to tailor general-purpose LLMs in developing and applying OSINT skills. 

We also extended this method by addressing two limitations revealed by our study context. First, our brainstorming sessions in a group setting, with a mix of co-located and online setups alleviated the limits of interpersonal engagement and creative expression among participants in a remote, individual study setting. Second, we used ChatGPT and Team-GPT as functional technology probes \cite{hutchinson_technology_2003}, compared to Wizard-of-Oz simulations, to help participants envision more diverse and concrete AI solutions.

\subsubsection{Extending the method for collaboration and learning}

Our goal was to support investigations of student analysts, as they are essential for the cybersecurity clinic model \cite{consortium_nodate}. Their needs, expertise, and skill levels differ significantly from expert practitioners, who are the primary stakeholders for previous matchmaking studies \cite{liu2024humancenterednlpfactcheckingcodesigning, Visser2005ContextmappingEF}. Our adaptations of the Matchmaking for AI method cater to learners and the clinic model, characterized by achieving learning goals and performing collaborative real-world investigations. 

Chin Jr. et al. highlight collaborative analysis as a critical approach for intelligence analysis, and we report on its evolution over time by observing AI interactions and teamwork across workshops and real investigations \cite{chin2009exploring}. In Study 2 (Section~\ref{collab-study2}), we investigate new challenges and design ideas that emerge as the team collaborates and incorporates generative AI tools over time. Our longitudinal and cyclical study design promotes fluidity between the phases of co-design and helps to study collaboration.

As Stahl suggests, understanding groupware requires groups to use it in real-world conditions over time \cite{stahl2005groups}. In Study 3 (Section~\ref{study3}), we enhance the matchmaking process by validating design ideas in a natural setting. Team-GPT along with other content collaboration tools like Google Docs and Mattermost serve as a research tool to observe team behaviors and use of generative AI in ongoing investigations. The insights gained from real-world investigations help prioritize the design ideas, providing a deeper understanding of how to integrate AI tools effectively into collaborative OSINT practices. Our study goes beyond the typical use of AI for data collection and analysis \cite{browne_systematic_2024}, focusing on how student teams learn new skills and close skill gaps. Our study design refines the matchmaking process to support collaboration and real-world needs, making it more flexible and useful for areas like crowdsourcing complex tasks \cite{venkatagiri2021crowdsolve, retelny2017noworkflow} and co-creation \cite{briggs2020brainstorming}.

{\strike{\textbf{Recommendations for streamlining collaborative OSINT investigations}}

\strike{\subsubsection{Summary of co-designed ideas}}

\strike{OSINT investigators work with limited resources, in terms of funds, time, and training or software skills to improve their workflows \cite{glassman2012intelligence, pastor2020not}. In our work, we bridge the translational gap in HCI research by describing how to leverage generative AI in OSINT-based vulnerability assessments. The learning goals (Table~\ref{tab:challenges-table}) reflect desired outcomes and the design ideas for achieving them can be useful for OSINT practitioners. The challenges from the perspective of students will resonate with novices in the OSINT community, which has been characterized as a community of practice \cite{belghith_compete_2022}. Prior work has described multiple roles played by ChatGPT in the OSINT workflows of experts: (1) research assistance, (2) efficiencies, and (3) technical enablement \cite{poulter_osint_2023, flashpoint_generative_2024}. Though these findings are relevant, our characterization is based on the particular needs of amateurs as they learn OSINT skills, use generative AI to support their collaborative investigations, and deliver high-quality results to clients. Overall, generative AI was useful in training and skill enhancement, scaffolding investigations, and reporting on findings.}

\strike{\paragraph{Training and skill enhancement}}

\strike{During OSINT investigations, participants used generative AI to understand ethical guidelines and create sock puppet accounts (LG1). They overcame skill gaps in data collection through guidance on effective tools, such as Google dorking and identifying breached data sources (LG3). ChatGPT also supported verification efforts by suggesting alternate sources to participants for cross-referencing information (LG6). AI-assisted interpretation of tool outputs enabled participants to better identify vulnerabilities, enhancing their analytical skills (LG7). Similar benefits like getting new insights and unstuck \cite{tholander_design_2023} have been demonstrated in applications of group ideation \cite{he_ai_2024}, coding \cite{peng_impact_2023}, and general creativity \cite{doshi_generative_2023}.}

\strike{\paragraph{Scaffolding investigations}}

\strike{ChatGPT helped teams get started and stay organized. It helped participants break down complex OSINT tasks into specific subtasks (LG2) and divide them based on preferences. ChatGPT-generated templates were used to coordinate data collection efforts, reducing redundant work and enhancing communication among team members (LG4). These results contribute to research on using AI to structure complex tasks \cite{bhattacharjee_understanding_2023, subramonyam_bridging_2023}.}

\strike{\paragraph{Reporting on findings}}

\strike{Participants used AI probes to structure raw data into tables and lists, making it easier for team members to follow (LG5). The documentation helped present supporting information for findings in the report. AI was also used to generate actionable recommendations based on identified vulnerabilities, tailoring advice to specific client needs (LG8). Participants iteratively refined their results by simulating client interactions to ensure that recommendations were clear and relevant (LG9). These improvements demonstrate the role of generative AI in the articulation work involved in sensemaking tasks \cite{schmidt_cooperative_2013}.}

\strike{\subsubsection{Collaboration and leadership on generative AI platforms}}

\strike{We found that Team-GPT supported three different types of collaboration --- sharing, functional, and content --- first mentioned in Kang et al.'s longitudinal study on collaboration in intelligence analysis \cite{kang_characterizing_2014}. The shared workspace in Team-GPT made successful use of AI visible and helped avoid duplication. Functional collaboration was enhanced through features like commenting and forking.}
\strike{Content collaboration was supported by the prompt library and saved chats, which helped create a repository of useful prompts that could be adapted to various contexts. These features can be explored further in the context of team-AI collaboration across other domains \cite{hendriks_exploring_2024}.}

\strike{Leadership behaviors like planning and problem-solving were supported by ChatGPT. Leaders frequently used the prompt library on Team-GPT to create templates that guided the team's work and structured their presentations. Participants emphasized that AI should complement, not replace, direct team interactions to maintain effective collaboration, aligning with findings from previous studies \cite{kim_ai_2021, he_ai_2024}.}


\subsection{\newtext{Role of Generative AI in Learning and Collaboration for Investigations}} 

\newtext{Studies 1 and 2 helped identify and achieve a set of learning goals (Table~\ref{tab:challenges-table}) for performing vulnerability assessments as a part of the OSINT Clinic. Generative AI, particularly ChatGPT, was instrumental in helping students plan investigations (LG2) and select effective OSINT tools (LG3). For example, AI simplified the process of breaking down complex tasks into structured subtasks tailored to the team’s capabilities. This scaffolding allowed students to navigate the vast array of OSINT techniques without feeling overwhelmed, supporting novice learners in developing systematic approaches to investigations. These results contribute to research on using AI to structure complex tasks \cite{bhattacharjee_understanding_2023, subramonyam_bridging_2023}.}

\newtext{In terms of data collection and verification (LG4, LG6), AI-generated templates provided consistent structures for organizing collected data, reducing redundancy, and enhancing collaboration. AI’s ability to suggest alternative sources for verification encouraged students to adopt rigorous cross-referencing practices. However, the AI’s generalization tendencies and outdated information posed challenges, requiring students to manually verify AI-suggested tools and sources. AI-assisted interpretation of tool outputs enabled participants to better identify vulnerabilities, enhancing their analytical skills (LG7). By simplifying technical findings into clear, understandable language, AI helped students translate complex data into actionable insights. This was particularly beneficial in learning how to present findings to non-technical clients (LG9). Similar benefits like getting new insights and support \cite{tholander_design_2023} have been demonstrated in applications of group ideation \cite{he_ai_2024}, coding \cite{peng_impact_2023}, and general creativity \cite{doshi_generative_2023}.}

\newtext{We found that Team-GPT supported sharing and content collaboration, which are collaborative behaviors first identified in Kang et al.'s longitudinal study of teams of intelligence analysts~\cite{kang_characterizing_2014}. Teams used the shared workspace to finalize templates and structure workflows, which minimized duplication of effort and streamlined task division. Content collaboration was supported by the prompt library and saved chats, and led to the development of a repository of useful prompts (see Appendix~\ref{appendix_prompts}). However, challenges emerged around maintaining consistency in formatting and style across different team members' outputs. This required students to develop stronger communication and coordination skills, reflecting real-world team dynamics in cybersecurity investigations \cite{riebe_values_2023}.}

\newtext{While generative AI supported the development of OSINT skills and collaboration, challenges such as privacy concerns, generalization, and inconsistent outputs underscore the need for careful integration. The students’ ability to critically engage with AI, verify its suggestions, and adapt their workflows reflects a key outcome: they not only learned OSINT techniques but also developed meta-skills in evaluating AI outputs and incorporating them responsibly.}


\strike{\textbf{Design requirements for generative AI-powered systems for supporting OSINT investigations}}

\strike{Previous research has highlighted several challenges in using generative AI, such as misalignments in understanding its capabilities \cite{subramonyam_bridging_2023, tankelevitch_metacognitive_2024}, prompting techniques \cite{zamfirescu-pereira_why_2023, subramonyam_bridging_2023}, and evaluating its outputs \cite{kulkarni_word_2023, subramonyam_bridging_2023}. Chris Poulter's blog discusses issues with ChatGPT for OSINT investigations, including operational security, data retention, storage, accuracy, and sourcing \cite{poulter_osint_2023}. Building on this research, we aimed to identify outstanding challenges within an AI-augmented workflow. Some of the challenges mentioned in the results have been alleviated with recent updates in ChatGPT like providing links to websites as references, formatting content into tables, and generating downloadable files. Here, we outline the key difficulties investigators faced in developing skills and conducting real-world investigations and offer implications for human-AI collaboration research to better adapt foundational models to OSINT-specific applications.}

\strike{\subsubsection{Strengthening privacy measures}}

\strike{Participants were cautious about sharing potentially sensitive public information about companies due to unclear data practices on platforms like ChatGPT and Team-GPT. To mitigate these concerns, they enabled private search \cite{openAI_data_nodate} and made general, scenario-based queries without specifics. OSINT experts have suggested that small, fine-tuned models that can run locally are a promising alternative, given their growing capabilities \cite{gururangan_dont_2020, privategpt_nodate}. Additionally, software systems can enhance security by adding checkpoints before sending information to models via APIs. For instance, we experimented with running a PII scrubber script locally to sanitize documents containing collected data before using them with ChatGPT.}

\strike{\subsubsection{Integrating existing tools for data collection}}

\strike{There is a lack of support for collecting OSINT information with general-purpose models, even though this is the most time-consuming part of the investigation. Agentic workflows could help automate data collection by orchestrating the use of reputable tools \cite{noauthor_openaiassistants_nodate, dennis_ai_2023, flathmann_fostering_2021}. These workflows can follow a chain-of-thought reasoning process to document the data collection and analysis based on examples of successful investigations \cite{jin_learn_2022}. Based on design guidelines, improving trust in results can be achieved by integrating well-known and reliable tools \cite{weisz_design_2024}. With trustworthy AI workflows in place, analysts can focus on verifying information and identifying areas that require deeper investigation.}

\strike{\subsubsection{Improving control over model output}}

\strike{Participants faced challenges with the overwhelming amount of information generated by GPT models, making it difficult to identify the most relevant insights. They highlighted the need for greater control over the output, such as filtering options and creating data visualizations. These features relate to the need for human control over AI-infused systems \cite{shneiderman_ensuring_nodate, yin_can_2024}. Models can be fine-tuned and augmented with knowledge bases to provide the required trustworthy and updated references while suggesting tools and creating templates \cite{gururangan_dont_2020, lewis_retrieval-augmented_2020}.} 
 
\strike{\subsubsection{Developing prompt templates collaboratively}}

\strike{Participants found it challenging to create effective prompts and wanted to learn from others’ successful examples but struggled to track them. They highlighted the difficulty in achieving consistent results due to significant changes in response with small changes in prompts \cite{jiang_promptmaker_2022}. Identifying reusable prompts was crucial for enhancing the efficiency of their work and getting around "guardrails" imposed by ChatGPT. Maintaining awareness of how AI is being used and helping team members leverage the output correctly is vital for teamwork \cite{porsdam_mann_generative_2023}. To enable better team awareness while using generative AI tools, there needs to be improved documentation of prompts, their context, and sources of any external information used.}

\strike{\subsubsection{Enhancing leadership support}}

\strike{Effective leadership was crucial for managing investigations and breaking down information silos that formed around individual tasks.} 
\strike{To facilitate this, tools should support summarizing ongoing work, identifying areas where help is needed like additional verification, and offering solutions \cite{hepenstal2021analysis}. This could involve prompting other team members or automatically generating suggestions based on OSINT best practices and guidelines. Personalization based on user behavior and expertise is expected to improve collaborative processes such as planning and monitoring progress, by leveraging the unique perspectives of team members \cite{wang_towards_2021, bhattacharjee_understanding_2023}.}


\subsection{\newtext{Design Considerations for Using Generative AI in Real-World Investigations}}

\newtext{During the real-world investigations conducted in Study 3, generative AI tools like ChatGPT and Team-GPT played a supportive role across multiple phases of the Intelligence Cycle. Students used generative AI to create templates for structuring data collection, generate actionable recommendations for clients, and simplify technical findings into more digestible reports. Leaders in the team relied on AI to break down complex tasks, assign subtasks, and scaffold the investigation process, ensuring the team maintained a structured workflow. Generative AI also supported the verification process by suggesting alternative data sources, enhancing the rigor of the investigations. However, participants used these tools selectively, often choosing to work independently with OSINT tools and personal ChatGPT instances to maintain efficiency.} 

\newtext{The results underscored the need for balancing AI automation with human judgment, aligned with findings from previous studies \cite{kim_ai_2021, he_ai_2024}. While generative AI can handle routine tasks like template generation and summarization, more nuanced activities — such as verifying sensitive information or providing context-aware leadership — still require human expertise. These results build on previous research that has highlighted several challenges in using generative AI, such as misalignments in understanding its capabilities \cite{subramonyam_bridging_2023, tankelevitch_metacognitive_2024}, prompting techniques \cite{zamfirescu-pereira_why_2023, subramonyam_bridging_2023}, and evaluating its outputs \cite{kulkarni_word_2023, subramonyam_bridging_2023}. Integrating generative AI into real-world OSINT workflows revealed design considerations related to privacy, workflow integration, and collaboration. We combined participants' design ideas with recent technological advancements to propose future directions and mitigation strategies across these challenges.}

\subsubsection{\newtext{Privacy Challenges}}

\newtext{Privacy concerns were a significant limitation in fully leveraging generative AI tools like ChatGPT and Team-GPT. Although clients did not specifically mention concerns about AI tools or privacy, our NDA explicitly ensured that all data would remain within the clinic and not be shared externally. Participants were cautious about feeding sensitive client data into AI platforms due to uncertainty about how the data might be stored or used by these tools. This caution was more pronounced during the later stages of the investigations, particularly when analysis and dissemination involved sensitive and identifiable information. Even though participants attempted to anonymize data or use generalized scenarios, the fear of exposing confidential information remained prevalent. While a Python script was provided for this purpose, the extra setup was a barrier to its adoption.}

\newtext{The reliance on commercially available, cloud-based AI platforms creates a tension between efficiency and privacy. This issue becomes more critical with OSINT clinics as students may be more prone to taking "shortcuts" or inadvertently breaching confidentiality. This tension highlights the need for privacy-preserving AI frameworks that can support OSINT investigations without risking client confidentiality. For instance, AI models that run locally on secure machines, rather than cloud-based platforms, could allow for sensitive data processing without risking data leakage \cite{gururangan_dont_2020, privategpt_nodate}. Features that let users toggle privacy modes, disable chat history, or automatically redact sensitive information can help build trust in these tools. Educational guidelines and training for students on responsible AI use can reduce the likelihood of shortcuts that compromise privacy.}

\subsubsection{\newtext{Integration with Existing Workflows}}

\newtext{Integrating generative AI into the existing workflows of student analysts presented practical challenges. There is a lack of support for collecting OSINT information with general-purpose models, even though this is the most time-consuming part of the investigation. Analysts also faced workflow disruptions due to context-switching between multiple tools and platforms (ChatGPT, Team-GPT, Google Docs, Mattermost, browser, command line tools).} 

\newtext{To address this, future design efforts should focus on creating seamless integrations between generative AI tools and commonly used collaboration platforms. Features like embedded AI prompts within investigation tools or plugin-based extensions could streamline workflows and reduce context-switching. Agentic workflows could help automate data collection by orchestrating the use of reputable tools \cite{noauthor_openaiassistants_nodate, dennis_ai_2023, flathmann_fostering_2021}. Trust in results can be achieved by integrating well-known and reliable OSINT tools \cite{weisz_design_2024, jin_learn_2022}.}

\subsubsection{\newtext{Collaboration and Leadership Issues}}

\newtext{Generative AI tools introduced both opportunities and challenges for collaboration and leadership within the student teams. Team-GPT features facilitated initial brainstorming and task organization but led to fragmented workflows when used excessively. Forking conversations and managing multiple active chats often created silos, reducing the efficiency of team coordination. As a result, participants preferred to use personal ChatGPT instances to maintain focus and avoid cluttered workspaces.}

\newtext{Effective leadership was crucial in mitigating these challenges. Leaders used AI to structure tasks and generate templates, but monitoring team progress remained difficult. Here, AI can assist leaders by suggesting task assignments based on team members’ skills and past performance \cite{wang_towards_2021, bhattacharjee_understanding_2023}. Features that maintain group awareness, for example, real-time dashboards showing team progress, task ownership, and AI interactions could help leaders and members stay synchronized \cite{porsdam_mann_generative_2023}.}

\newtext{Moreover, while AI facilitated some aspects of task-oriented leadership, it lacked the contextual understanding needed for decision-making. This reinforces the importance of human oversight and interpersonal skills in leadership roles, suggesting that AI should be seen as a complement rather than a replacement for leadership behaviors \cite{shneiderman_ensuring_nodate, yin_can_2024}.}

\subsection{\newtext{Limitations and Future Work}}

\newtext{A key limitation of this study is the small participant size of six undergraduate students, which may limit the generalizability of the results. However, this is consistent with prior HCI and intelligence analysis research, where studies often rely on small groups due to the sensitive nature of the work \cite{hepenstal2021analysis, riebe_values_2023, chin2009exploring, kang_characterizing_2014, hayes_open-source_2018, gerber_how_2016, wong_how_2015}. This challenge reflects broader issues in accessing cybersecurity professionals for research. As Sundaramurthy et al. found, barriers such as intensive workloads, sensitive work environments, and mistrust of outsiders make it difficult to recruit larger, more diverse groups \cite{sundaramurthy_anthropological_2014}.}

\newtext{Another limitation is that the findings are primarily reflective of a small group with limited professional experience. While only half of our student participants were computer science majors, the group represented diverse academic backgrounds, including business and national security majors. Despite this limitation, the insights from this group remain valuable. The participants’ status as novice OSINT analysts mirrors the demographics of many cybersecurity clinics and educational programs, where students are the primary workforce. Their challenges, skill gaps, and learning goals offer a critical lens for understanding how generative AI can support early-stage analysts, who represent the future of the OSINT and cybersecurity workforce. Addressing the needs of novices can lead to the development of AI tools that are accessible, educational, and scalable for broader use in training environments and real-world applications.}

\newtext{The current study provides initial findings on the integration of generative AI in OSINT clinics. The developed prompt library, along with examples of generative AI use across the different phases of OSINT investigations, offers a starting point for refining design ideas in future research. Expanding this work to include larger and more varied participant pools—such as students from diverse backgrounds, professional analysts, or interdisciplinary teams -- will help assess the broader applicability and effectiveness of these AI-driven approaches. Including students from various disciplines (e.g., social sciences, law, journalism) could provide a more holistic understanding of how different backgrounds influence the adoption and effectiveness of AI tools in OSINT. Additionally, longitudinal studies that follow participants into professional settings could reveal how AI-supported skills translate to real-world practice.}


\section{Conclusion}

This paper introduced the OSINT clinic, which was aimed at providing students with training and practical experience in conducting cybersecurity vulnerability assessments using publicly available data, addressing the scalability challenge of helping small businesses protect themselves against cyber attacks. Through a three-phase co-design study, we identified key challenges, such as information management, skill gaps, and collaboration inefficiencies. We developed design ideas to address these gaps by integrating generative AI into various stages of the OSINT investigation process, including planning, data collection, processing, analysis, and dissemination. The design ideas can benefit novices and experts to leverage large language models like ChatGPT and collaborative AI platforms like Team-GPT in their workflows. A pilot with three small businesses demonstrated the benefits of using generative AI to streamline OSINT investigations, as clients found value in the resulting vulnerability assessment reports. It also revealed challenges, including privacy concerns and difficulties in monitoring progress. We propose future research on human-AI collaboration for OSINT tasks, informed by our design requirements and recommendations.

\bibliographystyle{ACM-Reference-Format}
\bibliography{sample-base,discussion,Co-design_methods,Vulenrability_Assessment_OSINT,Opportunities_AI_OSINT,introduction,sample-base-ORS,Teamwork_Leadership, OSINT-tools-websites, discussion2, Revision}


\begin{thebibliography}{114}


\ifx \showCODEN    \undefined \def \showCODEN     #1{\unskip}     \fi
\ifx \showDOI      \undefined \def \showDOI       #1{#1}\fi
\ifx \showISBNx    \undefined \def \showISBNx     #1{\unskip}     \fi
\ifx \showISBNxiii \undefined \def \showISBNxiii  #1{\unskip}     \fi
\ifx \showISSN     \undefined \def \showISSN      #1{\unskip}     \fi
\ifx \showLCCN     \undefined \def \showLCCN      #1{\unskip}     \fi
\ifx \shownote     \undefined \def \shownote      #1{#1}          \fi
\ifx \showarticletitle \undefined \def \showarticletitle #1{#1}   \fi
\ifx \showURL      \undefined \def \showURL       {\relax}        \fi
\providecommand\bibfield[2]{#2}
\providecommand\bibinfo[2]{#2}
\providecommand\natexlab[1]{#1}
\providecommand\showeprint[2][]{arXiv:#2}

\bibitem[noa(2020)]%
        {noauthor_nist_2020}
 \bibinfo{year}{2020}\natexlab{}.
\newblock \showarticletitle{{NIST} {SP} 800-115}.
\newblock \bibinfo{journal}{\emph{NIST}} (\bibinfo{date}{Jan.} \bibinfo{year}{2020}).
\newblock
\urldef\tempurl%
\url{https://www.nist.gov/privacy-framework/nist-sp-800-115}
\showURL{%
Retrieved 2024-09-13 from \tempurl}
\newblock
\shownote{Last Modified: 2021-04-23T09:13-04:00}.


\bibitem[HRC(2021)]%
        {HRC_Berkeley}
 \bibinfo{year}{2021}\natexlab{}.
\newblock \bibinfo{booktitle}{\emph{J298 OSINT Seminar --- Open Source Investigations}}.
\newblock
\urldef\tempurl%
\url{https://journalism.berkeley.edu/course-section/j298-human-rights-center-seminar-f21/}
\showURL{%
Retrieved 2024-09-10 from \tempurl}


\bibitem[new(2022)]%
        {news}
 \bibinfo{year}{2022}\natexlab{}.
\newblock \bibinfo{title}{Join us in pushing back on misinformation.}
\newblock
\newblock
\urldef\tempurl%
\url{https://web.archive.org/web/20200602035809/https://our.news/}
\showURL{%
Retrieved 2024-09-10 from \tempurl}


\bibitem[ver(2024)]%
        {verizon_2019_nodate}
 \bibinfo{year}{2024}\natexlab{}.
\newblock \bibinfo{title}{2019 {Data Breach Investigations Report} {Summary} of {Findings}}.
\newblock
\newblock
\urldef\tempurl%
\url{https://enterprise.verizon.com/resources/reports/dbir/2019/summary-of-findings/}
\showURL{%
Retrieved 2024-09-08 from \tempurl}


\bibitem[cit(2024)]%
        {citizen_lab_about_nodate}
 \bibinfo{year}{2024}\natexlab{}.
\newblock \bibinfo{booktitle}{\emph{About the {Citizen} {Lab}}}.
\newblock \bibinfo{type}{{T}echnical {R}eport}. \bibinfo{institution}{Citizen Lab, University of Toronto}.
\newblock
\urldef\tempurl%
\url{https://citizenlab.ca/about/}
\showURL{%
Retrieved 2024-12-10 from \tempurl}
\newblock
\shownote{Section: Uncategorized}.


\bibitem[bui(2024)]%
        {builtwith_2024_nodate}
 \bibinfo{year}{2024}\natexlab{}.
\newblock \bibinfo{title}{{BuiltWith}}.
\newblock
\newblock
\urldef\tempurl%
\url{https://builtwith.com/}
\showURL{%
Retrieved 2024-09-08 from \tempurl}


\bibitem[cen(2024)]%
        {censys_exposure_nodate}
 \bibinfo{year}{2024}\natexlab{}.
\newblock \bibinfo{title}{Censys: Exposure {Management} and {Threat} {Hunting} {Solutions}}.
\newblock
\newblock
\urldef\tempurl%
\url{https://censys.com/}
\showURL{%
Retrieved 2024-09-08 from \tempurl}


\bibitem[cis(2024)]%
        {cis_nodate}
 \bibinfo{year}{2024}\natexlab{}.
\newblock \bibinfo{title}{{CIS} {Controls} {Navigator} v8.1}.
\newblock
\newblock
\urldef\tempurl%
\url{https://www.cisecurity.org/controls/cis-controls-navigator/}
\showURL{%
Retrieved 2024-09-08 from \tempurl}


\bibitem[cli(2024a)]%
        {clinic_commonly_2024}
 \bibinfo{year}{2024}\natexlab{a}.
\newblock \bibinfo{title}{Commonly {Asked} {Questions} {About} {Cybersecurity} {Clinics} – {Consortium} of {Cybersecurity} {Clinics}}.
\newblock
\newblock
\urldef\tempurl%
\url{https://cybersecurityclinics.org/blog/commonly-asked-questions-about-cybersecurity-clinics/}
\showURL{%
Retrieved 2024-09-12 from \tempurl}


\bibitem[con(2024)]%
        {consortium_nodate}
 \bibinfo{year}{2024}\natexlab{}.
\newblock \bibinfo{title}{Consortium of {Cybersecurity} {Clinics}}.
\newblock
\newblock
\urldef\tempurl%
\url{https://cybersecurityclinics.org/}
\showURL{%
Retrieved 2024-09-07 from \tempurl}


\bibitem[goo(2024)]%
        {googledorking_nodate}
 \bibinfo{year}{2024}\natexlab{}.
\newblock \bibinfo{title}{Google {Dorking}: {An} {Introduction} for {Cybersecurity} {Professionals}}.
\newblock
\newblock
\urldef\tempurl%
\url{https://www.splunk.com/en_us/blog/learn/google-dorking.html}
\showURL{%
Retrieved 2024-09-08 from \tempurl}


\bibitem[hav(2024)]%
        {haveibeenpwned_nodate}
 \bibinfo{year}{2024}\natexlab{}.
\newblock \bibinfo{title}{Have {I} {Been} {Pwned}: {Check} if your email has been compromised in a data breach}.
\newblock
\newblock
\urldef\tempurl%
\url{https://haveibeenpwned.com/}
\showURL{%
Retrieved 2024-09-08 from \tempurl}


\bibitem[noa(2024a)]%
        {noauthor_intelligence_nodate}
 \bibinfo{year}{2024}\natexlab{a}.
\newblock \bibinfo{title}{Intelligence {X}}.
\newblock
\newblock
\urldef\tempurl%
\url{https://intelx.io/}
\showURL{%
Retrieved 2024-09-11 from \tempurl}


\bibitem[cha(2024)]%
        {chatgpt_plus}
 \bibinfo{year}{2024}\natexlab{}.
\newblock \bibinfo{title}{Introducing ChatGPT Plus}.
\newblock
\newblock
\urldef\tempurl%
\url{https://openai.com/index/chatgpt-plus/}
\showURL{%
Retrieved 10-09-2024 from \tempurl}


\bibitem[nes(2024)]%
        {nessus_nodate}
 \bibinfo{year}{2024}\natexlab{}.
\newblock \bibinfo{title}{Nessus {Vulnerability} {Scanner}: {Network} {Security} {Solution} {\textbar} {Tenable}®}.
\newblock
\newblock
\urldef\tempurl%
\url{https://www.tenable.com/products/nessus}
\showURL{%
Retrieved 2024-09-08 from \tempurl}


\bibitem[nma(2024)]%
        {nmap_nodate}
 \bibinfo{year}{2024}\natexlab{}.
\newblock \bibinfo{title}{Nmap: the {Network} {Mapper} - {Free} {Security} {Scanner}}.
\newblock
\newblock
\urldef\tempurl%
\url{https://nmap.org/}
\showURL{%
Retrieved 2024-09-08 from \tempurl}


\bibitem[noa(2024b)]%
        {noauthor_openaiassistants_nodate}
 \bibinfo{year}{2024}\natexlab{b}.
\newblock \bibinfo{title}{{OpenAI} {Platform}}.
\newblock
\newblock
\urldef\tempurl%
\url{https://platform.openai.com}
\showURL{%
Retrieved 2024-09-11 from \tempurl}


\bibitem[pri(2024)]%
        {privategpt_nodate}
 \bibinfo{year}{2024}\natexlab{}.
\newblock \bibinfo{title}{{PrivateGPT}: {Use} {Cases} in {OSINT}}.
\newblock
\newblock
\urldef\tempurl%
\url{https://www.osintcombine.com/post/privategpt-use-cases-in-osint}
\showURL{%
Retrieved 2024-09-07 from \tempurl}


\bibitem[noa(2024c)]%
        {noauthor_shodan_nodate}
 \bibinfo{year}{2024}\natexlab{c}.
\newblock \bibinfo{title}{Shodan}.
\newblock
\newblock
\urldef\tempurl%
\url{https://www.shodan.io}
\showURL{%
Retrieved 2024-09-08 from \tempurl}


\bibitem[cli(2024b)]%
        {clinic_uc_nodate}
 \bibinfo{year}{2024}\natexlab{b}.
\newblock \bibinfo{title}{{UC} {Berkeley} {Cybersecurity} {Clinic}}.
\newblock
\newblock
\urldef\tempurl%
\url{https://cltc.berkeley.edu/program/cybersecurity-clinic/}
\showURL{%
Retrieved 2024-12-10 from \tempurl}


\bibitem[who(2024)]%
        {whoiscom_nodate}
 \bibinfo{year}{2024}\natexlab{}.
\newblock \bibinfo{title}{Whois.com - {Free} {Whois} {Lookup}}.
\newblock
\newblock
\urldef\tempurl%
\url{https://www.whois.com/whois/}
\showURL{%
Retrieved 2024-09-08 from \tempurl}


\bibitem[Ackerman(2000)]%
        {ackerman2000intellectual}
\bibfield{author}{\bibinfo{person}{Mark~S Ackerman}.} \bibinfo{year}{2000}\natexlab{}.
\newblock \showarticletitle{The intellectual challenge of CSCW: the gap between social requirements and technical feasibility}.
\newblock \bibinfo{journal}{\emph{Human--Computer Interaction}} \bibinfo{volume}{15}, \bibinfo{number}{2-3} (\bibinfo{year}{2000}), \bibinfo{pages}{179--203}.
\newblock


\bibitem[Akhgar et~al\mbox{.}(2017)]%
        {akhgar2017open}
\bibfield{author}{\bibinfo{person}{Babak Akhgar}, \bibinfo{person}{P~Saskia Bayerl}, {and} \bibinfo{person}{Fraser Sampson}.} \bibinfo{year}{2017}\natexlab{}.
\newblock \bibinfo{booktitle}{\emph{Open source intelligence investigation: from strategy to implementation}}.
\newblock \bibinfo{publisher}{Springer}.
\newblock


\bibitem[Alcaidinho et~al\mbox{.}(2017)]%
        {alcaidinho_2017_investigation}
\bibfield{author}{\bibinfo{person}{Joelle Alcaidinho}, \bibinfo{person}{Larry Freil}, \bibinfo{person}{Taylor Kelly}, \bibinfo{person}{Kayla Marland}, \bibinfo{person}{Chunhui Wu}, \bibinfo{person}{Bradley Wittenbrook}, \bibinfo{person}{Giancarlo Valentin}, {and} \bibinfo{person}{Melody Jackson}.} \bibinfo{year}{2017}\natexlab{}.
\newblock \showarticletitle{Mobile Collaboration for Human and Canine Police Explosive Detection Teams}. In \bibinfo{booktitle}{\emph{Proceedings of the 2017 ACM Conference on Computer Supported Cooperative Work and Social Computing}} (Portland, Oregon, USA) \emph{(\bibinfo{series}{CSCW '17})}. \bibinfo{publisher}{Association for Computing Machinery}, \bibinfo{address}{New York, NY, USA}, \bibinfo{pages}{925–933}.
\newblock
\showISBNx{9781450343350}
\urldef\tempurl%
\url{https://doi.org/10.1145/2998181.2998271}
\showDOI{\tempurl}


\bibitem[Alharthi et~al\mbox{.}(2021)]%
        {alharthi_2021}
\bibfield{author}{\bibinfo{person}{Sultan~A. Alharthi}, \bibinfo{person}{Nicolas~James LaLone}, \bibinfo{person}{Hitesh~Nidhi Sharma}, \bibinfo{person}{Igor Dolgov}, {and} \bibinfo{person}{Z~O. Toups}.} \bibinfo{year}{2021}\natexlab{}.
\newblock \showarticletitle{An Activity Theory Analysis of Search and; Rescue Collective Sensemaking and Planning Practices}. In \bibinfo{booktitle}{\emph{Proceedings of the 2021 CHI Conference on Human Factors in Computing Systems}} (Yokohama, Japan) \emph{(\bibinfo{series}{CHI '21})}. \bibinfo{publisher}{Association for Computing Machinery}, \bibinfo{address}{New York, NY, USA}, Article \bibinfo{articleno}{146}, \bibinfo{numpages}{20}~pages.
\newblock
\showISBNx{9781450380966}
\urldef\tempurl%
\url{https://doi.org/10.1145/3411764.3445272}
\showDOI{\tempurl}


\bibitem[Aliprandi et~al\mbox{.}(2014)]%
        {aliprandi_caper_nodate}
\bibfield{author}{\bibinfo{person}{Carlo Aliprandi}, \bibinfo{person}{Juan~Arraiza Irujo}, \bibinfo{person}{Montse Cuadros}, \bibinfo{person}{Sebastian Maier}, \bibinfo{person}{Felipe Melero}, {and} \bibinfo{person}{Matteo Raffaelli}.} \bibinfo{year}{2014}\natexlab{}.
\newblock \showarticletitle{{CAPER}: {Collaborative} {Information}, {Acquisition}, {Processing}, {Exploitation} and {Reporting} for the {Prevention} of {Organised} {Crime}}.
\newblock \bibinfo{journal}{\emph{Communications in Computer and Information Science}} (\bibinfo{year}{2014}), \bibinfo{pages}{6}.
\newblock


\bibitem[Bansal et~al\mbox{.}(2021)]%
        {bansal_is_2021}
\bibfield{author}{\bibinfo{person}{Gagan Bansal}, \bibinfo{person}{Besmira Nushi}, \bibinfo{person}{Ece Kamar}, \bibinfo{person}{Eric Horvitz}, {and} \bibinfo{person}{Daniel~S. Weld}.} \bibinfo{year}{2021}\natexlab{}.
\newblock \bibinfo{title}{Is the {Most} {Accurate} {AI} the {Best} {Teammate}? {Optimizing} {AI} for {Teamwork}}.
\newblock
\newblock
\urldef\tempurl%
\url{http://arxiv.org/abs/2004.13102}
\showURL{%
\tempurl}
\newblock
\shownote{arXiv:2004.13102 [cs]}.


\bibitem[Belghith et~al\mbox{.}(2022)]%
        {belghith_compete_2022}
\bibfield{author}{\bibinfo{person}{Yasmine Belghith}, \bibinfo{person}{Sukrit Venkatagiri}, {and} \bibinfo{person}{Kurt Luther}.} \bibinfo{year}{2022}\natexlab{}.
\newblock \showarticletitle{Compete, {Collaborate}, {Investigate}: {Exploring} the {Social} {Structures} of {Open} {Source} {Intelligence} {Investigations}}. In \bibinfo{booktitle}{\emph{Proceedings of the 2022 {CHI} {Conference} on {Human} {Factors} in {Computing} {Systems}}} \emph{(\bibinfo{series}{{CHI} '22})}. \bibinfo{publisher}{Association for Computing Machinery}, \bibinfo{address}{New York, NY, USA}, \bibinfo{pages}{1--18}.
\newblock
\showISBNx{978-1-4503-9157-3}
\urldef\tempurl%
\url{https://doi.org/10.1145/3491102.3517526}
\showDOI{\tempurl}


\bibitem[Bhattacharjee et~al\mbox{.}(2023)]%
        {bhattacharjee_understanding_2023}
\bibfield{author}{\bibinfo{person}{Ananya Bhattacharjee}, \bibinfo{person}{Yuchen Zeng}, \bibinfo{person}{Sarah~Yi Xu}, \bibinfo{person}{Dana Kulzhabayeva}, \bibinfo{person}{Minyi Ma}, \bibinfo{person}{Rachel Kornfield}, \bibinfo{person}{Syed~Ishtiaque Ahmed}, \bibinfo{person}{Alex Mariakakis}, \bibinfo{person}{Mary~P. Czerwinski}, \bibinfo{person}{Anastasia Kuzminykh}, \bibinfo{person}{Michael Liut}, {and} \bibinfo{person}{Joseph~Jay Williams}.} \bibinfo{year}{2023}\natexlab{}.
\newblock \bibinfo{title}{Understanding the {Role} of {Large} {Language} {Models} in {Personalizing} and {Scaffolding} {Strategies} to {Combat} {Academic} {Procrastination}}.
\newblock
\newblock
\urldef\tempurl%
\url{http://arxiv.org/abs/2312.13581}
\showURL{%
\tempurl}
\newblock
\shownote{arXiv:2312.13581 [cs]}.


\bibitem[Bly and Churchill(1999)]%
        {bly_design_1999}
\bibfield{author}{\bibinfo{person}{Sara Bly} {and} \bibinfo{person}{Elizabeth~F. Churchill}.} \bibinfo{year}{1999}\natexlab{}.
\newblock \showarticletitle{Design through matchmaking: technology in search of users}.
\newblock \bibinfo{journal}{\emph{Interactions}} \bibinfo{volume}{6}, \bibinfo{number}{2} (\bibinfo{date}{March} \bibinfo{year}{1999}), \bibinfo{pages}{23--31}.
\newblock
\showISSN{1072-5520, 1558-3449}
\urldef\tempurl%
\url{https://doi.org/10.1145/296165.296174}
\showDOI{\tempurl}


\bibitem[Braun and Clarke(2006)]%
        {braun2006thematic}
\bibfield{author}{\bibinfo{person}{Virginia Braun} {and} \bibinfo{person}{Victoria Clarke}.} \bibinfo{year}{2006}\natexlab{}.
\newblock \showarticletitle{Using thematic analysis in psychology}.
\newblock \bibinfo{journal}{\emph{Qualitative research in psychology}} \bibinfo{volume}{3}, \bibinfo{number}{2} (\bibinfo{date}{Jan.} \bibinfo{year}{2006}), \bibinfo{pages}{77--101}.
\newblock
\showISSN{1478-0887}
\urldef\tempurl%
\url{https://doi.org/10.1191/1478088706qp063oa}
\showDOI{\tempurl}
\newblock
\shownote{Publisher: Routledge \_eprint: https://www.tandfonline.com/doi/pdf/10.1191/1478088706qp063oa}.


\bibitem[Briggs et~al\mbox{.}(2020)]%
        {briggs2020brainstorming}
\bibfield{author}{\bibinfo{person}{Shannon Briggs}, \bibinfo{person}{Matthew Peveler}, \bibinfo{person}{Jaimie Drozdal}, \bibinfo{person}{Lilit Balagyozyan}, \bibinfo{person}{Jonas Braasch}, {and} \bibinfo{person}{Hui Su}.} \bibinfo{year}{2020}\natexlab{}.
\newblock \showarticletitle{Brainstorming for sensemaking in a multimodal, multiuser cognitive environment}. In \bibinfo{booktitle}{\emph{HCI International 2020-Late Breaking Papers: User Experience Design and Case Studies: 22nd HCI International Conference, HCII 2020, Copenhagen, Denmark, July 19--24, 2020, Proceedings 22}}. Springer, \bibinfo{pages}{66--83}.
\newblock


\bibitem[Browne et~al\mbox{.}(2024a)]%
        {browne_systematic_2024}
\bibfield{author}{\bibinfo{person}{Thomas~Oakley Browne}, \bibinfo{person}{Mohammad Abedin}, {and} \bibinfo{person}{Mohammad Jabed~Morshed Chowdhury}.} \bibinfo{year}{2024}\natexlab{a}.
\newblock \showarticletitle{A systematic review on research utilising artificial intelligence for open source intelligence ({OSINT}) applications}.
\newblock \bibinfo{journal}{\emph{International Journal of Information Security}} (\bibinfo{date}{June} \bibinfo{year}{2024}).
\newblock
\showISSN{1615-5270}
\urldef\tempurl%
\url{https://doi.org/10.1007/s10207-024-00868-2}
\showDOI{\tempurl}


\bibitem[Browne et~al\mbox{.}(2024b)]%
        {browne2024systematic}
\bibfield{author}{\bibinfo{person}{Thomas~Oakley Browne}, \bibinfo{person}{Mohammad Abedin}, {and} \bibinfo{person}{Mohammad Jabed~Morshed Chowdhury}.} \bibinfo{year}{2024}\natexlab{b}.
\newblock \showarticletitle{A systematic review on research utilising artificial intelligence for open source intelligence (OSINT) applications}.
\newblock \bibinfo{journal}{\emph{International Journal of Information Security}} (\bibinfo{year}{2024}), \bibinfo{pages}{1--28}.
\newblock


\bibitem[Chen(2021)]%
        {chen2021supporting}
\bibfield{author}{\bibinfo{person}{Dong Chen}.} \bibinfo{year}{2021}\natexlab{}.
\newblock \bibinfo{booktitle}{\emph{Supporting Collaborative Information Analysis Using Interactive Visualization: A Design Research}}.
\newblock \bibinfo{publisher}{The Pennsylvania State University}.
\newblock


\bibitem[Chen et~al\mbox{.}(2023)]%
        {chen_next_2023}
\bibfield{author}{\bibinfo{person}{Xiang~'Anthony' Chen}, \bibinfo{person}{Jeff Burke}, \bibinfo{person}{Ruofei Du}, \bibinfo{person}{Matthew~K. Hong}, \bibinfo{person}{Jennifer Jacobs}, \bibinfo{person}{Philippe Laban}, \bibinfo{person}{Dingzeyu Li}, \bibinfo{person}{Nanyun Peng}, \bibinfo{person}{Karl D.~D. Willis}, \bibinfo{person}{Chien-Sheng Wu}, {and} \bibinfo{person}{Bolei Zhou}.} \bibinfo{year}{2023}\natexlab{}.
\newblock \bibinfo{title}{Next {Steps} for {Human}-{Centered} {Generative} {AI}: {A} {Technical} {Perspective}}.
\newblock
\newblock
\urldef\tempurl%
\url{http://arxiv.org/abs/2306.15774}
\showURL{%
\tempurl}
\newblock
\shownote{arXiv:2306.15774 [cs]}.


\bibitem[Chin~Jr et~al\mbox{.}(2009)]%
        {chin2009exploring}
\bibfield{author}{\bibinfo{person}{George Chin~Jr}, \bibinfo{person}{Olga~A Kuchar}, {and} \bibinfo{person}{Katherine~E Wolf}.} \bibinfo{year}{2009}\natexlab{}.
\newblock \showarticletitle{Exploring the analytical processes of intelligence analysts}. In \bibinfo{booktitle}{\emph{Proceedings of the SIGCHI Conference on Human Factors in Computing Systems}}. \bibinfo{pages}{11--20}.
\newblock


\bibitem[Dailey and Starbird(2015)]%
        {dailey_dispersants_2015}
\bibfield{author}{\bibinfo{person}{Dharma Dailey} {and} \bibinfo{person}{Kate Starbird}.} \bibinfo{year}{2015}\natexlab{}.
\newblock \showarticletitle{"It's Raining Dispersants": Collective Sensemaking of Complex Information in Crisis Contexts}. In \bibinfo{booktitle}{\emph{Proceedings of the 18th ACM Conference Companion on Computer Supported Cooperative Work and; Social Computing}} (Vancouver, BC, Canada) \emph{(\bibinfo{series}{CSCW'15 Companion})}. \bibinfo{publisher}{Association for Computing Machinery}, \bibinfo{address}{New York, NY, USA}, \bibinfo{pages}{155–158}.
\newblock
\showISBNx{9781450329460}
\urldef\tempurl%
\url{https://doi.org/10.1145/2685553.2698995}
\showDOI{\tempurl}


\bibitem[Deng et~al\mbox{.}(2024)]%
        {deng2024pentestgpt}
\bibfield{author}{\bibinfo{person}{Gelei Deng}, \bibinfo{person}{Yi Liu}, \bibinfo{person}{V{\'\i}ctor Mayoral-Vilches}, \bibinfo{person}{Peng Liu}, \bibinfo{person}{Yuekang Li}, \bibinfo{person}{Yuan Xu}, \bibinfo{person}{Tianwei Zhang}, \bibinfo{person}{Yang Liu}, \bibinfo{person}{Martin Pinzger}, {and} \bibinfo{person}{Stefan Rass}.} \bibinfo{year}{2024}\natexlab{}.
\newblock \showarticletitle{{PentestGPT}: Evaluating and Harnessing Large Language Models for Automated Penetration Testing}. In \bibinfo{booktitle}{\emph{33rd USENIX Security Symposium (USENIX Security 24)}}. \bibinfo{publisher}{USENIX Association}, \bibinfo{address}{Philadelphia, PA}, \bibinfo{pages}{847--864}.
\newblock
\showISBNx{978-1-939133-44-1}
\urldef\tempurl%
\url{https://www.usenix.org/conference/usenixsecurity24/presentation/deng}
\showURL{%
\tempurl}


\bibitem[Dennis et~al\mbox{.}(2023)]%
        {dennis_ai_2023}
\bibfield{author}{\bibinfo{person}{Alan~R. Dennis}, \bibinfo{person}{Akshat Lakhiwal}, {and} \bibinfo{person}{Agrim Sachdeva}.} \bibinfo{year}{2023}\natexlab{}.
\newblock \showarticletitle{{AI} {Agents} as {Team} {Members}: {Effects} on {Satisfaction}, {Conflict}, {Trustworthiness}, and {Willingness} to {Work} {With}}.
\newblock \bibinfo{journal}{\emph{Journal of Management Information Systems}} \bibinfo{volume}{40}, \bibinfo{number}{2} (\bibinfo{date}{April} \bibinfo{year}{2023}), \bibinfo{pages}{307--337}.
\newblock
\showISSN{0742-1222}
\urldef\tempurl%
\url{https://doi.org/10.1080/07421222.2023.2196773}
\showDOI{\tempurl}
\newblock
\shownote{Publisher: Routledge \_eprint: https://doi.org/10.1080/07421222.2023.2196773}.


\bibitem[Doroudi et~al\mbox{.}(2016)]%
        {doroudi_toward_2016}
\bibfield{author}{\bibinfo{person}{Shayan Doroudi}, \bibinfo{person}{Ece Kamar}, \bibinfo{person}{Emma Brunskill}, {and} \bibinfo{person}{Eric Horvitz}.} \bibinfo{year}{2016}\natexlab{}.
\newblock \showarticletitle{Toward a {Learning} {Science} for {Complex} {Crowdsourcing} {Tasks}}. In \bibinfo{booktitle}{\emph{Proceedings of the 2016 {CHI} {Conference} on {Human} {Factors} in {Computing} {Systems}}} \emph{(\bibinfo{series}{{CHI} '16})}. \bibinfo{publisher}{Association for Computing Machinery}, \bibinfo{address}{New York, NY, USA}, \bibinfo{pages}{2623--2634}.
\newblock
\showISBNx{978-1-4503-3362-7}
\urldef\tempurl%
\url{https://doi.org/10.1145/2858036.2858268}
\showDOI{\tempurl}


\bibitem[Dorton and Hall(2021)]%
        {Dorton2021CollaborativeHS}
\bibfield{author}{\bibinfo{person}{Stephen~L. Dorton} {and} \bibinfo{person}{Rob~A. Hall}.} \bibinfo{year}{2021}\natexlab{}.
\newblock \showarticletitle{Collaborative Human-AI Sensemaking for Intelligence Analysis}. In \bibinfo{booktitle}{\emph{Interacci{\'o}n}}.
\newblock
\urldef\tempurl%
\url{https://api.semanticscholar.org/CorpusID:236150803}
\showURL{%
\tempurl}


\bibitem[Doshi and Hauser(2023)]%
        {doshi_generative_2023}
\bibfield{author}{\bibinfo{person}{Anil~R. Doshi} {and} \bibinfo{person}{Oliver Hauser}.} \bibinfo{year}{2023}\natexlab{}.
\newblock \bibinfo{title}{Generative artificial intelligence enhances creativity but reduces the diversity of novel content}.
\newblock
\newblock
\urldef\tempurl%
\url{https://doi.org/10.2139/ssrn.4535536}
\showDOI{\tempurl}


\bibitem[Dow et~al\mbox{.}(2012)]%
        {dow2012shepherding}
\bibfield{author}{\bibinfo{person}{Steven Dow}, \bibinfo{person}{Anand Kulkarni}, \bibinfo{person}{Scott Klemmer}, {and} \bibinfo{person}{Bj\"{o}rn Hartmann}.} \bibinfo{year}{2012}\natexlab{}.
\newblock \showarticletitle{Shepherding the Crowd Yields Better Work}. In \bibinfo{booktitle}{\emph{Proceedings of the ACM 2012 Conference on Computer Supported Cooperative Work}} (Seattle, Washington, USA) \emph{(\bibinfo{series}{CSCW ’12})}. \bibinfo{publisher}{Association for Computing Machinery}, \bibinfo{address}{New York, NY, USA}, \bibinfo{pages}{1013–1022}.
\newblock
\showISBNx{9781450310864}
\urldef\tempurl%
\url{https://doi.org/10.1145/2145204.2145355}
\showDOI{\tempurl}


\bibitem[Edwards et~al\mbox{.}(2017)]%
        {edwards_panning_2017}
\bibfield{author}{\bibinfo{person}{Matthew Edwards}, \bibinfo{person}{Robert Larson}, \bibinfo{person}{Benjamin Green}, \bibinfo{person}{Awais Rashid}, {and} \bibinfo{person}{Alistair Baron}.} \bibinfo{year}{2017}\natexlab{}.
\newblock \showarticletitle{Panning for gold: {Automatically} analysing online social engineering attack surfaces}.
\newblock \bibinfo{journal}{\emph{Computers \& Security}}  \bibinfo{volume}{69} (\bibinfo{date}{Aug.} \bibinfo{year}{2017}), \bibinfo{pages}{18--34}.
\newblock
\showISSN{0167-4048}
\urldef\tempurl%
\url{https://doi.org/10.1016/j.cose.2016.12.013}
\showDOI{\tempurl}


\bibitem[Fisher et~al\mbox{.}(2012)]%
        {fisher2012distributed}
\bibfield{author}{\bibinfo{person}{Kristie Fisher}, \bibinfo{person}{Scott Counts}, {and} \bibinfo{person}{Aniket Kittur}.} \bibinfo{year}{2012}\natexlab{}.
\newblock \showarticletitle{Distributed Sensemaking: Improving Sensemaking by Leveraging the Efforts of Previous Users}. In \bibinfo{booktitle}{\emph{Proceedings of the {SIGCHI} Conference on Human Factors in Computing Systems}} (New York, {NY}, {USA}) \emph{(\bibinfo{series}{{CHI} '12})}. \bibinfo{publisher}{{ACM}}, \bibinfo{address}{New York, NY, USA}, \bibinfo{pages}{247--256}.
\newblock
\showISBNx{978-1-4503-1015-4}
\urldef\tempurl%
\url{https://doi.org/10.1145/2207676.2207711}
\showDOI{\tempurl}


\bibitem[{Flashpoint}(2024)]%
        {flashpoint_generative_2024}
\bibfield{author}{\bibinfo{person}{{Flashpoint}}.} \bibinfo{year}{2024}\natexlab{}.
\newblock \bibinfo{title}{Generative {AI} for {OSINT}: {Next} {Level} {Techniques} for {ChatGPT} and {Beyond}}.
\newblock
\newblock
\urldef\tempurl%
\url{https://www.youtube.com/watch?v=7zzubjKEUW4}
\showURL{%
Retrieved 2024-09-08 from \tempurl}


\bibitem[Flathmann et~al\mbox{.}(2021)]%
        {flathmann_fostering_2021}
\bibfield{author}{\bibinfo{person}{Christopher Flathmann}, \bibinfo{person}{Beau~G. Schelble}, {and} \bibinfo{person}{Nathan~J. McNeese}.} \bibinfo{year}{2021}\natexlab{}.
\newblock \showarticletitle{Fostering {Human}-{Agent} {Team} {Leadership} by {Leveraging} {Human} {Teaming} {Principles}}. In \bibinfo{booktitle}{\emph{2021 {IEEE} 2nd {International} {Conference} on {Human}-{Machine} {Systems} ({ICHMS})}}. \bibinfo{pages}{1--6}.
\newblock
\urldef\tempurl%
\url{https://doi.org/10.1109/ICHMS53169.2021.9582649}
\showDOI{\tempurl}


\bibitem[Freed et~al\mbox{.}(2019)]%
        {freed2019my}
\bibfield{author}{\bibinfo{person}{Diana Freed}, \bibinfo{person}{Sam Havron}, \bibinfo{person}{Emily Tseng}, \bibinfo{person}{Andrea Gallardo}, \bibinfo{person}{Rahul Chatterjee}, \bibinfo{person}{Thomas Ristenpart}, {and} \bibinfo{person}{Nicola Dell}.} \bibinfo{year}{2019}\natexlab{}.
\newblock \showarticletitle{" Is My Phone Hacked?" Analyzing Clinical Computer Security Interventions With Survivors of Intimate Partner Violence}.
\newblock \bibinfo{journal}{\emph{Proceedings of the ACM on Human-Computer Interaction}} \bibinfo{volume}{3}, \bibinfo{number}{CSCW} (\bibinfo{year}{2019}), \bibinfo{pages}{1--24}.
\newblock


\bibitem[Gerber et~al\mbox{.}(2016)]%
        {gerber_how_2016}
\bibfield{author}{\bibinfo{person}{Matylda Gerber}, \bibinfo{person}{B.L.~William Wong}, {and} \bibinfo{person}{Neesha Kodagoda}.} \bibinfo{year}{2016}\natexlab{}.
\newblock \showarticletitle{How {Analysts} {Think}: {Intuition}, {Leap} of {Faith} and {Insight}}.
\newblock \bibinfo{journal}{\emph{Proceedings of the Human Factors and Ergonomics Society Annual Meeting}} \bibinfo{volume}{60}, \bibinfo{number}{1} (\bibinfo{date}{Sept.} \bibinfo{year}{2016}), \bibinfo{pages}{173--177}.
\newblock
\showISSN{1071-1813}
\urldef\tempurl%
\url{https://doi.org/10.1177/1541931213601039}
\showDOI{\tempurl}
\newblock
\shownote{Publisher: SAGE Publications Inc}.


\bibitem[Gibson(2016)]%
        {gibson2016acquisition}
\bibfield{author}{\bibinfo{person}{Helen Gibson}.} \bibinfo{year}{2016}\natexlab{}.
\newblock \showarticletitle{Acquisition and preparation of data for OSINT investigations}.
\newblock \bibinfo{journal}{\emph{Open source intelligence investigation: From strategy to implementation}} (\bibinfo{year}{2016}), \bibinfo{pages}{69--93}.
\newblock


\bibitem[Glassman and Kang(2012)]%
        {glassman2012intelligence}
\bibfield{author}{\bibinfo{person}{Michael Glassman} {and} \bibinfo{person}{Min~Ju Kang}.} \bibinfo{year}{2012}\natexlab{}.
\newblock \showarticletitle{Intelligence in the internet age: The emergence and evolution of Open Source Intelligence (OSINT)}.
\newblock \bibinfo{journal}{\emph{Computers in Human Behavior}} \bibinfo{volume}{28}, \bibinfo{number}{2} (\bibinfo{date}{March} \bibinfo{year}{2012}), \bibinfo{pages}{673--682}.
\newblock
\showISSN{0747-5632}
\urldef\tempurl%
\url{https://doi.org/10.1016/j.chb.2011.11.014}
\showDOI{\tempurl}


\bibitem[Govardhan et~al\mbox{.}(2023)]%
        {govardhan_key_2023}
\bibfield{author}{\bibinfo{person}{Devu Govardhan}, \bibinfo{person}{Grandhi Guna Sai~Hari Krishna}, \bibinfo{person}{V. Charan}, \bibinfo{person}{Sribhashyam Venkata~Anantha Sai}, {and} \bibinfo{person}{Radhika~Rani Chintala}.} \bibinfo{year}{2023}\natexlab{}.
\newblock \showarticletitle{Key {Challenges} and {Limitations} of the {OSINT} {Framework} in the {Context} of {Cybersecurity}}. In \bibinfo{booktitle}{\emph{2023 2nd {International} {Conference} on {Edge} {Computing} and {Applications} ({ICECAA})}}. \bibinfo{pages}{236--243}.
\newblock
\urldef\tempurl%
\url{https://doi.org/10.1109/ICECAA58104.2023.10212168}
\showDOI{\tempurl}


\bibitem[Gururangan et~al\mbox{.}(2020)]%
        {gururangan_dont_2020}
\bibfield{author}{\bibinfo{person}{Suchin Gururangan}, \bibinfo{person}{Ana Marasović}, \bibinfo{person}{Swabha Swayamdipta}, \bibinfo{person}{Kyle Lo}, \bibinfo{person}{Iz Beltagy}, \bibinfo{person}{Doug Downey}, {and} \bibinfo{person}{Noah~A. Smith}.} \bibinfo{year}{2020}\natexlab{}.
\newblock \bibinfo{title}{Don't {Stop} {Pretraining}: {Adapt} {Language} {Models} to {Domains} and {Tasks}}.
\newblock
\newblock
\urldef\tempurl%
\url{https://doi.org/10.48550/arXiv.2004.10964}
\showDOI{\tempurl}
\newblock
\shownote{arXiv:2004.10964 [cs]}.


\bibitem[Hassan et~al\mbox{.}(2018)]%
        {hassan2018evolution}
\bibfield{author}{\bibinfo{person}{Nihad~A Hassan}, \bibinfo{person}{Rami Hijazi}, \bibinfo{person}{Nihad~A Hassan}, {and} \bibinfo{person}{Rami Hijazi}.} \bibinfo{year}{2018}\natexlab{}.
\newblock \showarticletitle{The evolution of open source intelligence}.
\newblock \bibinfo{journal}{\emph{Open Source Intelligence Methods and Tools: A Practical Guide to Online Intelligence}} (\bibinfo{year}{2018}), \bibinfo{pages}{1--20}.
\newblock


\bibitem[Havron et~al\mbox{.}(2019)]%
        {havron_clinical_2019}
\bibfield{author}{\bibinfo{person}{Sam Havron}, \bibinfo{person}{Diana Freed}, \bibinfo{person}{Rahul Chatterjee}, \bibinfo{person}{Damon McCoy}, \bibinfo{person}{Nicola Dell}, {and} \bibinfo{person}{Thomas Ristenpart}.} \bibinfo{year}{2019}\natexlab{}.
\newblock \showarticletitle{Clinical {Computer} {Security} for {Victims} of {Intimate} {Partner} {Violence}}. \bibinfo{pages}{105--122}.
\newblock
\showISBNx{978-1-939133-06-9}
\urldef\tempurl%
\url{https://www.usenix.org/conference/usenixsecurity19/presentation/havron}
\showURL{%
\tempurl}


\bibitem[Hayes and Cappa(2018)]%
        {hayes_open-source_2018}
\bibfield{author}{\bibinfo{person}{Darren~R. Hayes} {and} \bibinfo{person}{Francesco Cappa}.} \bibinfo{year}{2018}\natexlab{}.
\newblock \showarticletitle{Open-source intelligence for risk assessment}.
\newblock \bibinfo{journal}{\emph{Business Horizons}} \bibinfo{volume}{61}, \bibinfo{number}{5} (\bibinfo{date}{Sept.} \bibinfo{year}{2018}), \bibinfo{pages}{689--697}.
\newblock
\showISSN{0007-6813}
\urldef\tempurl%
\url{https://doi.org/10.1016/j.bushor.2018.02.001}
\showDOI{\tempurl}


\bibitem[He et~al\mbox{.}(2024)]%
        {he_ai_2024}
\bibfield{author}{\bibinfo{person}{Jessica He}, \bibinfo{person}{Stephanie Houde}, \bibinfo{person}{Gabriel~E. Gonzalez}, \bibinfo{person}{Darío~Andrés Silva~Moran}, \bibinfo{person}{Steven~I. Ross}, \bibinfo{person}{Michael Muller}, {and} \bibinfo{person}{Justin~D. Weisz}.} \bibinfo{year}{2024}\natexlab{}.
\newblock \showarticletitle{{AI} and the {Future} of {Collaborative} {Work}: {Group} {Ideation} with an {LLM} in a {Virtual} {Canvas}}. In \bibinfo{booktitle}{\emph{Proceedings of the 3rd {Annual} {Meeting} of the {Symposium} on {Human}-{Computer} {Interaction} for {Work}}}. \bibinfo{publisher}{ACM}, \bibinfo{address}{Newcastle upon Tyne United Kingdom}, \bibinfo{pages}{1--14}.
\newblock
\showISBNx{9798400710179}
\urldef\tempurl%
\url{https://doi.org/10.1145/3663384.3663398}
\showDOI{\tempurl}


\bibitem[Hepenstal et~al\mbox{.}(2021)]%
        {hepenstal2021analysis}
\bibfield{author}{\bibinfo{person}{Sam Hepenstal}, \bibinfo{person}{Leishi Zhang}, {and} \bibinfo{person}{BL~William Wong}.} \bibinfo{year}{2021}\natexlab{}.
\newblock \showarticletitle{An analysis of expertise in intelligence analysis to support the design of Human-Centered Artificial Intelligence}. In \bibinfo{booktitle}{\emph{2021 ieee international conference on systems, man, and cybernetics (smc)}}. IEEE, \bibinfo{pages}{107--112}.
\newblock


\bibitem[Huang et~al\mbox{.}(2015)]%
        {huang_2015_investigation}
\bibfield{author}{\bibinfo{person}{Y.~Linlin Huang}, \bibinfo{person}{Kate Starbird}, \bibinfo{person}{Mania Orand}, \bibinfo{person}{Stephanie~A. Stanek}, {and} \bibinfo{person}{Heather~T. Pedersen}.} \bibinfo{year}{2015}\natexlab{}.
\newblock \showarticletitle{Connected Through Crisis: Emotional Proximity and the Spread of Misinformation Online}. In \bibinfo{booktitle}{\emph{Proceedings of the 18th ACM Conference on Computer Supported Cooperative Work \& Social Computing}} (Vancouver, BC, Canada) \emph{(\bibinfo{series}{CSCW '15})}. \bibinfo{publisher}{Association for Computing Machinery}, \bibinfo{address}{New York, NY, USA}, \bibinfo{pages}{969–980}.
\newblock
\showISBNx{9781450329224}
\urldef\tempurl%
\url{https://doi.org/10.1145/2675133.2675202}
\showDOI{\tempurl}


\bibitem[Hutchinson et~al\mbox{.}(2003)]%
        {hutchinson_technology_2003}
\bibfield{author}{\bibinfo{person}{Hilary Hutchinson}, \bibinfo{person}{Wendy Mackay}, \bibinfo{person}{Bo Westerlund}, \bibinfo{person}{Benjamin~B. Bederson}, \bibinfo{person}{Allison Druin}, \bibinfo{person}{Catherine Plaisant}, \bibinfo{person}{Michel Beaudouin-Lafon}, \bibinfo{person}{Stéphane Conversy}, \bibinfo{person}{Helen Evans}, \bibinfo{person}{Heiko Hansen}, \bibinfo{person}{Nicolas Roussel}, {and} \bibinfo{person}{Björn Eiderbäck}.} \bibinfo{year}{2003}\natexlab{}.
\newblock \showarticletitle{Technology probes: inspiring design for and with families}. In \bibinfo{booktitle}{\emph{Proceedings of the {SIGCHI} {Conference} on {Human} {Factors} in {Computing} {Systems}}} \emph{(\bibinfo{series}{{CHI} '03})}. \bibinfo{publisher}{Association for Computing Machinery}, \bibinfo{address}{New York, NY, USA}, \bibinfo{pages}{17--24}.
\newblock
\showISBNx{978-1-58113-630-2}
\urldef\tempurl%
\url{https://doi.org/10.1145/642611.642616}
\showDOI{\tempurl}


\bibitem[Hwang et~al\mbox{.}(2022)]%
        {hwang2022current}
\bibfield{author}{\bibinfo{person}{Yong-Woon Hwang}, \bibinfo{person}{Im-Yeong Lee}, \bibinfo{person}{Hwankuk Kim}, \bibinfo{person}{Hyejung Lee}, {and} \bibinfo{person}{Donghyun Kim}.} \bibinfo{year}{2022}\natexlab{}.
\newblock \showarticletitle{Current status and security trend of osint}.
\newblock \bibinfo{journal}{\emph{Wireless Communications and Mobile Computing}} \bibinfo{volume}{2022}, \bibinfo{number}{1} (\bibinfo{year}{2022}), \bibinfo{pages}{1290129}.
\newblock


\bibitem[Jin et~al\mbox{.}(2022)]%
        {jin_learn_2022}
\bibfield{author}{\bibinfo{person}{Xisen Jin}, \bibinfo{person}{Bill~Yuchen Lin}, \bibinfo{person}{Mohammad Rostami}, {and} \bibinfo{person}{Xiang Ren}.} \bibinfo{year}{2022}\natexlab{}.
\newblock \bibinfo{title}{Learn {Continually}, {Generalize} {Rapidly}: {Lifelong} {Knowledge} {Accumulation} for {Few}-shot {Learning}}.
\newblock
\newblock
\urldef\tempurl%
\url{https://doi.org/10.48550/arXiv.2104.08808}
\showDOI{\tempurl}
\newblock
\shownote{arXiv:2104.08808 [cs]}.


\bibitem[João Rafael Gonçalves~Evangelista and Napolitano(2021)]%
        {evangelista2021systematic}
\bibfield{author}{\bibinfo{person}{Márcio~Romero João Rafael Gonçalves~Evangelista, Renato José~Sassi} {and} \bibinfo{person}{Domingos Napolitano}.} \bibinfo{year}{2021}\natexlab{}.
\newblock \showarticletitle{Systematic Literature Review to Investigate the Application of Open Source Intelligence (OSINT) with Artificial Intelligence}.
\newblock \bibinfo{journal}{\emph{Journal of Applied Security Research}} \bibinfo{volume}{16}, \bibinfo{number}{3} (\bibinfo{year}{2021}), \bibinfo{pages}{345--369}.
\newblock
\urldef\tempurl%
\url{https://doi.org/10.1080/19361610.2020.1761737}
\showDOI{\tempurl}
\showeprint{https://doi.org/10.1080/19361610.2020.1761737}


\bibitem[Kang and Stasko(2014)]%
        {kang_characterizing_2014}
\bibfield{author}{\bibinfo{person}{Youn-ah Kang} {and} \bibinfo{person}{John Stasko}.} \bibinfo{year}{2014}\natexlab{}.
\newblock \showarticletitle{Characterizing the intelligence analysis process through a longitudinal field study: {Implications} for visual analytics}.
\newblock \bibinfo{journal}{\emph{Information Visualization}} \bibinfo{volume}{13}, \bibinfo{number}{2} (\bibinfo{date}{April} \bibinfo{year}{2014}), \bibinfo{pages}{134--158}.
\newblock
\showISSN{1473-8716}
\urldef\tempurl%
\url{https://doi.org/10.1177/1473871612468877}
\showDOI{\tempurl}
\newblock
\shownote{Publisher: SAGE Publications}.


\bibitem[Kim et~al\mbox{.}(2021)]%
        {kim_ai_2021}
\bibfield{author}{\bibinfo{person}{Jihyun Kim}, \bibinfo{person}{Kelly Merrill~Jr.}, {and} \bibinfo{person}{Chad Collins}.} \bibinfo{year}{2021}\natexlab{}.
\newblock \showarticletitle{{AI} as a friend or assistant: {The} mediating role of perceived usefulness in social {AI} vs. functional {AI}}.
\newblock \bibinfo{journal}{\emph{Telematics and Informatics}}  \bibinfo{volume}{64} (\bibinfo{date}{Nov.} \bibinfo{year}{2021}), \bibinfo{pages}{101694}.
\newblock
\showISSN{0736-5853}
\urldef\tempurl%
\url{https://doi.org/10.1016/j.tele.2021.101694}
\showDOI{\tempurl}


\bibitem[Kulkarni et~al\mbox{.}(2023)]%
        {kulkarni_word_2023}
\bibfield{author}{\bibinfo{person}{Chinmay Kulkarni}, \bibinfo{person}{Stefania Druga}, \bibinfo{person}{Minsuk Chang}, \bibinfo{person}{Alex Fiannaca}, \bibinfo{person}{Carrie Cai}, {and} \bibinfo{person}{Michael Terry}.} \bibinfo{year}{2023}\natexlab{}.
\newblock \bibinfo{title}{A {Word} is {Worth} a {Thousand} {Pictures}: {Prompts} as {AI} {Design} {Material}}.
\newblock
\newblock
\urldef\tempurl%
\url{https://doi.org/10.48550/arXiv.2303.12647}
\showDOI{\tempurl}
\newblock
\shownote{arXiv:2303.12647 [cs]}.


\bibitem[Lederman({[n.\,d.]})]%
        {lederman_latest_nodate}
\bibfield{author}{\bibinfo{person}{Doug Lederman}.} \bibinfo{year}{[n.\,d.]}\natexlab{}.
\newblock \bibinfo{title}{The {Latest} {Campus} {Clinic}: {Cybersecurity}}.
\newblock
\newblock
\urldef\tempurl%
\url{https://www.insidehighered.com/news/tech-innovation/2023/06/27/cybersecurity-clinics-blend-hands-training-and-community-services}
\showURL{%
Retrieved 2024-09-11 from \tempurl}


\bibitem[Liu et~al\mbox{.}(2024)]%
        {liu2024humancenterednlpfactcheckingcodesigning}
\bibfield{author}{\bibinfo{person}{Houjiang Liu}, \bibinfo{person}{Anubrata Das}, \bibinfo{person}{Alexander Boltz}, \bibinfo{person}{Didi Zhou}, \bibinfo{person}{Daisy Pinaroc}, \bibinfo{person}{Matthew Lease}, {and} \bibinfo{person}{Min~Kyung Lee}.} \bibinfo{year}{2024}\natexlab{}.
\newblock \showarticletitle{Human-centered NLP Fact-checking: Co-Designing with Fact-checkers using Matchmaking for AI}.
\newblock \bibinfo{journal}{\emph{Proceedings of the ACM on Human-Computer Interaction}} \bibinfo{volume}{8}, \bibinfo{number}{CSCW2} (\bibinfo{year}{2024}), \bibinfo{pages}{1--44}.
\newblock


\bibitem[Luther et~al\mbox{.}(2013)]%
        {luther_redistributing_2013}
\bibfield{author}{\bibinfo{person}{Kurt Luther}, \bibinfo{person}{Casey Fiesler}, {and} \bibinfo{person}{Amy Bruckman}.} \bibinfo{year}{2013}\natexlab{}.
\newblock \showarticletitle{Redistributing leadership in online creative collaboration}. In \bibinfo{booktitle}{\emph{Proceedings of the 2013 conference on {Computer} supported cooperative work}} \emph{(\bibinfo{series}{{CSCW} '13})}. \bibinfo{publisher}{Association for Computing Machinery}, \bibinfo{address}{New York, NY, USA}, \bibinfo{pages}{1007--1022}.
\newblock
\showISBNx{978-1-4503-1331-5}
\urldef\tempurl%
\url{https://doi.org/10.1145/2441776.2441891}
\showDOI{\tempurl}


\bibitem[Martins and Medeiros(2022)]%
        {martins_generating_2022}
\bibfield{author}{\bibinfo{person}{Cláudio Martins} {and} \bibinfo{person}{Ibéria Medeiros}.} \bibinfo{year}{2022}\natexlab{}.
\newblock \showarticletitle{Generating {Quality} {Threat} {Intelligence} {Leveraging} {OSINT} and a {Cyber} {Threat} {Unified} {Taxonomy}}.
\newblock \bibinfo{journal}{\emph{ACM Transactions on Privacy and Security}} \bibinfo{volume}{25}, \bibinfo{number}{3} (\bibinfo{date}{Aug.} \bibinfo{year}{2022}), \bibinfo{pages}{1--39}.
\newblock
\showISSN{2471-2566, 2471-2574}
\urldef\tempurl%
\url{https://doi.org/10.1145/3530977}
\showDOI{\tempurl}


\bibitem[Mukhopadhyay et~al\mbox{.}(2024)]%
        {mukhopadhyay2024osint}
\bibfield{author}{\bibinfo{person}{Anirban Mukhopadhyay}, \bibinfo{person}{Sukrit Venkatagiri}, {and} \bibinfo{person}{Kurt Luther}.} \bibinfo{year}{2024}\natexlab{}.
\newblock \showarticletitle{OSINT Research Studios: A Flexible Crowdsourcing Framework to Scale Up Open Source Intelligence Investigations}.
\newblock \bibinfo{journal}{\emph{Proceedings of the ACM on Human-Computer Interaction}} \bibinfo{volume}{8}, \bibinfo{number}{CSCW1} (\bibinfo{year}{2024}), \bibinfo{pages}{1--38}.
\newblock


\bibitem[Muller and Weisz(2022)]%
        {muller2022frameworks}
\bibfield{author}{\bibinfo{person}{Michael Muller} {and} \bibinfo{person}{Justin Weisz}.} \bibinfo{year}{2022}\natexlab{}.
\newblock \showarticletitle{Frameworks for Collaborating Humans and AIs: Sequence and Sociality in Organizational Applications}. In \bibinfo{booktitle}{\emph{CHIWORK}}.
\newblock


\bibitem[of~Technology(2023)]%
        {MITCybersecurityClinic}
\bibfield{author}{\bibinfo{person}{Massachusetts~Institute of Technology}.} \bibinfo{year}{2023}\natexlab{}.
\newblock \bibinfo{title}{Urban Cyber Defense: Cybersecurity Clinic}.
\newblock
\newblock
\urldef\tempurl%
\url{http://urbancyberdefense.mit.edu/cybersecurityclinic}
\showURL{%
\tempurl}
\newblock
\shownote{Accessed: 2023-10-02}.


\bibitem[Pastor-Galindo et~al\mbox{.}(2020)]%
        {pastor2020not}
\bibfield{author}{\bibinfo{person}{Javier Pastor-Galindo}, \bibinfo{person}{Pantaleone Nespoli}, \bibinfo{person}{F{\'e}lix~G{\'o}mez M{\'a}rmol}, {and} \bibinfo{person}{Gregorio~Mart{\'\i}nez P{\'e}rez}.} \bibinfo{year}{2020}\natexlab{}.
\newblock \showarticletitle{The not yet exploited goldmine of OSINT: Opportunities, open challenges and future trends}.
\newblock \bibinfo{journal}{\emph{IEEE access}}  \bibinfo{volume}{8} (\bibinfo{year}{2020}), \bibinfo{pages}{10282--10304}.
\newblock


\bibitem[Peng et~al\mbox{.}(2023)]%
        {peng_impact_2023}
\bibfield{author}{\bibinfo{person}{Sida Peng}, \bibinfo{person}{Eirini Kalliamvakou}, \bibinfo{person}{Peter Cihon}, {and} \bibinfo{person}{Mert Demirer}.} \bibinfo{year}{2023}\natexlab{}.
\newblock \bibinfo{title}{The {Impact} of {AI} on {Developer} {Productivity}: {Evidence} from {GitHub} {Copilot}}.
\newblock
\newblock
\urldef\tempurl%
\url{https://doi.org/10.48550/arXiv.2302.06590}
\showDOI{\tempurl}
\newblock
\shownote{arXiv:2302.06590 [cs]}.


\bibitem[Pervez et~al\mbox{.}(2023)]%
        {pervez_towards_2023}
\bibfield{author}{\bibinfo{person}{Muhammad~Hasban Pervez}, \bibinfo{person}{Mert~Ilhan Ecevit}, \bibinfo{person}{Najiba~Zainab Naqvi}, \bibinfo{person}{Reiner Creutzburg}, {and} \bibinfo{person}{Hasan Dag}.} \bibinfo{year}{2023}\natexlab{}.
\newblock \showarticletitle{Towards {Better} {Cyber} {Security} {Consciousness}: {The} {Ease} and {Danger} of {OSINT} {Tools} in {Exposing} {Critical} {Infrastructure} {Vulnerabilities}}. In \bibinfo{booktitle}{\emph{2023 8th {International} {Conference} on {Computer} {Science} and {Engineering} ({UBMK})}}. \bibinfo{pages}{438--443}.
\newblock
\urldef\tempurl%
\url{https://doi.org/10.1109/UBMK59864.2023.10286573}
\showDOI{\tempurl}
\newblock
\shownote{ISSN: 2521-1641}.


\bibitem[Porsdam~Mann et~al\mbox{.}(2023)]%
        {porsdam_mann_generative_2023}
\bibfield{author}{\bibinfo{person}{Sebastian Porsdam~Mann}, \bibinfo{person}{Brian~D. Earp}, \bibinfo{person}{Sven Nyholm}, \bibinfo{person}{John Danaher}, \bibinfo{person}{Nikolaj Møller}, \bibinfo{person}{Hilary Bowman-Smart}, \bibinfo{person}{Joshua Hatherley}, \bibinfo{person}{Julian Koplin}, \bibinfo{person}{Monika Plozza}, \bibinfo{person}{Daniel Rodger}, \bibinfo{person}{Peter~V. Treit}, \bibinfo{person}{Gregory Renard}, \bibinfo{person}{John McMillan}, {and} \bibinfo{person}{Julian Savulescu}.} \bibinfo{year}{2023}\natexlab{}.
\newblock \showarticletitle{Generative {AI} entails a credit–blame asymmetry}.
\newblock \bibinfo{journal}{\emph{Nature Machine Intelligence}} \bibinfo{volume}{5}, \bibinfo{number}{5} (\bibinfo{date}{May} \bibinfo{year}{2023}), \bibinfo{pages}{472--475}.
\newblock
\showISSN{2522-5839}
\urldef\tempurl%
\url{https://doi.org/10.1038/s42256-023-00653-1}
\showDOI{\tempurl}
\newblock
\shownote{Publisher: Nature Publishing Group}.


\bibitem[Poulter(2023)]%
        {poulter_osint_2023}
\bibfield{author}{\bibinfo{person}{Chris Poulter}.} \bibinfo{year}{2023}\natexlab{}.
\newblock \bibinfo{title}{{OSINT} {Workflow} with {ChatGPT}: {Tips}, {Risks}, and {Benefits}}.
\newblock
\newblock
\urldef\tempurl%
\url{https://www.osintcombine.com/post/osint-workflow-with-chatgpt-tips-risks-and-benefits}
\showURL{%
\tempurl}


\bibitem[Retelny et~al\mbox{.}(2017)]%
        {retelny2017noworkflow}
\bibfield{author}{\bibinfo{person}{Daniela Retelny}, \bibinfo{person}{Michael~S. Bernstein}, {and} \bibinfo{person}{Melissa~A. Valentine}.} \bibinfo{year}{2017}\natexlab{}.
\newblock \showarticletitle{No Workflow Can Ever Be Enough: How Crowdsourcing Workflows Constrain Complex Work}.
\newblock \bibinfo{journal}{\emph{Proc. ACM Hum.-Comput. Interact.}} \bibinfo{volume}{1}, \bibinfo{number}{CSCW}, Article \bibinfo{articleno}{89} (\bibinfo{date}{Dec.} \bibinfo{year}{2017}), \bibinfo{numpages}{23}~pages.
\newblock
\urldef\tempurl%
\url{https://doi.org/10.1145/3134724}
\showDOI{\tempurl}


\bibitem[Riebe et~al\mbox{.}(2023)]%
        {riebe_values_2023}
\bibfield{author}{\bibinfo{person}{Thea Riebe}, \bibinfo{person}{Julian Bäumler}, \bibinfo{person}{Marc-André Kaufhold}, {and} \bibinfo{person}{Christian Reuter}.} \bibinfo{year}{2023}\natexlab{}.
\newblock \showarticletitle{Values and {Value} {Conflicts} in the {Context} of {OSINT} {Technologies} for {Cybersecurity} {Incident} {Response}: {A} {Value} {Sensitive} {Design} {Perspective}}.
\newblock \bibinfo{journal}{\emph{Computer Supported Cooperative Work (CSCW)}} (\bibinfo{date}{April} \bibinfo{year}{2023}).
\newblock
\showISSN{1573-7551}
\urldef\tempurl%
\url{https://doi.org/10.1007/s10606-022-09453-4}
\showDOI{\tempurl}


\bibitem[Sanders and Stappers(2008)]%
        {Sanders2008CocreationAT}
\bibfield{author}{\bibinfo{person}{Elizabeth B.-N. Sanders} {and} \bibinfo{person}{Pieter~Jan Stappers}.} \bibinfo{year}{2008}\natexlab{}.
\newblock \showarticletitle{Co-creation and the new landscapes of design}.
\newblock \bibinfo{journal}{\emph{CoDesign}}  \bibinfo{volume}{4} (\bibinfo{year}{2008}), \bibinfo{pages}{18 -- 5}.
\newblock
\urldef\tempurl%
\url{https://api.semanticscholar.org/CorpusID:111148589}
\showURL{%
\tempurl}


\bibitem[Sarkar et~al\mbox{.}(2023)]%
        {sarkar2023participatory}
\bibfield{author}{\bibinfo{person}{Advait Sarkar}, \bibinfo{person}{Ian Drosos}, \bibinfo{person}{Rob Deline}, \bibinfo{person}{Andrew~D Gordon}, \bibinfo{person}{Carina Negreanu}, \bibinfo{person}{Sean Rintel}, \bibinfo{person}{Jack Williams}, {and} \bibinfo{person}{Benjamin Zorn}.} \bibinfo{year}{2023}\natexlab{}.
\newblock \showarticletitle{Participatory prompting: a user-centric research method for eliciting AI assistance opportunities in knowledge workflows}.
\newblock \bibinfo{journal}{\emph{arXiv preprint arXiv:2312.16633}} (\bibinfo{year}{2023}).
\newblock


\bibitem[Shneiderman(2022)]%
        {shneiderman_ensuring_nodate}
\bibfield{author}{\bibinfo{person}{Ben Shneiderman}.} \bibinfo{year}{2022}\natexlab{}.
\newblock \bibinfo{title}{Ensuring {Human} {Control} over {AI}-{Infused} {Systems} {\textbar} {National} {Academies}}.
\newblock
\newblock
\urldef\tempurl%
\url{https://www.nationalacademies.org/news/2022/04/ensuring-human-control-over-ai-infused-systems}
\showURL{%
\tempurl}


\bibitem[Stahl(2005)]%
        {stahl2005groups}
\bibfield{author}{\bibinfo{person}{Gerry Stahl}.} \bibinfo{year}{2005}\natexlab{}.
\newblock \showarticletitle{Groups, group cognition and groupware}. In \bibinfo{booktitle}{\emph{International Conference on Collaboration and Technology}}. Springer, \bibinfo{pages}{1--16}.
\newblock


\bibitem[Subramonyam et~al\mbox{.}(2023)]%
        {subramonyam_bridging_2023}
\bibfield{author}{\bibinfo{person}{Hariharan Subramonyam}, \bibinfo{person}{Christopher~Lawrence Pondoc}, \bibinfo{person}{Colleen Seifert}, \bibinfo{person}{Maneesh Agrawala}, {and} \bibinfo{person}{Roy Pea}.} \bibinfo{year}{2023}\natexlab{}.
\newblock \bibinfo{title}{Bridging the {Gulf} of {Envisioning}: {Cognitive} {Design} {Challenges} in {LLM} {Interfaces}}.
\newblock
\newblock
\urldef\tempurl%
\url{http://arxiv.org/abs/2309.14459}
\showURL{%
\tempurl}
\newblock
\shownote{arXiv:2309.14459 [cs]}.


\bibitem[Suh et~al\mbox{.}(2021)]%
        {suh_ai_2021}
\bibfield{author}{\bibinfo{person}{Minhyang~(Mia) Suh}, \bibinfo{person}{Emily Youngblom}, \bibinfo{person}{Michael Terry}, {and} \bibinfo{person}{Carrie~J Cai}.} \bibinfo{year}{2021}\natexlab{}.
\newblock \showarticletitle{{AI} as {Social} {Glue}: {Uncovering} the {Roles} of {Deep} {Generative} {AI} during {Social} {Music} {Composition}}. In \bibinfo{booktitle}{\emph{Proceedings of the 2021 {CHI} {Conference} on {Human} {Factors} in {Computing} {Systems}}} \emph{(\bibinfo{series}{{CHI} '21})}. \bibinfo{publisher}{Association for Computing Machinery}, \bibinfo{address}{New York, NY, USA}, \bibinfo{pages}{1--11}.
\newblock
\showISBNx{978-1-4503-8096-6}
\urldef\tempurl%
\url{https://doi.org/10.1145/3411764.3445219}
\showDOI{\tempurl}


\bibitem[Sundaramurthy et~al\mbox{.}(2014)]%
        {sundaramurthy_anthropological_2014}
\bibfield{author}{\bibinfo{person}{Sathya~Chandran Sundaramurthy}, \bibinfo{person}{John McHugh}, \bibinfo{person}{Xinming~Simon Ou}, \bibinfo{person}{S.~Raj Rajagopalan}, {and} \bibinfo{person}{Michael Wesch}.} \bibinfo{year}{2014}\natexlab{}.
\newblock \showarticletitle{An {Anthropological} {Approach} to {Studying} {CSIRTs}}.
\newblock \bibinfo{journal}{\emph{IEEE Security \& Privacy}} \bibinfo{volume}{12}, \bibinfo{number}{5} (\bibinfo{date}{Sept.} \bibinfo{year}{2014}), \bibinfo{pages}{52--60}.
\newblock
\showISSN{1540-7993, 1558-4046}
\urldef\tempurl%
\url{https://doi.org/10.1109/MSP.2014.84}
\showDOI{\tempurl}


\bibitem[Suri et~al\mbox{.}(2024)]%
        {suri_use_2024}
\bibfield{author}{\bibinfo{person}{Siddharth Suri}, \bibinfo{person}{Scott Counts}, \bibinfo{person}{Leijie Wang}, \bibinfo{person}{Chacha Chen}, \bibinfo{person}{Mengting Wan}, \bibinfo{person}{Tara Safavi}, \bibinfo{person}{Jennifer Neville}, \bibinfo{person}{Chirag Shah}, \bibinfo{person}{Ryen~W. White}, \bibinfo{person}{Reid Andersen}, \bibinfo{person}{Georg Buscher}, \bibinfo{person}{Sathish Manivannan}, \bibinfo{person}{Nagu Rangan}, {and} \bibinfo{person}{Longqi Yang}.} \bibinfo{year}{2024}\natexlab{}.
\newblock \bibinfo{title}{The {Use} of {Generative} {Search} {Engines} for {Knowledge} {Work} and {Complex} {Tasks}}.
\newblock
\newblock
\urldef\tempurl%
\url{https://doi.org/10.48550/arXiv.2404.04268}
\showDOI{\tempurl}
\newblock
\shownote{arXiv:2404.04268 [cs]}.


\bibitem[Szymoniak and Foks(2024)]%
        {szymoniak_open_2024}
\bibfield{author}{\bibinfo{person}{Sabina Szymoniak} {and} \bibinfo{person}{Kacper Foks}.} \bibinfo{year}{2024}\natexlab{}.
\newblock \showarticletitle{Open {Source} {Intelligence} {Opportunities} and {Challenges}: a {Review}}.
\newblock \bibinfo{journal}{\emph{Advances in Science and Technology Research Journal}} \bibinfo{volume}{18}, \bibinfo{number}{3} (\bibinfo{date}{June} \bibinfo{year}{2024}), \bibinfo{pages}{123--139}.
\newblock
\showISSN{2080-4075, 2299-8624}
\urldef\tempurl%
\url{https://doi.org/10.12913/22998624/186036}
\showDOI{\tempurl}


\bibitem[Tabatabaei and Wells(2017)]%
        {tabatabaei2017osint}
\bibfield{author}{\bibinfo{person}{Fahimeh Tabatabaei} {and} \bibinfo{person}{Douglas Wells}.} \bibinfo{year}{2017}\natexlab{}.
\newblock \showarticletitle{OSINT in the Context of Cyber-Security}.
\newblock \bibinfo{journal}{\emph{Open Source Intelligence Investigation: From Strategy to Implementation}} (\bibinfo{year}{2017}), \bibinfo{pages}{213--231}.
\newblock


\bibitem[Tankelevitch et~al\mbox{.}(2024)]%
        {tankelevitch_metacognitive_2024}
\bibfield{author}{\bibinfo{person}{Lev Tankelevitch}, \bibinfo{person}{Viktor Kewenig}, \bibinfo{person}{Auste Simkute}, \bibinfo{person}{Ava~Elizabeth Scott}, \bibinfo{person}{Advait Sarkar}, \bibinfo{person}{Abigail Sellen}, {and} \bibinfo{person}{Sean Rintel}.} \bibinfo{year}{2024}\natexlab{}.
\newblock \showarticletitle{The {Metacognitive} {Demands} and {Opportunities} of {Generative} {AI}}. In \bibinfo{booktitle}{\emph{Proceedings of the {CHI} {Conference} on {Human} {Factors} in {Computing} {Systems}}} \emph{(\bibinfo{series}{{CHI} '24})}. \bibinfo{publisher}{Association for Computing Machinery}, \bibinfo{address}{New York, NY, USA}, \bibinfo{pages}{1--24}.
\newblock
\showISBNx{9798400703300}
\urldef\tempurl%
\url{https://doi.org/10.1145/3613904.3642902}
\showDOI{\tempurl}


\bibitem[Team-GPT(2024)]%
        {team-gpt_enterprise_nodate}
\bibfield{author}{\bibinfo{person}{Team-GPT}.} \bibinfo{year}{2024}\natexlab{}.
\newblock \bibinfo{title}{Enterprise {AI} software for teams between 2 and 5,000}.
\newblock
\newblock
\urldef\tempurl%
\url{https://team-gpt.com/}
\showURL{%
Retrieved 2024-09-08 from \tempurl}


\bibitem[Tholander and Jonsson(2023)]%
        {tholander_design_2023}
\bibfield{author}{\bibinfo{person}{Jakob Tholander} {and} \bibinfo{person}{Martin Jonsson}.} \bibinfo{year}{2023}\natexlab{}.
\newblock \showarticletitle{Design {Ideation} with {AI} - {Sketching}, {Thinking} and {Talking} with {Generative} {Machine} {Learning} {Models}}. In \bibinfo{booktitle}{\emph{Proceedings of the 2023 {ACM} {Designing} {Interactive} {Systems} {Conference}}} \emph{(\bibinfo{series}{{DIS} '23})}. \bibinfo{publisher}{Association for Computing Machinery}, \bibinfo{address}{New York, NY, USA}, \bibinfo{pages}{1930--1940}.
\newblock
\showISBNx{978-1-4503-9893-0}
\urldef\tempurl%
\url{https://doi.org/10.1145/3563657.3596014}
\showDOI{\tempurl}


\bibitem[Tseng et~al\mbox{.}(2024)]%
        {tseng2024data}
\bibfield{author}{\bibinfo{person}{Emily Tseng}, \bibinfo{person}{Rosanna Bellini}, \bibinfo{person}{Yeuk-Yu Lee}, \bibinfo{person}{Alana Ramjit}, \bibinfo{person}{Thomas Ristenpart}, {and} \bibinfo{person}{Nicola Dell}.} \bibinfo{year}{2024}\natexlab{}.
\newblock \showarticletitle{Data Stewardship in Clinical Computer Security: Balancing Benefit and Burden in Participatory Systems}.
\newblock \bibinfo{journal}{\emph{Proceedings of the ACM on Human-Computer Interaction}} \bibinfo{volume}{8}, \bibinfo{number}{CSCW1} (\bibinfo{year}{2024}), \bibinfo{pages}{1--29}.
\newblock


\bibitem[Tseng et~al\mbox{.}(2022)]%
        {tseng2022care}
\bibfield{author}{\bibinfo{person}{Emily Tseng}, \bibinfo{person}{Mehrnaz Sabet}, \bibinfo{person}{Rosanna Bellini}, \bibinfo{person}{Harkiran~Kaur Sodhi}, \bibinfo{person}{Thomas Ristenpart}, {and} \bibinfo{person}{Nicola Dell}.} \bibinfo{year}{2022}\natexlab{}.
\newblock \showarticletitle{Care infrastructures for digital security in intimate partner violence}. In \bibinfo{booktitle}{\emph{Proceedings of the 2022 CHI Conference on Human Factors in Computing Systems}}. \bibinfo{pages}{1--20}.
\newblock


\bibitem[Tundis et~al\mbox{.}(2022)]%
        {tundis2022feature}
\bibfield{author}{\bibinfo{person}{Andrea Tundis}, \bibinfo{person}{Samuel Ruppert}, {and} \bibinfo{person}{Max M{\"u}hlh{\"a}user}.} \bibinfo{year}{2022}\natexlab{}.
\newblock \showarticletitle{A feature-driven method for automating the assessment of osint cyber threat sources}.
\newblock \bibinfo{journal}{\emph{Computers \& Security}}  \bibinfo{volume}{113} (\bibinfo{year}{2022}), \bibinfo{pages}{102576}.
\newblock


\bibitem[Urban et~al\mbox{.}(2020)]%
        {urban_plenty_2020}
\bibfield{author}{\bibinfo{person}{Tobias Urban}, \bibinfo{person}{Matteo Große-Kampmann}, \bibinfo{person}{Dennis Tatang}, \bibinfo{person}{Thorsten Holz}, {and} \bibinfo{person}{Norbert Pohlmann}.} \bibinfo{year}{2020}\natexlab{}.
\newblock \showarticletitle{Plenty of {Phish} in the {Sea}: {Analyzing} {Potential} {Pre}-attack {Surfaces}}. In \bibinfo{booktitle}{\emph{Computer {Security} – {ESORICS} 2020}} \emph{(\bibinfo{series}{Lecture {Notes} in {Computer} {Science}})}, \bibfield{editor}{\bibinfo{person}{Liqun Chen}, \bibinfo{person}{Ninghui Li}, \bibinfo{person}{Kaitai Liang}, {and} \bibinfo{person}{Steve Schneider}} (Eds.). \bibinfo{publisher}{Springer International Publishing}, \bibinfo{address}{Cham}, \bibinfo{pages}{272--291}.
\newblock
\showISBNx{978-3-030-59013-0}
\urldef\tempurl%
\url{https://doi.org/10.1007/978-3-030-59013-0_14}
\showDOI{\tempurl}


\bibitem[van Dijk and Zimmerman(2019)]%
        {Dijk2019DiscoveringUF}
\bibfield{author}{\bibinfo{person}{Bart van Dijk} {and} \bibinfo{person}{John Zimmerman}.} \bibinfo{year}{2019}\natexlab{}.
\newblock \showarticletitle{Discovering Users for Technical Innovations through Systematic Matchmaking}.
\newblock \bibinfo{journal}{\emph{Extended Abstracts of the 2019 CHI Conference on Human Factors in Computing Systems}} (\bibinfo{year}{2019}).
\newblock
\urldef\tempurl%
\url{https://api.semanticscholar.org/CorpusID:144207663}
\showURL{%
\tempurl}


\bibitem[Venkatagiri et~al\mbox{.}(2021)]%
        {venkatagiri2021crowdsolve}
\bibfield{author}{\bibinfo{person}{Sukrit Venkatagiri}, \bibinfo{person}{Aakash Gautam}, {and} \bibinfo{person}{Kurt Luther}.} \bibinfo{year}{2021}\natexlab{}.
\newblock \showarticletitle{CrowdSolve: Managing Tensions in an Expert-Led Crowdsourced Investigation}.
\newblock \bibinfo{journal}{\emph{Proc. ACM Hum.-Comput. Interact.}} \bibinfo{volume}{5}, \bibinfo{number}{CSCW1}, Article \bibinfo{articleno}{118} (\bibinfo{date}{April} \bibinfo{year}{2021}), \bibinfo{numpages}{30}~pages.
\newblock
\urldef\tempurl%
\url{https://doi.org/10.1145/3449192}
\showDOI{\tempurl}


\bibitem[Venkatagiri et~al\mbox{.}(2023)]%
        {venkatagiri2023cosint}
\bibfield{author}{\bibinfo{person}{Sukrit Venkatagiri}, \bibinfo{person}{Anirban Mukhopadhyay}, \bibinfo{person}{David Hicks}, \bibinfo{person}{Aaron Brantly}, {and} \bibinfo{person}{Kurt Luther}.} \bibinfo{year}{2023}\natexlab{}.
\newblock \showarticletitle{CoSINT: Designing a Collaborative Capture the Flag Competition to Investigate Misinformation}. In \bibinfo{booktitle}{\emph{Proceedings of the 2023 ACM Designing Interactive Systems Conference}} (Pittsburgh, PA, USA) \emph{(\bibinfo{series}{DIS '23})}. \bibinfo{publisher}{Association for Computing Machinery}, \bibinfo{address}{New York, NY, USA}, \bibinfo{pages}{2551–2572}.
\newblock
\showISBNx{9781450398930}
\urldef\tempurl%
\url{https://doi.org/10.1145/3563657.3595997}
\showDOI{\tempurl}


\bibitem[Venkatagiri et~al\mbox{.}(2019)]%
        {venkatagiri2019groundtruth}
\bibfield{author}{\bibinfo{person}{Sukrit Venkatagiri}, \bibinfo{person}{Jacob Thebault-Spieker}, \bibinfo{person}{Rachel Kohler}, \bibinfo{person}{John Purviance}, \bibinfo{person}{Rifat~Sabbir Mansur}, {and} \bibinfo{person}{Kurt Luther}.} \bibinfo{year}{2019}\natexlab{}.
\newblock \showarticletitle{GroundTruth: Augmenting Expert Image Geolocation with Crowdsourcing and Shared Representations}.
\newblock \bibinfo{journal}{\emph{Proc. ACM Hum.-Comput. Interact.}} \bibinfo{volume}{3}, \bibinfo{number}{CSCW}, Article \bibinfo{articleno}{107} (\bibinfo{date}{Nov.} \bibinfo{year}{2019}), \bibinfo{numpages}{30}~pages.
\newblock
\urldef\tempurl%
\url{https://doi.org/10.1145/3359209}
\showDOI{\tempurl}


\bibitem[Visser et~al\mbox{.}(2005)]%
        {Visser2005ContextmappingEF}
\bibfield{author}{\bibinfo{person}{Froukje~Sleeswijk Visser}, \bibinfo{person}{Pieter~Jan Stappers}, \bibinfo{person}{Remko van~der Lugt}, {and} \bibinfo{person}{Elizabeth B.-N. Sanders}.} \bibinfo{year}{2005}\natexlab{}.
\newblock \showarticletitle{Contextmapping: experiences from practice}.
\newblock \bibinfo{journal}{\emph{CoDesign}}  \bibinfo{volume}{1} (\bibinfo{year}{2005}), \bibinfo{pages}{119 -- 149}.
\newblock
\urldef\tempurl%
\url{https://api.semanticscholar.org/CorpusID:279110}
\showURL{%
\tempurl}


\bibitem[Wan et~al\mbox{.}(2024)]%
        {wan_it_2024}
\bibfield{author}{\bibinfo{person}{Qian Wan}, \bibinfo{person}{Siying Hu}, \bibinfo{person}{Yu Zhang}, \bibinfo{person}{Piaohong Wang}, \bibinfo{person}{Bo Wen}, {and} \bibinfo{person}{Zhicong Lu}.} \bibinfo{year}{2024}\natexlab{}.
\newblock \showarticletitle{"{It} {Felt} {Like} {Having} a {Second} {Mind}": {Investigating} {Human}-{AI} {Co}-creativity in {Prewriting} with {Large} {Language} {Models}}.
\newblock \bibinfo{journal}{\emph{Proceedings of the ACM on Human-Computer Interaction}} \bibinfo{volume}{8}, \bibinfo{number}{CSCW1} (\bibinfo{date}{April} \bibinfo{year}{2024}), \bibinfo{pages}{1--26}.
\newblock
\showISSN{2573-0142}
\urldef\tempurl%
\url{https://doi.org/10.1145/3637361}
\showDOI{\tempurl}
\newblock
\shownote{arXiv:2307.10811 [cs]}.


\bibitem[Wang et~al\mbox{.}(2021)]%
        {wang_towards_2021}
\bibfield{author}{\bibinfo{person}{Qiaosi Wang}, \bibinfo{person}{Koustuv Saha}, \bibinfo{person}{Eric Gregori}, \bibinfo{person}{David Joyner}, {and} \bibinfo{person}{Ashok Goel}.} \bibinfo{year}{2021}\natexlab{}.
\newblock \showarticletitle{Towards {Mutual} {Theory} of {Mind} in {Human}-{AI} {Interaction}: {How} {Language} {Reflects} {What} {Students} {Perceive} {About} a {Virtual} {Teaching} {Assistant}}. In \bibinfo{booktitle}{\emph{Proceedings of the 2021 {CHI} {Conference} on {Human} {Factors} in {Computing} {Systems}}} \emph{(\bibinfo{series}{{CHI} '21})}. \bibinfo{publisher}{Association for Computing Machinery}, \bibinfo{address}{New York, NY, USA}, \bibinfo{pages}{1--14}.
\newblock
\showISBNx{978-1-4503-8096-6}
\urldef\tempurl%
\url{https://doi.org/10.1145/3411764.3445645}
\showDOI{\tempurl}


\bibitem[Weisz et~al\mbox{.}(2024)]%
        {weisz_design_2024}
\bibfield{author}{\bibinfo{person}{Justin~D. Weisz}, \bibinfo{person}{Jessica He}, \bibinfo{person}{Michael Muller}, \bibinfo{person}{Gabriela Hoefer}, \bibinfo{person}{Rachel Miles}, {and} \bibinfo{person}{Werner Geyer}.} \bibinfo{year}{2024}\natexlab{}.
\newblock \showarticletitle{Design {Principles} for {Generative} {AI} {Applications}}. In \bibinfo{booktitle}{\emph{Proceedings of the {CHI} {Conference} on {Human} {Factors} in {Computing} {Systems}}} \emph{(\bibinfo{series}{{CHI} '24})}. \bibinfo{publisher}{Association for Computing Machinery}, \bibinfo{address}{New York, NY, USA}, \bibinfo{pages}{1--22}.
\newblock
\showISBNx{9798400703300}
\urldef\tempurl%
\url{https://doi.org/10.1145/3613904.3642466}
\showDOI{\tempurl}


\bibitem[Williams and Blum(2018)]%
        {williams_defining_2018}
\bibfield{author}{\bibinfo{person}{Heather~J. Williams} {and} \bibinfo{person}{Ilana Blum}.} \bibinfo{year}{2018}\natexlab{}.
\newblock \bibinfo{booktitle}{\emph{Defining {Second} {Generation} {Open} {Source} {Intelligence} ({OSINT}) for the {Defense} {Enterprise}}}.
\newblock \bibinfo{type}{{T}echnical {R}eport}. \bibinfo{institution}{RAND Corporation}.
\newblock
\urldef\tempurl%
\url{https://www.rand.org/pubs/research_reports/RR1964.html}
\showURL{%
\tempurl}


\bibitem[Williams et~al\mbox{.}(2016)]%
        {williams2016axis}
\bibfield{author}{\bibinfo{person}{Joseph~Jay Williams}, \bibinfo{person}{Juho Kim}, \bibinfo{person}{Anna Rafferty}, \bibinfo{person}{Samuel Maldonado}, \bibinfo{person}{Krzysztof~Z Gajos}, \bibinfo{person}{Walter~S Lasecki}, {and} \bibinfo{person}{Neil Heffernan}.} \bibinfo{year}{2016}\natexlab{}.
\newblock \showarticletitle{Axis: Generating explanations at scale with learnersourcing and machine learning}. In \bibinfo{booktitle}{\emph{Proceedings of the Third (2016) ACM Conference on Learning@ Scale}}. \bibinfo{pages}{379--388}.
\newblock


\bibitem[Wong and Kodagoda(2015)]%
        {wong_how_2015}
\bibfield{author}{\bibinfo{person}{B.L.~William Wong} {and} \bibinfo{person}{Neesha Kodagoda}.} \bibinfo{year}{2015}\natexlab{}.
\newblock \showarticletitle{How {Analysts} {Think}: {Inference} {Making} {Strategies}}.
\newblock \bibinfo{journal}{\emph{Proceedings of the Human Factors and Ergonomics Society Annual Meeting}} \bibinfo{volume}{59}, \bibinfo{number}{1} (\bibinfo{date}{Sept.} \bibinfo{year}{2015}), \bibinfo{pages}{269--273}.
\newblock
\showISSN{1071-1813}
\urldef\tempurl%
\url{https://doi.org/10.1177/1541931215591055}
\showDOI{\tempurl}
\newblock
\shownote{Publisher: SAGE Publications Inc}.


\bibitem[Xu et~al\mbox{.}(2015)]%
        {xu2015classroom}
\bibfield{author}{\bibinfo{person}{Anbang Xu}, \bibinfo{person}{Huaming Rao}, \bibinfo{person}{Steven~P Dow}, {and} \bibinfo{person}{Brian~P Bailey}.} \bibinfo{year}{2015}\natexlab{}.
\newblock \showarticletitle{A classroom study of using crowd feedback in the iterative design process}. In \bibinfo{booktitle}{\emph{Proceedings of the 18th ACM conference on computer supported cooperative work \& social computing}}. \bibinfo{pages}{1637--1648}.
\newblock


\bibitem[Yadav et~al\mbox{.}(2023)]%
        {yadav2023open}
\bibfield{author}{\bibinfo{person}{Ashok Yadav}, \bibinfo{person}{Atul Kumar}, {and} \bibinfo{person}{Vrijendra Singh}.} \bibinfo{year}{2023}\natexlab{}.
\newblock \showarticletitle{Open-source intelligence: a comprehensive review of the current state, applications and future perspectives in cyber security}.
\newblock \bibinfo{journal}{\emph{Artificial Intelligence Review}} \bibinfo{volume}{56}, \bibinfo{number}{11} (\bibinfo{year}{2023}), \bibinfo{pages}{12407--12438}.
\newblock


\bibitem[Yin et~al\mbox{.}(2024)]%
        {yin_can_2024}
\bibfield{author}{\bibinfo{person}{Meng Yin}, \bibinfo{person}{Shiyao Jiang}, {and} \bibinfo{person}{Xiongying Niu}.} \bibinfo{year}{2024}\natexlab{}.
\newblock \showarticletitle{Can {AI} really help? {The} double-edged sword effect of {AI} assistant on employees’ innovation behavior}.
\newblock \bibinfo{journal}{\emph{Computers in Human Behavior}}  \bibinfo{volume}{150} (\bibinfo{date}{Jan.} \bibinfo{year}{2024}), \bibinfo{pages}{107987}.
\newblock
\showISSN{0747-5632}
\urldef\tempurl%
\url{https://doi.org/10.1016/j.chb.2023.107987}
\showDOI{\tempurl}


\bibitem[Zamfirescu-Pereira et~al\mbox{.}(2023)]%
        {zamfirescu-pereira_why_2023}
\bibfield{author}{\bibinfo{person}{J.D. Zamfirescu-Pereira}, \bibinfo{person}{Richmond Wong}, \bibinfo{person}{Bjoern Hartmann}, {and} \bibinfo{person}{Qian Yang}.} \bibinfo{year}{2023}\natexlab{}.
\newblock \showarticletitle{Why {Johnny} {Can}'t {Prompt}: {How} {Non}-{AI} {Experts} {Try} (and {Fail}) to {Design} {LLM} {Prompts}}.
\newblock
\urldef\tempurl%
\url{https://doi.org/10.1145/3544548.3581388}
\showDOI{\tempurl}


\bibitem[Zhang et~al\mbox{.}(2023)]%
        {zhang2023investigating}
\bibfield{author}{\bibinfo{person}{Rui Zhang}, \bibinfo{person}{Wen Duan}, \bibinfo{person}{Christopher Flathmann}, \bibinfo{person}{Nathan McNeese}, \bibinfo{person}{Guo Freeman}, {and} \bibinfo{person}{Alyssa Williams}.} \bibinfo{year}{2023}\natexlab{}.
\newblock \showarticletitle{Investigating AI teammate communication strategies and their impact in human-AI teams for effective teamwork}.
\newblock \bibinfo{journal}{\emph{Proceedings of the ACM on Human-Computer Interaction}} \bibinfo{volume}{7}, \bibinfo{number}{CSCW2} (\bibinfo{year}{2023}), \bibinfo{pages}{1--31}.
\newblock


\end{thebibliography}

\appendix
\section{Appendix}
\label{appendix}
\subsection{Prompt templates for OSINT investigsations for cybersecurity vulnerability assessment}
\label{appendix_prompts}
\begin{table*}[]
\caption{Curated prompt templates for Planning in Team-GPT. Users are prompted to fill in \{\{\}\} placeholders before running the prompts}
\label{tab:prompts-table1}
\begin{tabular}{@{}p{0.07\linewidth}|p{0.19\linewidth}|p{0.64\linewidth}@{}}
\hline
\textbf{Phase}           & \textbf{Prompt name and purpose}                                                                                                                                                                                                                                                                                                                                                              & \textbf{Prompt details}                                                                                                                                                                                                                                                                                                                                                                                                                                                                                                                                                                                                                                                                                                                                                                                                                                                                                                                                                                                                                                                                                                                                                                                                                                                                                                                                                                                                                                                                                                                                                                                                                                                                                                                                                                                                                                                                                                                                                                                                                                                                                                                                                                                                                                                                                                      \\ \hline
Planning        & Division of work: Takes the team size and divides up tasks to perform network discovery, website vulnerabilities, physical assets, third-party services, employee public information, brand monitoring, and breach data in two weeks as a team.                                                                                                                                      & Create a planning template for an OSINT cybersecurity investigation involving 7 services: network discovery, website vulnerabilities, physical assets, third-party services, employee public information, brand monitoring, and breach data. Description of tasks are as follows: Network Discovery - Identify all digital assets (websites, email, servers, webcams) associated with the business. Website Vulnerabilities - Discover vulnerabilities on the business’s website such as outdated software, insecure admin panels / logins, files that should not be public, etc. Physical Assets - Identify physical assets like buildings, properties, or vehicles from online resources or images. Third Party Services - Document third party services (vendors, etc.) that the client does business with to identify possible supply chain vulnerabilities. Employee Public Information - Identify key employees from a company using social media platforms, gathering information like job titles, email addresses, or sensitive personal data accidentally shared online. Brand Monitoring - Find instances of negative mentions, disinformation, inappropriate content, legal filings and settlements related to the business. Breach Data - Search for a company’s leaked passwords and data in breach data dumps and on the dark web. My team of \{\{team\_size\}\} is looking to perform the 7 tasks. Divide our workflow into the most efficient subtasks between the \{\{team\_size\}\} of us. Create a timeline for two weeks for a team of \{\{team\_size\}\} to work on. The first week is for data collection and preliminary analysis to identify areas to dig deeper. The second week should focus on completing the analysis, incorporating any new information and feedback from reviewers.  At the end of the second week, the goal is to generate a set of vulnerabilities and recommendations to present to the small business. Take into account the member's expertise while dividing up the task. Member 1 has preference for the following tasks \{\{Member\_1\_preference\}\}. Member 2 has preference for the following tasks \{\{Member\_2\_preference\}\}. Member 3 has preference for the following tasks \{\{Member\_3\_preference\}\}. Create internal deadlines for the tasks. \\ \cline{2-3} 
                & Templates for tasks: Create template for data collection for a particular OSINT task/subtask                                                                                                                                                                                                                                                                                         & You are an OSINT investigator specializing in cybersecurity vulnerability assessments. I am developing a template for an Open Source Intelligence (OSINT) investigation focused on identifying cybersecurity risks within an organization. The investigation requires answering the following OSINT question: \{\{task\_description\_or\_guiding\_question\}\}. Create a comprehensive template that outlines the steps for data collection, analysis, and documentation related to these questions. The template should include sections for different data sources, how to use the data sources, methods for analyzing the information, risk assessment techniques, and protocols for reporting findings. Additionally, please suggest OSINT tools and resources that can aid in collecting and analyzing information effectively. Format the template and information so that it is within a table.                                                                                                                                                                                                                                                                                                                                                                                                                                                                                                                                                                                                                                                                                                                                                                                                                                                                                                                                                                                                                                                                                                                                                                                                                                                                                                                                                                                                              \\ \hline

\end{tabular}%
\end{table*}

\begin{table*}[]
\caption{Curated prompt templates for Data Collection in Team-GPT. Users are prompted to fill in \{\{\}\} placeholders before running the prompts}
\label{tab:prompts-table2}
\begin{tabular}{@{}p{0.07\linewidth}|p{0.19\linewidth}|p{0.64\linewidth}@{}}

\hline
\textbf{Phase}           & \textbf{Prompt name and purpose}                                                                                                                                                                                                                                                                                                                                                              & \textbf{Prompt details}                                                                                                                                                                                                                                       \\ \hline
Data Collection & OSINT Tools and Techniques: Learning about related tools by providing a known tool                                                                                                                                                                                                                                                                                                   & What are other examples of OSINT tools like \{\{tool/website\}\} ? Provide them in a list format with a description of each tool/website and how to utilize them for the \{\{OSINT\_task\}\}?                                                                                                                                                                                                                                                                                                                                                                                                                                                                                                                                                                                                                                                                                                                                                                                                                                                                                                                                                                                                                                                                                                                                                                                                                                                                                                                                                                                                                                                                                                                                                                                                                                                                                                                                                                                                                                                                                                                                                                                                                                                                                                                       \\ \cline{2-3} 
                & Process for collecting data: Given a subtask, the prompt generates a plan to collect relevant data                                                                                                                                                                                                                                                                                   & Develop a step-by-step guide for leveraging OSINT (Open Source Intelligence) techniques to gather data on \{\{OSINT\_subtask\}\}. Include methods for identifying relevant sources and analyzing the gathered data.                                                                                                                                                                                                                                                                                                                                                                                                                                                                                                                                                                                                                                                                                                                                                                                                                                                                                                                                                                                                                                                                                                                                                                                                                                                                                                                                                                                                                                                                                                                                                                                                                                                                                                                                                                                                                                                                                                                                                                                                                                                                                                 \\ \cline{2-3} 
                & Data collection templates: Each task has a detailed description based on tools and how to document that data                                                                                                                                                                                                                                                                         & Example for data breach: Develop a detailed template for analyzing and documenting a small business’s exposure in data breaches. Focus on searching public breach databases and the dark web for leaked data related to the company. The template should include sections for tool usage (e.g., Have I Been Pwned, BuiltWith, Censys), specific techniques for finding and verifying breaches, and the handling of sensitive information. Describe the process for updating the report with new data and organizing findings (including raw data descriptions and archived links) for effective communication with clients.                                                                                                                                                                                                                                                                                                                                                                                                                                                                                                                                                                                                                                                                                                                                                                                                                                                                                                                                                                                                                                                                                                                                                                                                                                                                                                                                                                                                                                                                                                                                                                                                                                                                                         \\ \hline

\end{tabular}%
\end{table*}

\begin{table*}[]
\caption{Curated prompt templates for Data Processing and Analysis in Team-GPT. Users are prompted to fill in \{\{\}\} placeholders before running the prompts}
\label{tab:prompts-table3}
\begin{tabular}{@{}p{0.07\linewidth}|p{0.19\linewidth}|p{0.64\linewidth}@{}}

\hline
\textbf{Phase}           & \textbf{Prompt name and purpose}                                                                                                                                                                                                                                                                                                                                                              & \textbf{Prompt details}                                                                                                                                                                                                                                       \\ \hline
Data Processing & Formatting raw data: Put the raw data into a table based on template                                                                                                                                                                                                                                                                                                                 & Create a table using this template \{\{template\_for\_organizing\_data\}\} and organize the information I collected \{\{raw\_data\}\}                                                                                                                                                                                                                                                                                                                                                                                                                                                                                                                                                                                                                                                                                                                                                                                                                                                                                                                                                                                                                                                                                                                                                                                                                                                                                                                                                                                                                                                                                                                                                                                                                                                                                                                                                                                                                                                                                                                                                                                                                                                                                                                                                                               \\ \cline{2-3} 
                & Finding next steps based on raw information: Enter the broad task like network discovery, website vulnerabilities, physical assets, third-party services, employee public information, brand monitoring, and breach data. Then input the raw data from different tools with a heading for what the data is about. The results describe the information and find areas to dig deeper. & You are an OSINT investigator specializing in cybersecurity vulnerability assessments. Accept redacted information for cybersecurity vulnerability assessment tasks using OSINT. The tasks include network discovery, website vulnerabilities, physical assets, third-party services, employee public information, brand monitoring, and breach data. The description of the tasks are: Network Discovery - Identify all digital assets (websites, email, servers, webcams) associated with the business. Website Vulnerabilities- Discover vulnerabilities on the business’s website such as outdated software, insecure admin panels / logins, files that should not be public, etc. Physical Assets -Identify physical assets like buildings, properties, or vehicles from online resources or images. Third Party Services -Document third party services (vendors, etc.) that the client does business with to identify possible supply chain vulnerabilities. Employee Public Information- Identify key employees from a company using social media platforms, gathering information like job titles, email addresses, or sensitive personal data accidentally shared online. Brand Monitoring - Find instances of negative mentions, disinformation, inappropriate content, legal filings and settlements related to the business. Breach Data - Search for a company’s leaked passwords and data in breach data dumps and on the dark web. The particular task is \{\{OSINT task\}\}. The raw information is \{\{raw\_information\}\}. Use raw information to find other sources of relevant OSINT information and describe data collection methods that can enhance current results for the task.  Describe the data explaining the significant aspects. What are some interesting insights that might be useful for the business client who wants to know about their cybersecurity vulnerabilities. Finally pinpoint areas to look into for future research for that task.                                                                                                                                                                                                                                                                                                                               \\ \hline
Data Analysis   & Interpret results from tools: Describe the results and identify possible vulnerabilities by providing the tool name and results from the tool                                                                                                                                                                                                                                        & Given the \{\{result\_from\_tool\}\} from tool \{\{tool\_name\}\}, can you explain the concepts and significance of each heading? What are some vulnerabilities that can be reported based on this data?                                                                                                                                                                                                                                                                                                                                                                                                                                                                                                                                                                                                                                                                                                                                                                                                                                                                                                                                                                                                                                                                                                                                                                                                                                                                                                                                                                                                                                                                                                                                                                                                                                                                                                                                                                                                                                                                                                                                                                                                                                                                                                            \\ \cline{2-3} 
                & Identifying vulnerabilities from data collected: Provide results from tools and raw data to identify possible vulnerabilities                                                                                                                                                                                                                                                        & Interpret the redacted OSINT data \{\{collected\_data\}\} to document how this information can be used to identify vulnerabilities. What vulnerabilities can be presented to a small business based on this information?                                                                                                                                                                                                                                                                                                                                                                                                                                                                                                                                                                                                                                                                                                                                                                                                                                                                                                                                                                                                                                                                                                                                                                                                                                                                                                                                                                                                                                                                                                                                                                                                                                                                                                                                                                                                                                                                                                                                                                                                                                                                                            \\ \hline

\end{tabular}%
\end{table*}

\begin{table*}[]
\caption{Curated prompt templates for Dissemination in Team-GPT. Users are prompted to fill in \{\{\}\} placeholders before running the prompts}
\label{tab:prompts-table4}
\begin{tabular}{@{}p{0.07\linewidth}|p{0.19\linewidth}|p{0.64\linewidth}@{}}

\hline
\textbf{Phase}           & \textbf{Prompt name and purpose}                                                                                                                                                                                                                                                                                                                                                              & \textbf{Prompt details}                                                                                                                                                                                                                                       \\ \hline
Dissemi- nation   & Develop recommendations from vulnerabilities: Provide the list of vulnerabilities to generate recommendations for the business                                                                                                                                                                                                                                                       & Given the list of website vulnerabilities :\{\{list\_of\_vulnerabilities\_or\_key\_findings\}\}, what are some recommendations you would make as a cybersecurity professional? Remember the client is a small business with no dedicated personnel for cybersecurity and operating on a small budget. Add a page at the end for summarized vulnerabilities and corresponding recommendations. Provide a timeline and more detailed explanations on these recommendations. Also add any other suggestions/disclaimers you would put in a document for an OSINT investigator who is giving this to a stakeholder.                                                                                                                                                                                                                                                                                                                                                                                                                                                                                                                                                                                                                                                                                                                                                                                                                                                                                                                                                                                                                                                                                                                                                                                                                                                                                                                                                                                                                                                                                                                                                                                                                                                                                                     \\ \cline{2-3} 
                & Iterating on recommendations: Simulate feedback from a small business on the recommendations                                                                                                                                                                                                                                                                                         & Act as a small business owner responding to the information I just gave them as a cybersecurity analyst they hired to look into vulnerabilities in their business. Here's a set of vulnerabilities and recommendations: \{\{vulnerabilities\_and\_recommendations\}\}. Provide feedback that can help with making the recommendations more actionable and complete.                                                                                                                                                                                                                                                                                                                                                                                                                                                                                                                                                                                                                                                                                                                                                                                                                                                                                                                                                                                                                                                                                                                                                                                                                                                                                                                                                                                                                                                                                                                                                                                                                                                                                                                                                                                                                                                                                                                                                 \\ \hline

\end{tabular}%
\end{table*}

\subsection{FigJam canvases showing categories and ideas during workshops}

\begin{figure*}[]
\includegraphics[width=15cm]{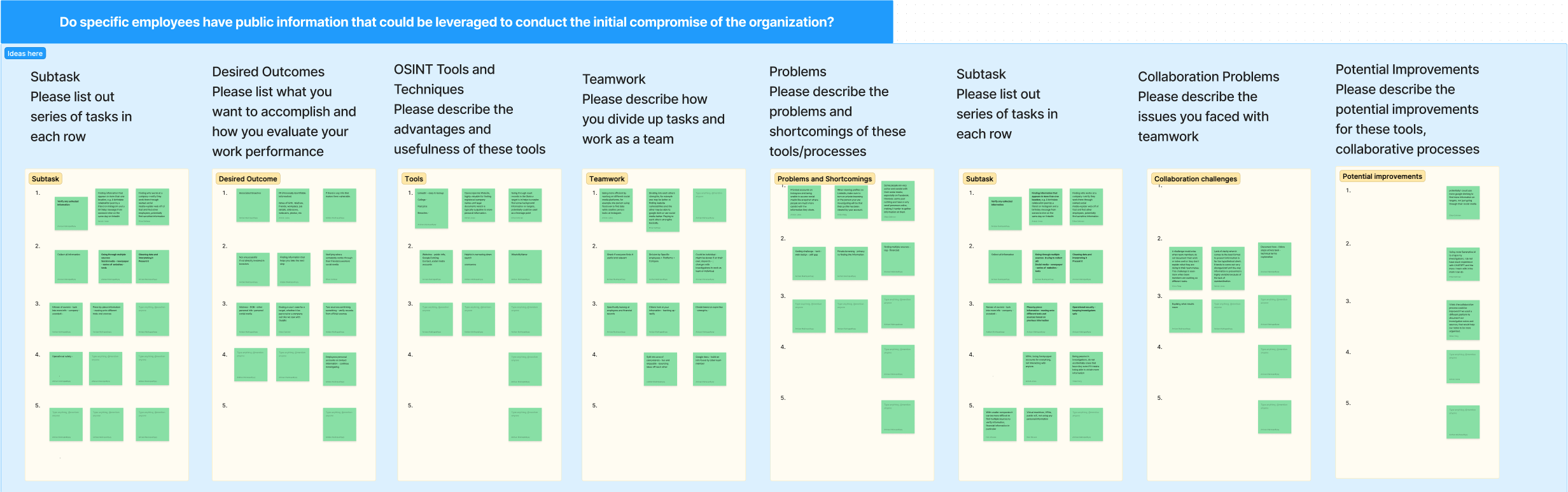}
\caption{FigJam board showing the canvas for brainstorming during Design Workshop 1}
\label{fig:figjam-study1}
\Description{The image presents a brainstorming canvas titled "Do specific employees have public information that could be leveraged to conduct the initial compromise of the organization?" This canvas contains several columns filled with sticky notes. The first column, "Subtask," lists a series of tasks required for the brainstorming process. The "Desired Outcomes" column specifies what each task aims to accomplish and the methods for evaluating work performance. The "OSINT Tools and Techniques" column discusses the advantages and usefulness of various tools employed in the tasks. The "Teamwork" column describes how tasks are divided and managed within a team. The "Problems and Shortcomings" section highlights the challenges and limitations associated with the tools and processes. There is also a column titled "Collaboration Problems," which describes the issues faced during teamwork. The final column, "Potential Improvements," outlines possible enhancements for the tools and collaborative processes. The description was generated using ChatGPT.}
\end{figure*}

\begin{figure*}[]
\includegraphics[width=15cm]{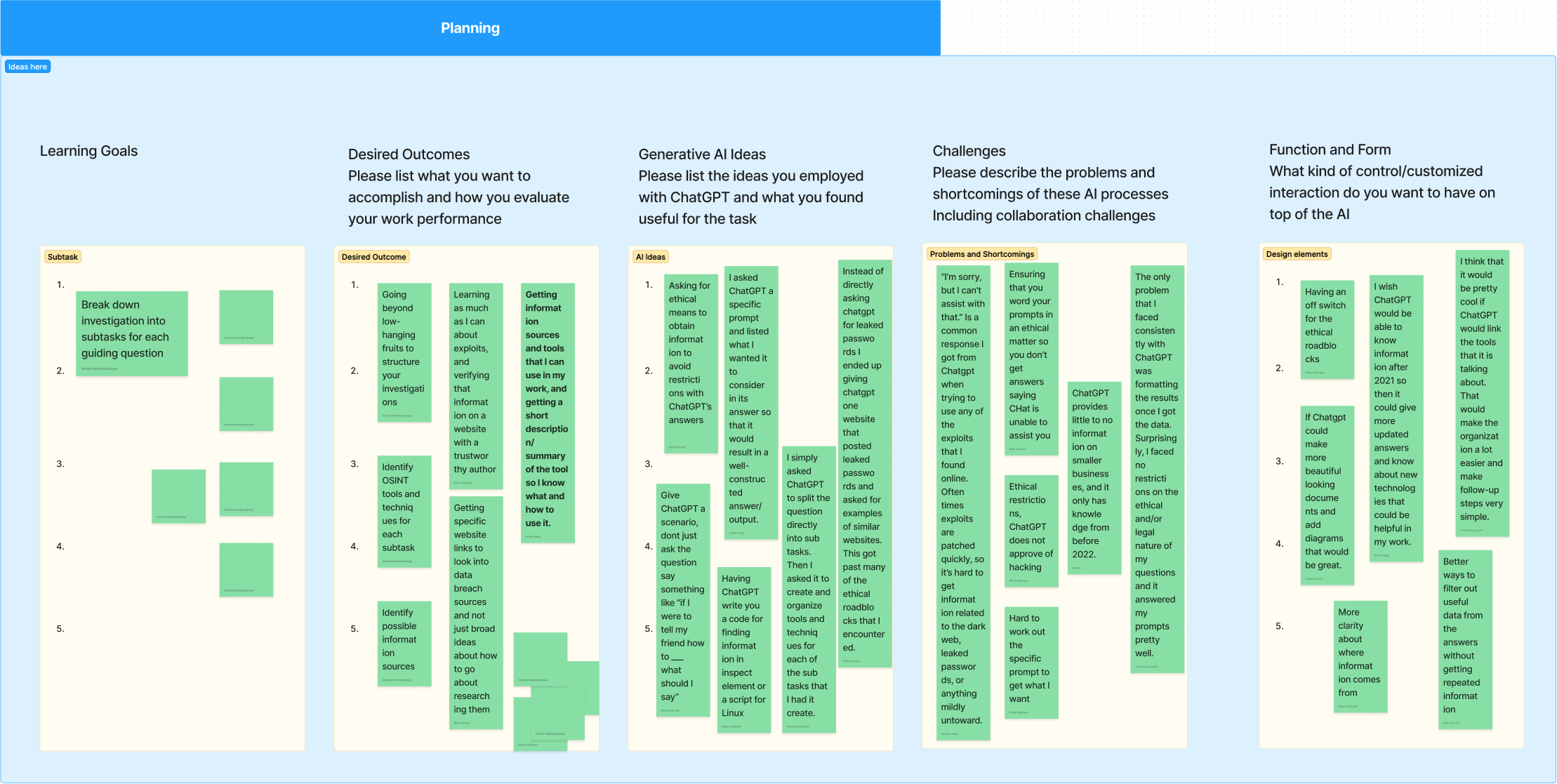}
\caption{FigJam board showing the canvas and results for Learning Goal 2 during Design Workshop 2}
\label{fig:figjam-study2}
\Description{The image is a planning canvas that lays out the framework for Study 2. The canvas begins with a section on "Learning Goals," which outlines the goals for breaking down an investigation into manageable subtasks based on guiding questions. The "Desired Outcomes" section lists specific objectives and the methods employed to evaluate performance. The "Generative AI Ideas" section details various ideas utilized with ChatGPT and assesses their usefulness for the given tasks. The "Challenges" section describes the problems and shortcomings encountered with AI processes, including issues related to collaboration. Finally, the "Function and Form" section outlines the desired levels of control and customization that users want to implement in their AI interactions. The description was generated using ChatGPT.}
\end{figure*}


\end{document}